\documentclass[twocolumn,showpacs,preprintnumbers,amsmath,amssymb, citeautoscript, superscriptaddress, prx]{revtex4}
\usepackage{rotating}
\usepackage{lscape}
\usepackage{rotating}
\usepackage{hyperref}
\usepackage{graphicx}
\usepackage{dcolumn}
\usepackage{bm}

\usepackage{ulem}
\renewcommand{\emph}{\textit}

\usepackage{color}
\definecolor{red}{rgb}{1,0,0}

\definecolor{blue}{rgb}{0,0,1}

\definecolor{green}{rgb}{0,1,0}

\newcommand{\Eq}[1]{Eq.\,(\ref{#1})}

\newcommand{\Fig}[1]{Fig.\,\ref{#1}}

\newcommand{\Ref}[1]{Ref.\mbox{\,}\cite{#1}}
\newcommand{\Refs}[1]{Refs.\mbox{\,}\cite{#1}}

\begin{document}

\preprint{APS}
\author {Mariano de Souza}
\email{mariano@rc.unesp.br; Current address: Institute of Semiconductor and Solid State Physics, Johannes Kepler University Linz, 4040 Linz, Austria}
\affiliation{Instituto de Geoci\^encias e Ci\^encias Exatas - IGCE, Unesp - Univ  Estadual Paulista, Departamento de
F\'isica, Cx.\,Postal
178, 13506-900 Rio Claro (SP), Brazil}
\affiliation{Physikalisches Institut, Goethe-Universit\"{a}t, Max-von-Laue-Str. 1, 60438 Frankfurt am Main, Germany}

\author {Lorenz Bartosch}
\email{lb@itp.uni-frankfurt.de}
\affiliation{Institut f\"{u}r Theoretische Physik, Goethe-Universit\"{a}t,
Max-von-Laue-Str. 1,
60438 Frankfurt am Main, Germany}

\title{Probing the Mott Physics in $\kappa$-(BEDT-TTF)$_2$X Salts via Thermal Expansion}

\begin{abstract}
In the field of interacting electron systems the Mott metal-to-insulator (MI) transition
represents one of the pivotal issues. The role played by lattice
degrees of freedom for the Mott MI
transition
and the Mott criticality
in a variety of materials
are current topics under debate.
In this context, molecular conductors of the
$\kappa$-(BEDT-TTF)$_2$X type constitute a class of
materials for unraveling several aspects of the Mott physics. In this review,
we present a synopsis of literature results with
focus on recent expansivity measurements probing the
Mott  MI transition in this class of materials.
Progress in the description of the Mott critical behavior is also addressed.
\end{abstract}

\pacs{72.15.Eb, 72.80.-r, 72.80.Le, 74.70.Kn}

\maketitle

\date{\today}

\section{Introduction}\label{Introduction}
A tiny change in physical parameters can sometimes have dramatic consequences. This is particularly true in correlated condensed matter systems in which a small alteration in pressure, chemical composition, or temperature can tune a system across a phase transition. \textcolor{black}{In this context}, the Mott
metal-to-insulator (MI) transition represents one of the most remarkable examples \textcolor{black}{of electronic correlation phenomena}.
In contrast to conventional band insulators in which all bands are either filled or empty, the Mott insulating state is driven by interactions and emerges once the ratio of the electron-electron interaction $U$ to the
hopping matrix element (kinetic energy)
$t$ exceeds a critical value.
It is essentially this ratio which is tuned in many experiments. \textcolor{black}{Molecular conductors of the}
$\kappa$-phase of (BEDT-TTF)$_2$X (where BEDT-TTF refers to
bisethylenedithio-tetrathiafulvalene, i.e.\ C$_{10}$S$_8$H$_8$,
and X to a monovalent \textcolor{black}{counter} anion), \textcolor{black}{usually called charge-transfer salts,}  have been recognized as model systems
to investigate electronic correlations in two dimensions. In
particular, single crystals of \textcolor{black}{wholly} deuterated \textcolor{black}{charge-transfer salts of
the $\kappa$ phase with X = Cu[N(CN)$_2$]Br} have been attracting
broad interest due to fact that they lie in the vicinity of a
first-order phase transition line in the phase diagram
\cite{Kawamoto1997-20}, which separates the metallic from the
insulating state, thus \textcolor{black}{giving the possibility of investigating the Mott MI transition using temperature as a tuning parameter}. One result to be discussed  in detail in the frame of this paper refers
to the first experimental observation of the role \textcolor{black}{played by}
lattice degrees of freedom for the Mott MI transition in the
above-mentioned materials \cite{Mariano20f1}. Both the discontinuity
and the anisotropy of the lattice parameters observed via
high-resolution thermal expansion experiments indicate \textcolor{black}{a complex}
role of the lattice \textcolor{black}{effects} at the Mott \textcolor{black}{MI} transition for
the $\kappa$-(BEDT-TTF)$_2$X \textcolor{black}{charge-transfer} salts,
which cannot be \textcolor{black}{interpreted simply by taking into account} a purely 2D
electronic model. Furthermore, in the frame of the present work, a
model based on the rigid-unit modes scenario is proposed to describe
the negative thermal expansion above the so-called glass-like
transition temperature $T_g$ $\simeq$ 77\,K in \textcolor{black}{deuterated charge-transfer}
salts of $\kappa$-(BEDT-TTF)$_2$Cu[N(CN)$_2$]Br and parent
compounds.

From a theoretical point of view,
perturbative approaches and their generalizations are restricted to the regimes
$U \ll t$ and $t \ll U$ and have difficulties in addressing phase transitions. In fact, they
entirely fail in addressing the critical behavior in the \textcolor{black}{immediate} vicinity of a putative critical point.
Instead, it is advantageous to start from the critical point itself and use concepts of scaling to describe its vicinity.
We will review here how some aspects of the lattice response observed in \Refs{Mariano20f1} and \mbox{\,}\onlinecite{Thesis} can be understood in terms of such a scaling theory \cite{Bartosch,Zacharias}.

\vspace*{0.4cm}

The compound $\kappa$-(BEDT-TTF)$_2$Cu$_2$(CN)$_3$ is another
material which attracted great interest  in the last few years, see,
\textcolor{black}{for instance} \Ref{Yamashita2008} and references \textcolor{black}{cited} therein. This system has an
almost perfect quasi-2D frustrated triangular lattice
with a ratio of hopping integrals close to $t'/t \simeq 1$.
Interestingly enough, while the ratio $U$/$t$ shows a monotonic temperature dependence, the degree of frustration $t'/t$  as a function of temperature presents a non-monotonic behavior and becomes more pronounced at low temperatures \cite{Harald}.

Up to now, \textcolor{black}{no clear experimental evidence of magnetic ordering in $\kappa$-(BEDT-TTF)$_2$Cu$_2$(CN)$_3$ has been reported in the literature} \cite{Shimzu2003}. Thus, this system has been
recognized as a candidate for the realization of a spin-liquid\cite{Balents10}. At
$T$ = 6\,K, a crossover from a \textcolor{black}{paramagnetic Mott insulator to a genuine
\emph{quantum spin-liquid}} has been proposed
\cite{Yamashita2008}. Nevertheless, the \textcolor{black}{physical} nature of such
a crossover, sometimes called \emph{hidden ordering}, is still
controversial. Expansivity measurements performed on this material
revealed a pronounced phase-transition-like-anomaly at $T$ = 6\,K, which
coincides nicely with specific heat results reported in the
literature \cite{Yamashita2008}. These results provide strong
evidence that this lattice instability at $T$ = 6\,K is directly
linked to the proposed spin-liquid phase.
\newline

The electronic properties of $\kappa$-(BEDT-TTF)$_2$X \textcolor{black}{charge-transfer salts have been reported in various review articles, see, for instance} \Refs{Lang20a,Lang2004-9c,Ishiguro20,Jerome1991-20,Mckenzie1998-20,
Wosnitza2000-20,Singleton2002-20,Myagawa2004-20,Fukuyama2006-20,Mori2006-20,Kanoda2006-20,Powell2006-20,Lang4,Kanoda11,Powell11}.
Here we focus on topics linked to dilatometric studies of the $\kappa$-(BEDT-TTF)$_2$X salts.
Some parts of this review are based on the Ph.D.~thesis of one of us \cite{Thesis}.
Before giving a detailed outline of the main body of this review article at the end of this introductory chapter, let us first briefly review some general properties of the $\kappa$-(BEDT-TTF)$_2$X salts and discuss their crystallographic structure in Subsection \ref{Crystal-Structure}, their phase diagram in Subsection \ref{State of the Art
kappa-ET} and the most relevant experimental results from the literature related to this work in Subsection \ref{Key experimental results}.


\subsection{The (BEDT-TTF) molecule and crystallographic structure of the $\kappa$-ET salts}\label{Crystal-Structure}

\begin{figure}
\begin{center}
\includegraphics[angle=0,width=0.48\textwidth]{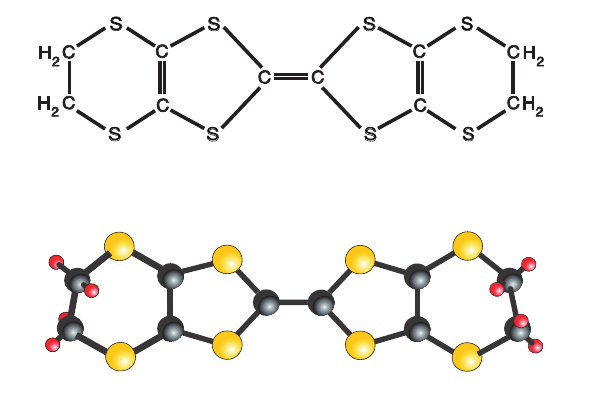}
\end{center}
\caption{\small (BEDT-TTF), or simply ET, molecule, which is the
basic entity that furnishes the quasi-2D (BEDT-TTF)-based
charge-transfer salts. Picture after
\Ref{Wosnitza2000-20}.}\label{ET1}
\end{figure}

The basic entity which furnishes the structure of the present class
of organic conductors is the (BEDT-TTF) molecule, usually
abbreviated simply to ET.
The structure of the ET molecule is shown in \Fig{ET1}. \textcolor{black}{As a result of}  the two \textcolor{black}{distinct}
possible configurations of the ethylene end groups in their extremes
(known as staggered and eclipsed configurations), the ET molecule is
not completely planar. It  \textcolor{black}{has been proposed} that the degrees of freedom of
the ethylene \textcolor{black}{end} groups can influence the electronic properties of the
salts of the $\kappa$-family, cf.\,\Ref{Jens2002-20} and references \textcolor{black}{cited}
therein.  \textcolor{black}{Employing infra-red imaging spectroscopy, the authors of \Ref{Sasaki2004-20} reported on the sensitivity of the vibration mode of the central C=C bonds, the so-called  $a_g(\nu_3)$ mode,
upon going from the metallic to the insulating state.}

In the following, let us briefly discuss the various phases in which ET charge-transfer
salts can crystallize.

\begin{figure}[!htb]
\begin{center}
\includegraphics[angle=0,width=0.50\textwidth]{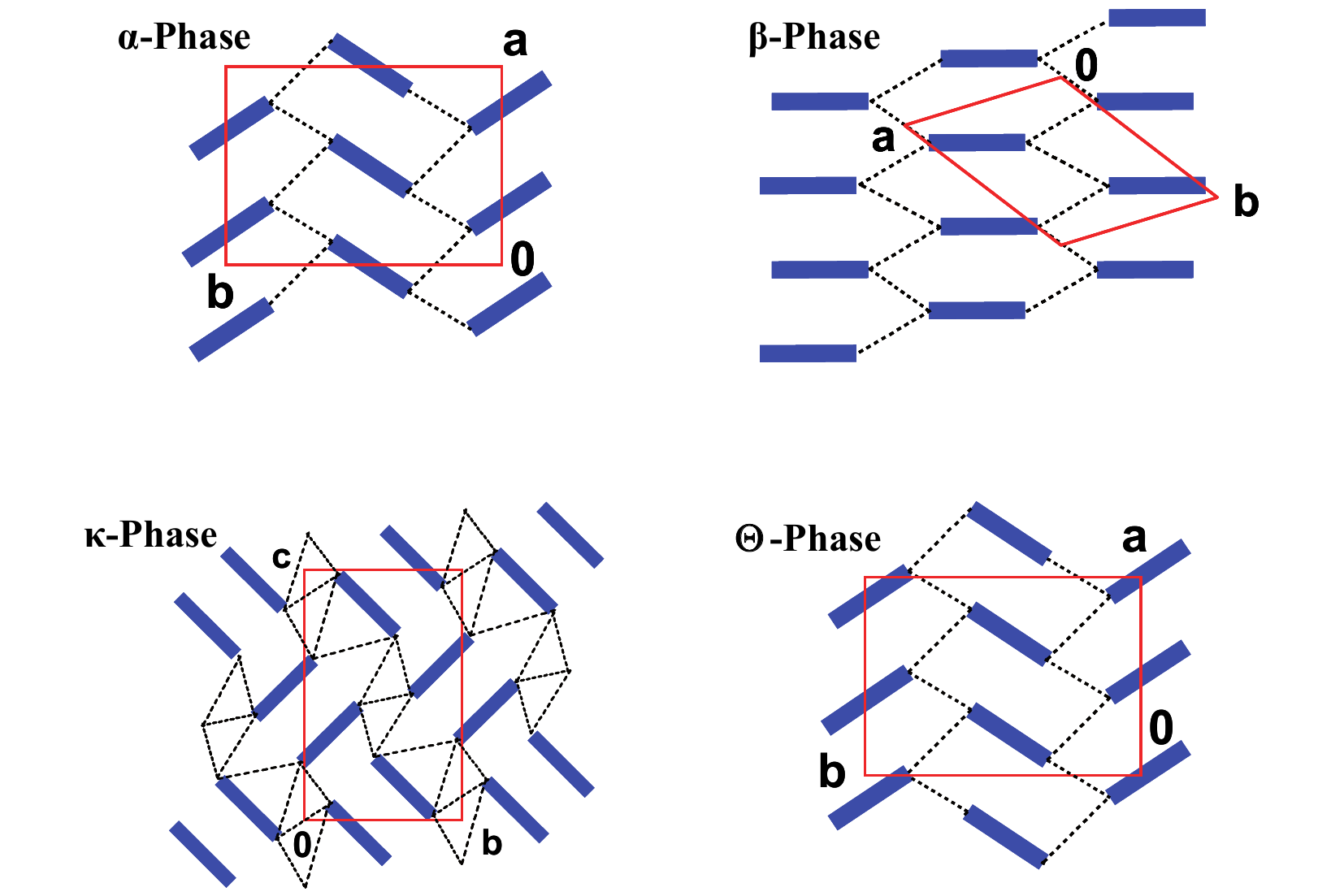}
\end{center}
\caption{\small Schematic top view, i.e.\ view along \textcolor{black}{the ET molecule} long axis of the $\alpha$-, $\beta$-, $\kappa$- and
$\theta$-phase of the ET-based organic conductors. The ET
molecules and the unit cells are, respectively, illustrated by solid blue and open red
rectangles. The dashed lines indicate the
$\pi$-orbital overlaps. Picture after
\Ref{Belal2005-20}.}\label{ET_Phases}
\end{figure}

During the crystallization process of the (ET)$_2$X salts,
the ET molecules can assume
different spatial arrangements, giving rise to different phases.
Several thermodynamically stable phases, usually referred to by the
Greek \textcolor{black}{letters} $\alpha$, $\beta$, $\theta$, $\lambda$, and
$\kappa$, are known,
see \Fig{ET_Phases} for a schematic
representation of the main known phases.
This review is focussed on salts of the
$\kappa$-phase family. Interestingly enough, different crystalline
arrangements imply different band fillings and consequently different
physical properties. A detailed discussion about each phase is
presented in \Refs{Lang4, Powell2006-20}.

\textcolor{black}{The packing
pattern of the $\kappa$-phase differs distinctly from the others in
the sense that it consists of two face-to-face ET molecules, cf.\,\Fig{ET_Phases}}. The two
molecules interact with each other via overlap between the highest
occupied molecular orbitals (HOMOS), as the HOMOS of each molecule
are partially occupied \cite{Powell2006-20}.

\begin{figure}[h]
\begin{center}
\includegraphics[angle=0,width=0.48\textwidth]{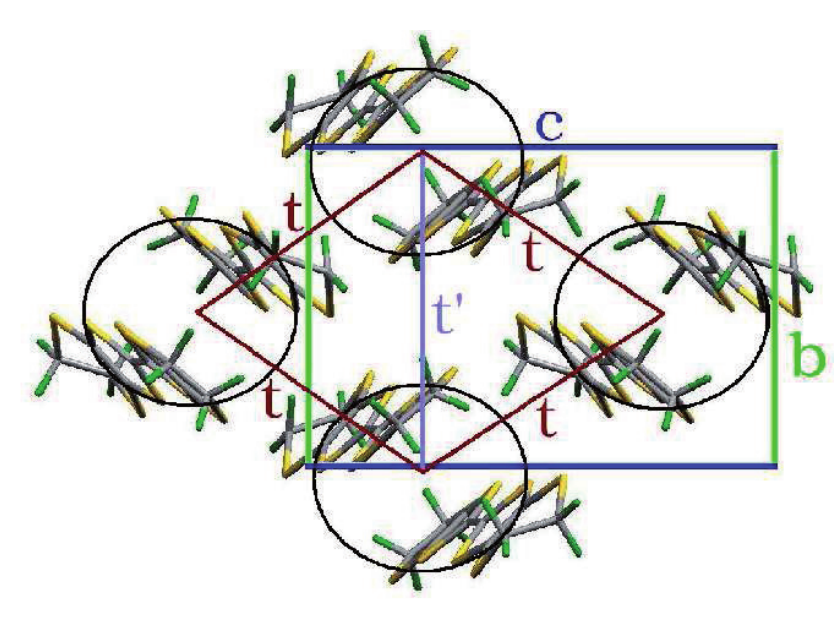}
\end{center}
\caption{\small \textcolor{black}{Crystal structure of the
$\kappa$-(ET)$_2$X salts (top view)}. The largest hopping terms between the
dimers are indicated by  $t$ and $t'$. Note that the hopping terms
$t$, $t'$ and $t$ form a triangular lattice.
It should be noted that the axes labeling refers to the monoclinic lattices, e.g. the salts with
X = Cu(NCS)$_{2}$ or Cu$_{2}$(CN)$_{3}$.
Picture taken from
\Ref{Powell2006-20}.}\label{Dimer Model}
\end{figure}

Due to the intrinsic strong dimerization in the $\kappa$-(ET)$_{2}$X
family, pairs of ET molecules can be treated as a dimer unity. As a result of such an assumption, an anisotropic triangular lattice is built by such dimer units as shown in \Fig{Dimer Model} and the system can be
described by
the Hubbard model. The counter anion X governs
 the ratio between the inter-dimer transfer integrals $t$ and $t'$.
For example, for the $\kappa$-(ET)$_2$Cu$_2$(CN)$_3$ salt
$t'/t = 0.83$ \cite{Kandpal09,Nakamura09}, suggesting a
strongly frustrated triangular lattice in this compound, to be
discussed in more detail below.

It is well to note that, more recently, a new phase, called mixed
phase, with alternating  $\alpha$- and $\kappa$-type arrangements of
the ET molecules, has been synthesized \cite{Schlueter20c}. In
addition, a dual-layered superconductor with alternating $\alpha'$-
and $\kappa$-type packing has been reported in the literature
\cite{JACS}.

\begin{figure}[!htb]
\begin{center}
\includegraphics[angle=0,width=0.42\textwidth]{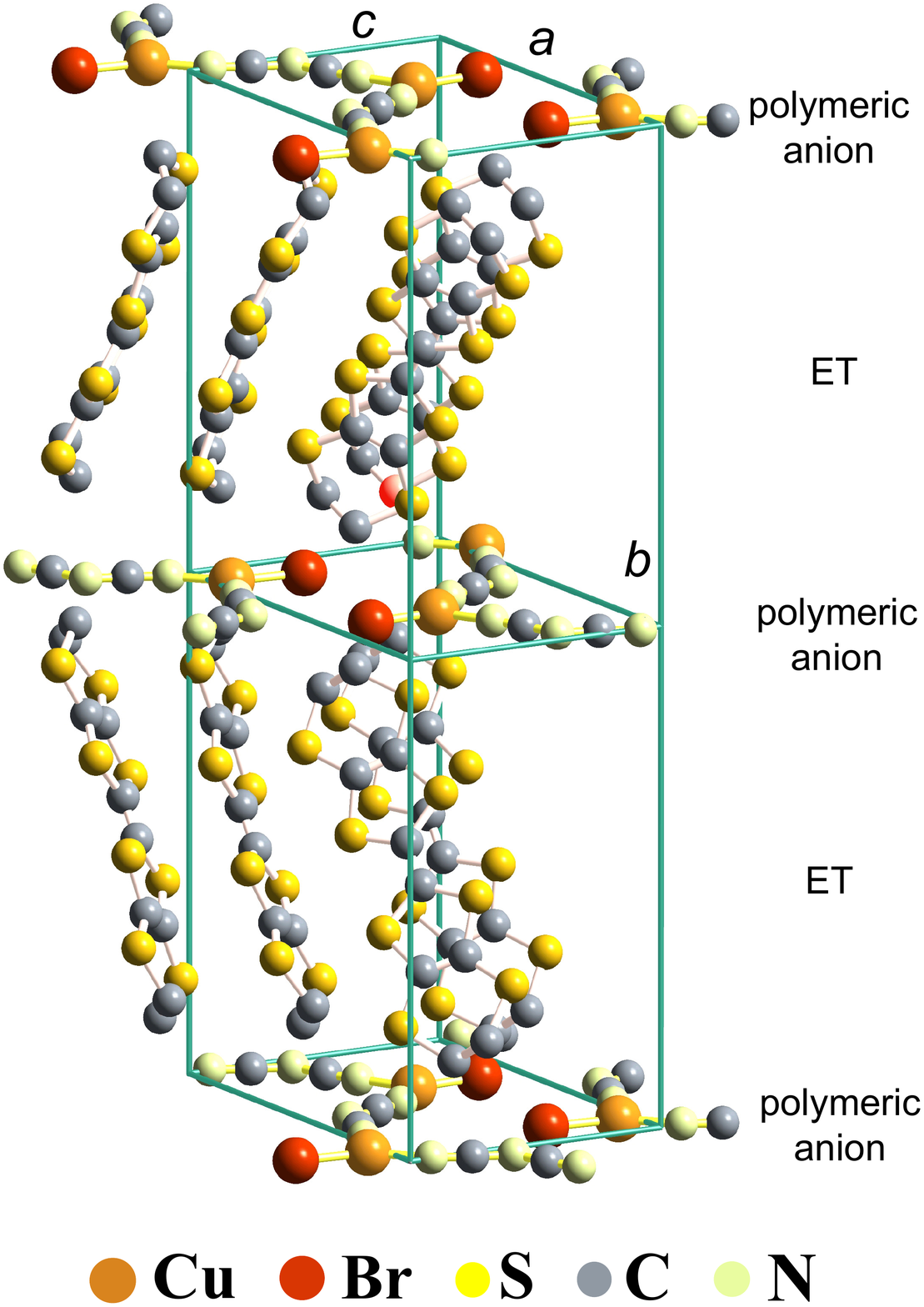}
\end{center}
\caption{\small The crystallographic structure of
$\kappa$-(ET)$_2$Cu[N(CN)$_2$]Br. While \emph{a}  and \emph{c} are the
in-plane axes,
\emph{b} is along the out-of-plane direction.
``ET'' and ``polymeric anion'' indicate layers of ET molecules and anions,
respectively. The various atoms are represented by different
colors, as indicated in the label. For clarity, protons at the end
of the ET molecules are omitted. Picture taken from \Ref{Thesis}.}\label{Structure}
\end{figure}

The crystallographic structure of the $\kappa$-(ET)$_{2}$Cu[N(CN)$_2$]Br
salt \cite{Kini20c1} is displayed in \Fig{Structure}.
The structure
of the latter substance is orthorhombic with the space group
\emph{Pnma} with four dimers and four anions in a unit cell. An
interesting aspect is that although the compounds
$\kappa$-(ET)$_{2}$Cu(NCS)$_2$ and $\kappa$-(ET)$_{2}$Cu$_2$(CN)$_3$
\cite{Geiser1991} also belong to the $\kappa$-phase family, their
crystallographic structure is monoclinic. Since  the counter
anion\linebreak X$^-$ = Cu[N(CN)$_2$]Br$^-$ assumes a closed-shell
configuration, metallic properties are observed only within the ET
layers, which lie in the crystallographic \emph{ac}-plane and
correspond to the large face of the crystal. \textcolor{black}{In the perpendicular
direction to the referred layers, i.e.\ the \emph{b}-direction
\cite{eixo}, the insulating anion layer partially blocks charge
transfer, so that the electrical conductivity along this direction
($\perp$ to the layers) is reduced by a factor of
100 -- 1000, depending on the counter anion.}
Due to this anisotropy, these materials are called
 quasi-two-dimensional (quasi-2D) conductors.

\begin{figure}[!htb]
\begin{center}
\includegraphics[angle=0,width=0.46\textwidth]{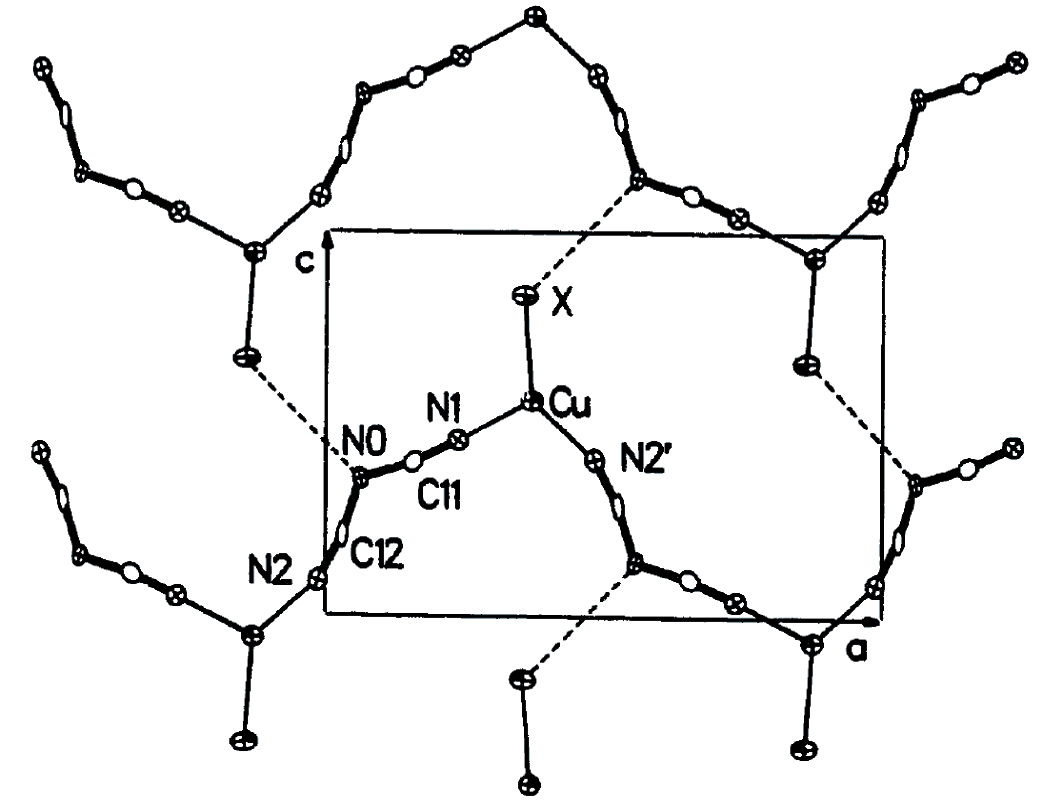}
\end{center}
\caption{\small \textcolor{black}{Layered structure formed by the Cu[N(CN)$_2$]X$^{-}$
polymeric anion in}
$\kappa$-(ET)$_2$Cu[N(CN)$_2$]X, view along the \emph{b}-axis.
Here, X refers to either Br or Cl.
Dashed lines highlight the X...N weak contacts between the anion
chains. The rectangle indicates the contour of the unit cell. Picture
taken from \Ref{Geiser1991a}.}\label{Anion}

\vspace*{0.5cm}
\includegraphics[angle=0,width=0.46\textwidth]{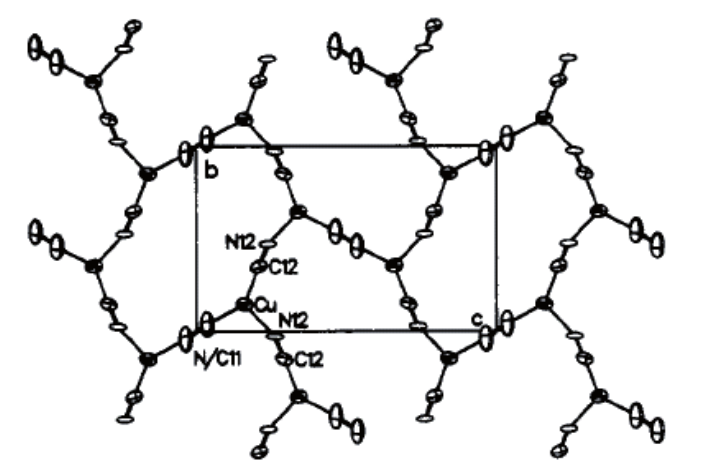}
\caption{\small \textcolor{black}{View along the out-of-plane direction (\emph{a}-axis) of the polymeric anion layer  of  Cu$_2$(CN)$_3^-$ in
$\kappa$-(ET)$_{2}$Cu$_2$(CN)$_3$. The
rectangle indicates the contour of the unit cell.
The N/C11 cyanide ion is located on an inversion center,
being thus crystallographically disordered.}
The ellipsoids therefore represent either a carbon or a nitrogen atom with a 50\% probability.
Picture
taken from \Ref{Geiser1991}.}\label{Anion-SL}
\end{figure}

As displayed in \Fig{Anion}, the Cu[N(CN)$_2$]Br$^-$
(or Cu[N(CN)$_2$]Cl$^-$)
polymeric anion structure is composed of  planar
triply-coordinated Cu(I) atoms with two bridging
dicyanamide
[(NC)N(CN)]$^{-}$ ligands, forming a zig-zag pattern chain along
the \emph{a}-axis. The terminal Br$^-$ ions, which overlap
slightly with the central N atoms of neighboring polymeric chains,
make the third bond at each Cu(I) atom \cite{Lang20a}.

\textcolor{black}{As shown in \Fig{Anion-SL}, the counter anion Cu$_2$(CN)$_3^-$
differs distinctly from the\linebreak Cu[N(CN)$_2$]Br$^-$ anion in
that it does not form chains, but a planar reticule of Cu(I) ions
coordinated in a triangular fashion and bridging cyanide groups. Notably,} one of the
cyanide groups (labeled N/C11 in \Fig{Anion-SL}) is located
on an inversion center and therefore must be crystallographically
disordered \cite{Geiser1991}. As will be discussed in
Section\,\ref{SpinLiquidSection}, this particular arrangement is
likely to be responsible for the absence of the so-called
glass-like transition in $\kappa$-(ET)$_{2}$Cu$_2$(CN)$_3$.

Directional-dependent thermal expansion experiments, to be discussed
in Section\,\ref{TEMeasurementsonkappaD8}, will show that the more
pronounced lattice effects, associated with the Mott MI transition
for $\kappa$-D8-Br, occur along the in-plane direction which is
parallel to the anion chains direction (\emph{a}-axis).
In addition,
we will propose a model based on the
collective vibration modes of the CN groups of the polymeric anion
chains
to explain the negative thermal expansion
along the \emph{a}-axis above $T_g$ $\simeq$ 77\,K.

For completeness, the lattice parameters of the $\kappa$-(ET)$_2$X
charge-transfer salts discussed in this work are listed in
Table\,\ref{LatticeParameters}.

\begin{table}[htb]
\centering \small
\begin{tabular}{|c|c|c|c|c|c|c|c|c|}
\hline\hline \textbf{Anion} & \textbf{\emph{a}({\AA})} & \textbf{\emph{b}({\AA})} & \textbf{\emph{c}({\AA})} & \textbf{V({\AA}$^3$)} & \textbf{$\alpha$} & \textbf{$\beta$} & \textbf{$\gamma$} & \textbf{Struc.}\\
\hline\hline
  D8-Br &   -   &    -  & - & -      & $\perp$ & $\perp$ & $\perp$ & Orth. \\ \hline
  H8-Br & 12.949  & 30.016 & 8.539 & 3317 & $\perp$ & $\perp$ & $\perp$ & Orth.\\ \hline
  H8-Cl & 12.977  & 29.977 & 8.480 & 3299 & $\perp$ & $\perp$ & $\perp$ & Orth.\\ \hline
  CuCN  & 16.117 & 8.5858 & 13.397 & 1701.2 & $\perp$ & 113.42$^\circ$ & $\perp$ & Monoc.\\ \hline\hline
\end{tabular}
\caption{\small Lattice parameters at room temperature and structure
of the investigated $\kappa$-(ET)$_2$X salts
\cite{Geiser1991a,Geiser1991}. CuCN refers to Cu$_2$(CN)$_3$. Data
for $\kappa$-D8-Br are not available in the literature. $\perp$
corresponds to 90$^\circ$.
Note that in contrast to the other compounds, for the Cu$_2$(CN)$_3$ salt the \emph{a}-direction is the out-of-plane direction.}\label{LatticeParameters}
\end{table}

\subsection{\textcolor{black}{Phase diagram of $\kappa$-(ET)$_2$X}}
\label{State of the Art
kappa-ET}

\begin{figure}[!htb]
\begin{center}
\includegraphics[angle=0,width=0.50\textwidth]{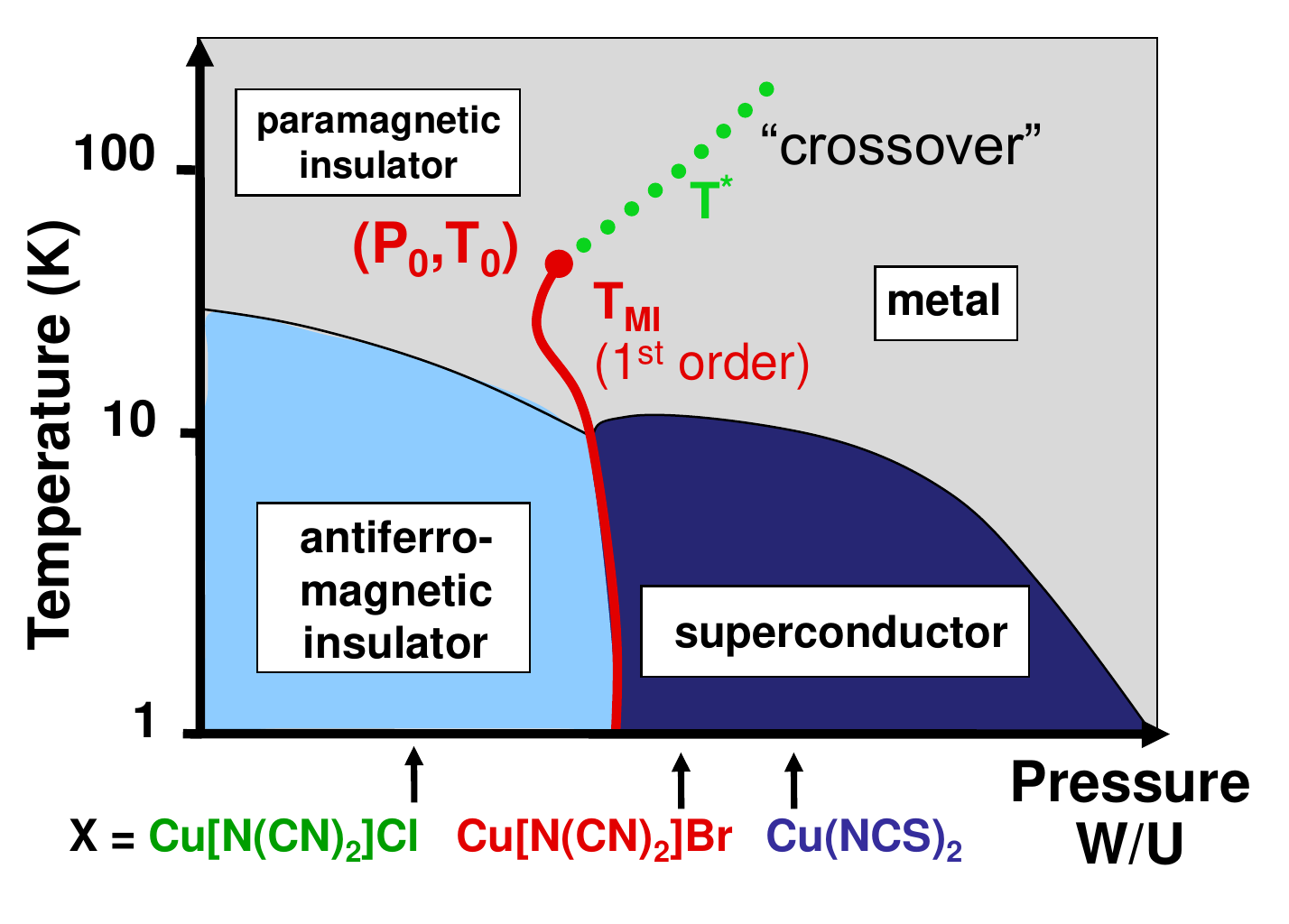}
\end{center}
\caption{\small \textcolor{black}{Conceptual pressure \emph{versus} temperature ($P$-$T$)  phase diagram for the
$\kappa$-(ET)$_2$X charge-transfer salts, after \Ref{Kanoda97-20}. Arrows
pinpoint the positions of the various salts with their counter anion X at
ambient pressure. The ratio of the on-site Coulomb repulsion relative
to the bandwidth, i.e.\ $U$/$W$, is reduced by application of hydrostatic pressure.}}\label{Schematic-PD}
\end{figure}

Charge-transfer salts of the $\kappa$-(ET)$_2$X family
have been recognized as strongly correlated electron systems.
Among other features, the latter statement is based on the unusual
metallic behavior inherent to these systems, as will be discussed
below. Band structure calculations suggest that these materials
should be metals down to low temperatures \cite{Powell2006-20}.
However, due to correlation effects, the phase diagram embodies
various phases, including the paramagnetic Mott insulator (PI), the antiferromagnetic Mott insulator (AF), the superconductor,
and the anomalous metal, see the phase diagram depicted in
\Fig{Schematic-PD}.

A remarkable feature is the competition between antiferromagnetic
ordering and superconductivity.
The
similarities between the present charge-transfer salts and
the high-$T_c$ cuprate superconductors have been discussed,
see e.g.\ \Refs{Mckenzie1998-20, Mackenzie20a, Degoto20b1}. The
major difference, however, between these two classes of materials is
that while for the  cuprate
superconductors one deals with a
\emph{doping-controlled} Mott transition, for the charge-transfer
salts to be discussed here one has a \emph{bandwidth-controlled}
Mott transition. In addition, \textcolor{black}{no spin-glass phase is observed in organic conductors
between the antiferromagnetic and superconducting phases}.
The symmetry of the superconducting order parameter in these materials
is still a
topic of debate, see e.g.\ \Ref{Taylor2007-20}
and references therein. From the experimental point of view, a
remarkable feature is the tunability of such systems. In contrast to
the cuprate
superconductors, where the ground states are tuned by
doping, i.e.\ band filling, and therefore disorder is unavoidable,
the ground states of organic conductors can be tuned by chemical
substitution, \textcolor{black}{namely counter anion substitution, and/or by} applying external pressure.
Pressure mainly induces strain which in turn leads to
modified hopping matrix elements $t$ and $t'$ and thereby modifies
the bandwidth $W$. As a consequence, applying pressure leads to a change of the ratio $W/U$.
A pressure of a few hundred bar, which is easily attainable
by employing the $^4$He-gas pressure technique, is enough to make a
wide sweep over relevant regions of the \textcolor{black}{pressure \emph{versus} temperature}
($P$-$T$) phase diagram \cite{pressao}.

\begin{figure}[!htb]
\begin{center}
\includegraphics[angle=0,width=0.48\textwidth]{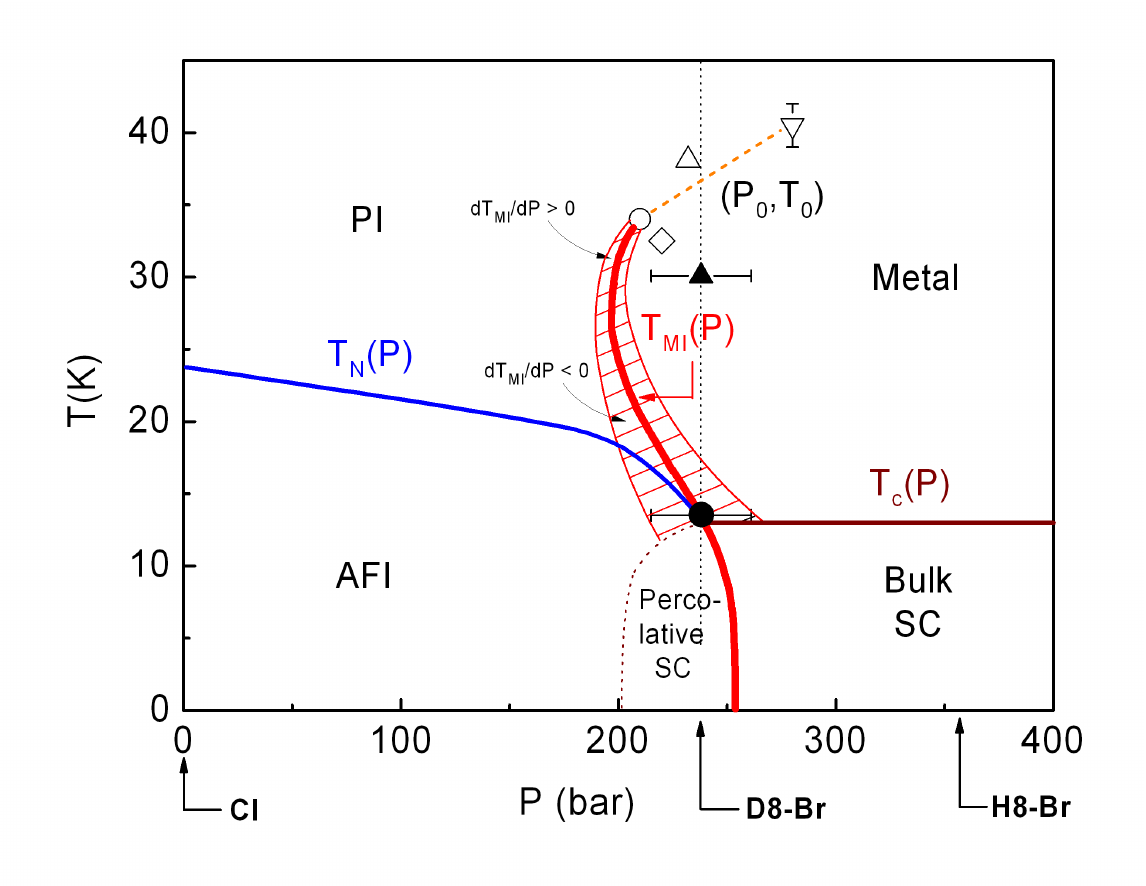}
\end{center}
\caption{\small \textcolor{black}{Pressure \emph{versus} temperature p}hase diagram of
$\kappa$-(ET)$_{2}$Cu[N(CN)$_{2}$]Z, with Z = Br or Cl. Lines
indicate the respective phase transitions. \textcolor{black}{D8-Br and H8-Br indicate the positions (estimated according to \Ref{Lang20f})} of
the $\kappa$-D8-Br and $\kappa$-H8-Br salts, respectively.
The orange broken line
indicates the
crossover from the metallic to the insulating regime above the critical temperature.
\textcolor{black}{Thin solid lines (red),
demarcating the hatched area, refer to the positions of the anomalies observed
via ultrasonic measurements (see \Fig{Fournier1}). The vertical
(black) dotted line indicates $T$ sweeps
of the $\kappa$-D8-Br salt,
depicting a crossing of
the \emph{S}-shaped first-order Mott MI phase transition line.
PI, AFI and SC denote the paramagnetic
insulator, the antiferromagnetic insulator and the superconducting phase,
respectively. The open circle refers to
the Mott critical end point.
At the point marked by a closed circle in the phase
diagram, \textit{T$_{MI}$} matches $T_N$,
cf.\ the discussion
in Section\,\ref{Entropy changes at the Mott transition}.}
As can be seen in the figure, the slope of $dT_{MI}$/$dP$
can be both positive and negative,
indicating
that upon crossing the MI line
the entropy can either increase or decrease.
While the entropy increases for $dT_{MI} / dP  > 0$ when crossing the MI line from the metallic to the PI state,
for $dT_{MI} / dP  < 0$, when going from the metallic to the AFI state, the
entropy associated with the metallic state is decreased due to
magnetic ordering. Picture taken from \Ref{Mariano20f1}.
}\label{Phase Diagram}
\end{figure}

The phase diagram shown in \Fig{Phase Diagram} has been mapped
by NMR and ac susceptibility \cite{Lefebvre7}, transport
\cite{Limelette2003-20} and ultra-sonic \cite{Fournier2003-20}
measurements under helium gas pressure on the salt with
X=Cu[N(CN)$_{2}$]Cl.  In this \emph{generic} phase diagram, the
antiferromagnetic transition line, not observed in ultra-sound
experiments \cite{Fournier2003-20}, was obtained from the NMR relaxation
rate \cite{Lefebvre7}. From the splitting of  NMR lines, the
estimated magnetic moment per dimer is (0.4 -- 1.0)\,$\mu_B$. The
superconducting transition line was determined from the ac susceptibility
\cite{Lefebvre7} and its fluctuations in the immediate vicinity of the Mott transition by means of Nernst effect measurements \cite{BlundNature}, whereas the \emph{S}-shaped first-order MI
transition line via ultra-sound \cite{Fournier2003-20}, transport
\cite{Limelette2003-20} and ac susceptibility \cite{Lefebvre7}
measurements.

\textcolor{black}{This first-order Mott MI transition line} ends in a critical point ($P_0, T_0$),
which has been studied by several groups \cite{Lefebvre7,
Limelette2003-20, Fournier2003-20, Kagawa2004-20, Kagawa2005_20}.
Among possible scenarios, this critical end-point has been discussed
in analogy with the liquid-gas transition, see e.g.\
\Ref{Papanikolaou2008_20}. In this scenario, above ($P_0,
T_0$) the metallic state cannot be distinguished from the
paramagnetic insulating state. \textcolor{black}{More recent studies suggest that
lattice effects are relevant close to the Mott
critical end-point \cite{Mariano20f1, Bartosch, Zacharias}.}

By means of resistivity measurements under pressure, the critical
behavior in the \textcolor{black}{proximity of the point ($P_0$, $T_0$) was investigated by
Kagawa \emph{et al.} \cite{Kagawa2005_20}.}
These authors found
critical exponents which do not fit in the most common universality
classes, and they assigned this to the quasi-2D structure of the
present substances. \textcolor{black}{Such a scenario shall be examined in more detail in
Section\,\ref{Criticality arount Tstar}. The possibility of tuning the
system from  the insulating to the metallic side of the $P$ \emph{versus} $T$ phase
diagram, i.e.\ crossing the first-order Mott MI line by application of
pressure, is  indicative of a bandwidth-controlled Mott MI transition.
Owing to the position of the salts with different counter anions X, as shown
in \Fig{Phase Diagram}, the fully hydrogenated salt with X =
Cu[N(CN)$_{2}$]Br superconducts below 12\,K, the
highest $T_c$ under ambient pressure among all organic conductors till the present date,
whereas the salt with counter anion X =
Cu[N(CN)$_{2}$]Cl is a Mott insulator  with $T_N$
$\simeq$ 24\,K.
It has very recently been shown that the antiferromagnetic N\'eel state
is accompanied by charge ordering, i.e.\ the salt
$\kappa$-(ET)$_{2}$Cu[N(CN)$_{2}$]Cl is in fact a multiferroic \cite{Lunke}.}

Interestingly enough, \textcolor{black}{fully deuteration
of the ethylene end groups of the ET molecules in the salt with X = Cu[N(CN)$_{2}$]Br, as described in detail above,
results in a shift of the latter
towards the boundary of the \emph{S}-shaped first-order Mott MI transition line. It is
due to their close position to this line \cite{Kawamoto1997-20}
that the deuterated salt with X = Cu[N(CN)$_{2}$]Br, hereafter named $\kappa$-D8-Br, has attracted particular
interest. As a matter of fact, the C-H and C-D bond lengths are slightly different. It is well known that the deuterium nucleus has one proton and one neutron, while the hydrogen nucleus is formed simply by one proton. Thus deuterium has a higher mass than hydrogen and, as consequence C-D
chemical bonds have a smaller eigen-frequency than the C-H bonds. Hence,
the displacement of the deuterium atoms from their equilibrium positions during the vibration process is smaller than that estimated for the hydrogen atoms. Thus the C-D chemical bonds are slightly smaller than the C-H bonds.}
Furthermore, the contact between the C-D$_2$ end groups and the
anions is weaker than that between the C-H$_2$ end groups and the
anions. As a consequence, the lattice of the fully deuterated salt
is softer than the lattice of the fully hydrogenated salt.
This process of exchanging atoms
is usually called applying ``chemical pressure''.  In fact, application of
pressure (chemical or external) \textcolor{black}{changes the distance
between the ET molecules so that the overlap between $\pi$-orbitals' is increased and, as a consequence,  the bandwidth
\emph{(W)} is changed. The ratio $W/U$ is considered the pivotal
parameter which defines the various phases of these substances
\cite{Kanoda97-20}.}

\begin{figure}[!htb]
\begin{center}
\includegraphics[angle=0,width=0.46\textwidth]{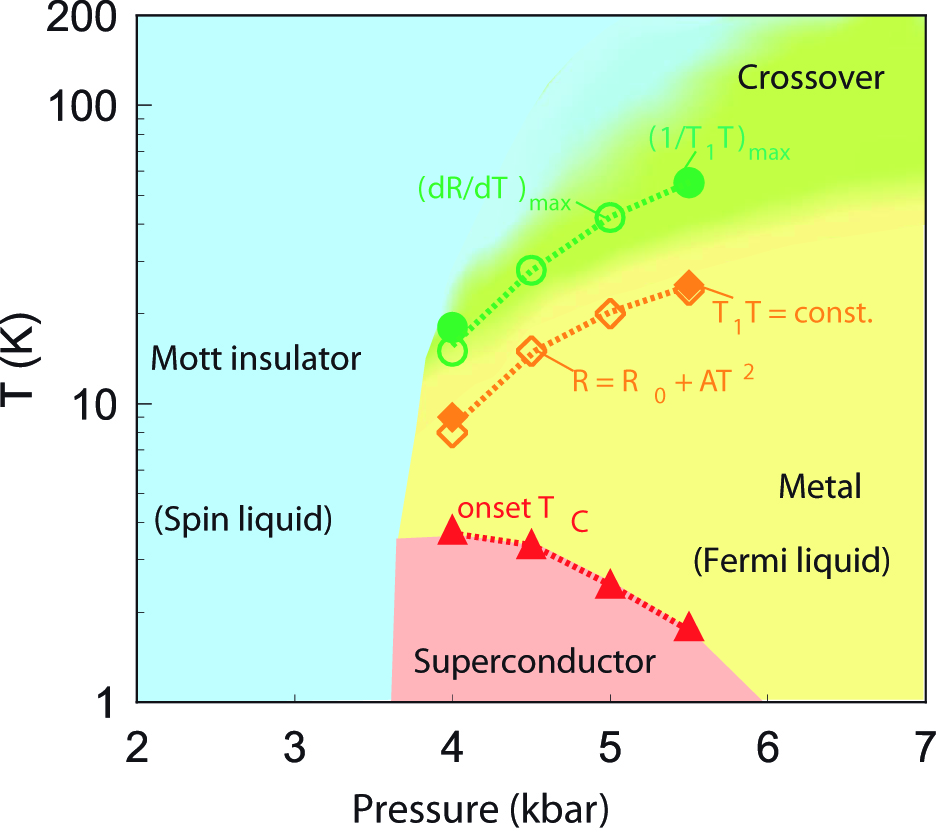}
\end{center}
\caption{\small $P$-$T$ phase diagram of the
$\kappa$-(ET)$_2$Cu$_2$(CN)$_3$ salt, obtained from resistance and
NMR measurements under pressure using an oil pressure cell. \textcolor{black}{The Mott MI
transition and crossover lines are associated with the temperature at
which the quantities spin-lattice relaxation rate
divided by temperature (1/$T_1 T$) and d$R$/d$T$ show a maximum, cf.\
labels indicated in the figure. The Fermi-liquid
regime (yellow  area)} was fixed by the temperature regime where the
resistance $R$ follows the $R_0$ + $AT^2$ relation and
(1/$T_1 T$) is constant.  The superconducting transition line was
obtained via in-plane resistance measurements.
Figure taken from
\Ref{Kurosaki2005}.}\label{SL-PD}
\end{figure}

Interestingly enough, for the salt with the anion X =
Cu$_2$(CN)$_3$, magnetic ordering is absent in the whole insulating
region. The \textcolor{black}{$P$ \emph{versus} $T$ phase diagram of the
$\kappa$-(ET)$_2$Cu$_2$(CN)$_3$ salt is depicted in
\Fig{SL-PD}.}

\subsection{Key experimental results}\label{Key experimental results}
\label{ET-LiteratureResults}

As mentioned above, the metallic state of
$\kappa$-(ET)$_2$X presents some distinct properties as compared to
those of conventional metals. For instance, in the $\kappa$-H8-Br
salt, the out-of-plane resistance as a function of temperature has a
non-monotonic behavior \cite{Strack20d} with a maximum around 90\,K,
which was observed to be sample-dependent.
Decreasing temperature, this maximum is followed
by a sudden drop and the resistivity sharply changes its slope
around $T^*$ $\simeq$ 40\,K.  Below  $T_c \simeq$ 11.5\,K
superconductivity is observed. On applying pressure, $T^*$ shifts to
higher temperatures
and disappears at roughly 2\,kbar.
From about
$T^*$ $\simeq$ 40\,K down to $T_c \simeq$ 11.5\,K the resistance
obeys a $T^2$ behavior which is frequently assigned to a coherent Fermi
liquid state. NMR studies revealed a peak in the spin-relaxation
rate divided by temperature $(1/T T_1)$ around $T^*$ for the present
compound \cite{Kawamoto95-20d}. The anomalous metallic behavior
around $T^*$ was also observed by thermal, magnetic, elastic, and
optical measurements, cf.\,\Ref{Strack20d} and references therein.
The origin of the $T^*$ anomaly is still controversial. Ultrasonic
velocity measurements under pressure on the compound
$\kappa$-(ET)$_2$Cu[N(CN)$_2$]Cl performed by Fournier \emph{et al.}
\cite{Fournier2003-20} revealed the existence of two new lines in
the phase diagram (not indicated in \Fig{Phase Diagram})
 which merge at the critical end point.
These lines indicate the softening of the lattice response and should therefore coincide with anomalies expected in future dilatometric studies under pressure.
It is important to keep in mind that the two lines observed in ultrasound experiments are different from the metal-to-insulator crossover line above $T_c$.

\begin{figure}[!htb]
\begin{center}
\includegraphics[angle=0,width=0.50\textwidth]{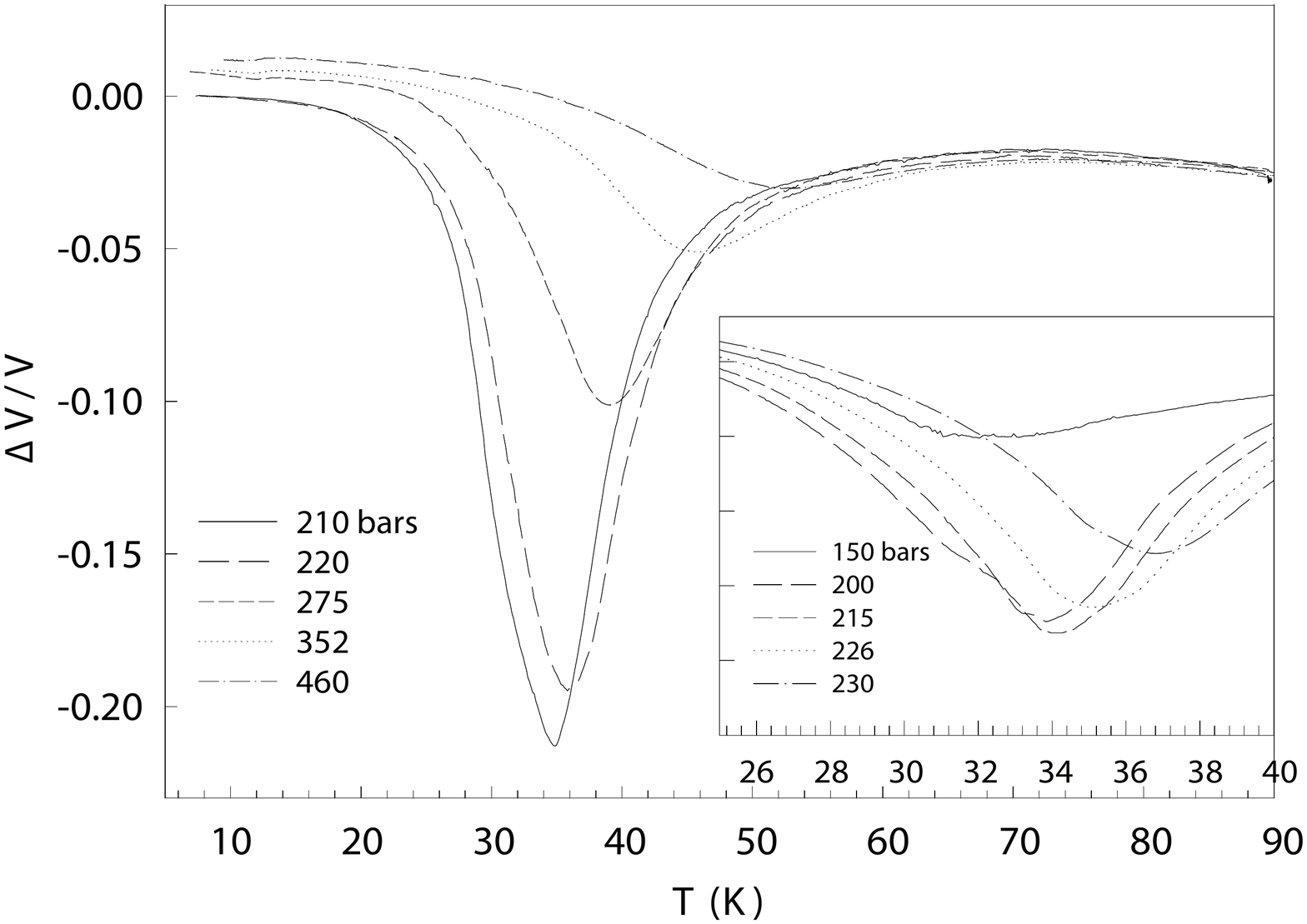}
\end{center}
\caption{\small Main panel: Relative \textcolor{black}{sound  velocity $\Delta V/V$ \emph{versus} $T$} under  pressure for the
$\kappa$-(ET)$_2$Cu[N(CN)$_2$]Cl salt. The various pressure values
are indicated in the label of the figure. Inset: Data for
pressures below 230\,bar. Picture extracted from
\Ref{Fournier2003-20}.}\label{Fournier1}
\end{figure}

\begin{figure}[!htb]
\begin{center}
\includegraphics[angle=0,width=0.50\textwidth]{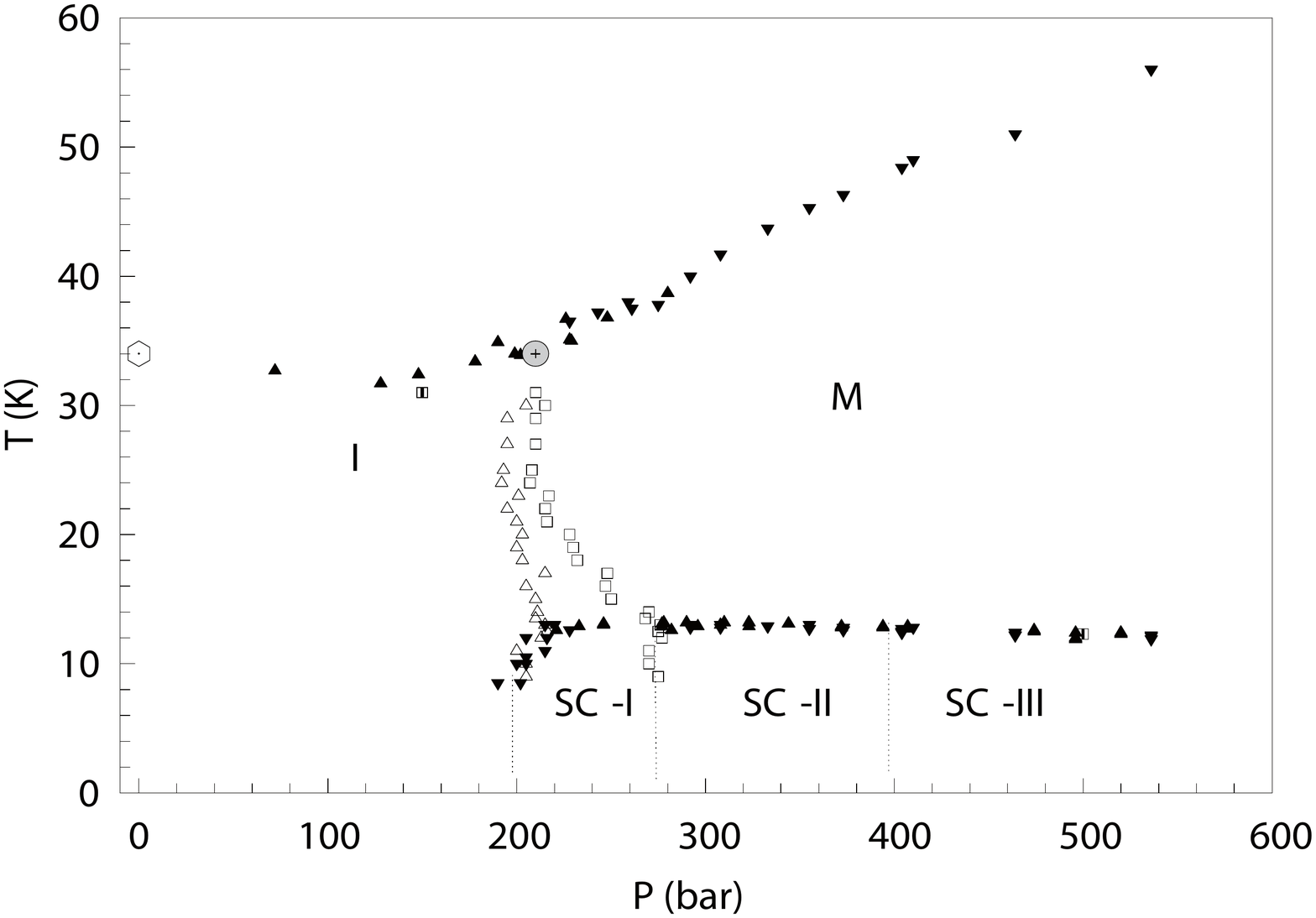}
\end{center}
\caption{\small Phase diagram obtained from ultrasound experiments
under pressure for the $\kappa$-(ET)$_2$Cu[N(CN)$_2$]Cl salt.
Different symbols refer to the various anomalies observed on three
different samples. The critical point ($P_0$, $T_0$) is indicated
by the gray circle. SC-I and -II indicate metastable
superconductivity, while SC-III indicates bulk superconductivity.
Dotted hexagon indicates the pressure point, obtained from
microwave resistivity measurement at ambient pressure. Picture
extracted from \Ref{Fournier2003-20}.}\label{Fournier2}
\end{figure}

The experimental findings obtained by Fournier \emph{et al.} are shown
in \Fig{Fournier1}. From this data set, pronounced anomalies,
depending on the pressure, are observed below around $40$\,K. The
maximum amplitude of the softening in the sound velocity (at $T\approx$
34\,K, $P\approx$ 210\,bar) marks the critical end point ($P_0$,
$T_0$) and corresponds to a softening of roughly 20\,\% of the sound
velocity.  On increasing the pressure, the peak position shifts its
position to higher temperatures. Below $P_0$ the anomaly becomes
gradually less pronounced and saturates at around 32\,K. The peak
position was thus used by the authors to draw the two
\emph{crossover} lines in the \textcolor{black}{$P$-$T$} phase diagram (see \Fig{Fournier2}).

\vspace*{0.4cm}

The large anomalies in the
sound velocity observed by Fournier \emph{et al.} were discussed in
terms of a diverging \textcolor{black}{electronic compressibility}. As a matter of fact, acoustic and lattice anomalies are,
according to dynamical mean-field theory (DMFT) calculations,
expected at the Mott MI transition as a reaction to the softening of
the \textcolor{black}{electronic channel}, cf.\,\Refs{Hassan2005-20,
Merino2000-20}. The explanation for this is that in the metallic
state the conduction electrons contribute more to the cohesion of
the solid than in the insulating one. Based on this argument,
according to \Ref{Hassan2005-20}, as an effect of the
localization, \textcolor{black}{abrupt changes of the lattice parameters are
expected at the Mott MI transition. We shall discuss the lattice effects for the present material
in Section\,\ref{TEMeasurementsonkappaD8}}.

\begin{figure}[!htb]
\begin{center}
\includegraphics[angle=0,width=0.50\textwidth]{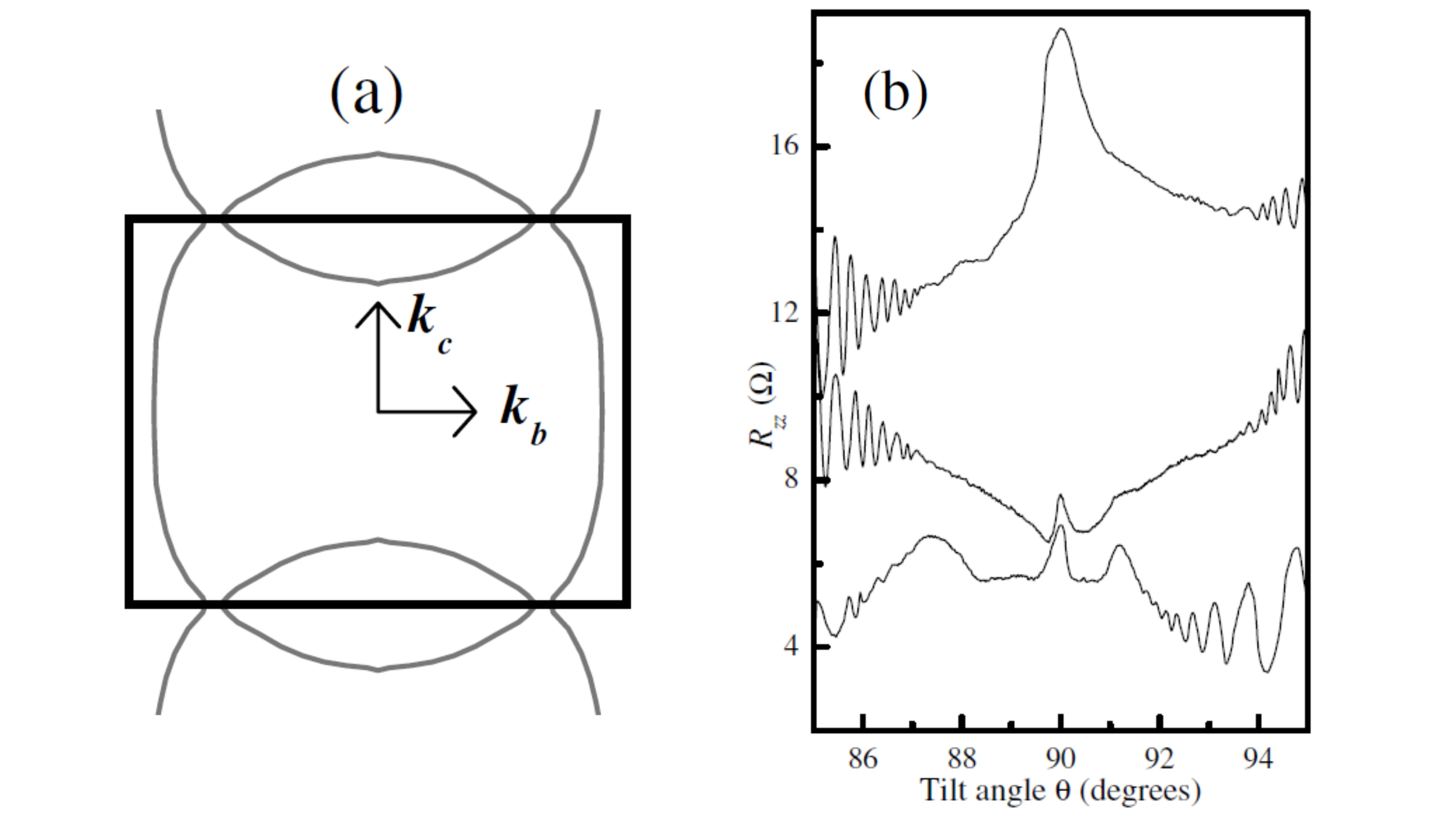}
\end{center}
\caption{\small {(a) Fermi surface cross section and Brillouin zone for $\kappa$-(ET)$_{2}$Cu(NCS)$_2$. (b) Interlayer magnetoresistance $R_{zz}$ for deuterated $\kappa$-(ET)$_{2}$Cu(NCS)$_2$ as a function of the tilt angle ($\theta \equiv$ angle formed between the applied magnetic field $B$ and the normal to the quasi-2D planes formed by the BEDT-TTF layers) taken under a static magnetic field $B$ = 45\,T at $T$ = 520\,mK. The central peak (around $\theta$ = 90$^0$) is associated with the closed orbits shown in (a), while the features in the edges are due to angle-dependent magnetoresistance oscillations (AMRO). Pictures taken from Ref.\cite{Single}.}}\label{FS}
\end{figure}

Regarding the Fermi surface, the weak interlayer transfer integrals were nicely estimated from magnetoresistance measurements for deuterated $\kappa$-(ET)$_{2}$Cu(NCS)$_2$ (Fig.\,\ref{FS}).

From a more fundamental point of view, the determination of the interlayer transfer integral in a quasi-2D system is quite helpful, for instance, to understand the validity of the criterion usually adopted to classify whether the interlayer charge transport in quasi-2D systems is incoherent or coherent.
For a detailed discussion, see e.g.\,Ref.\,\cite{Single} and references cited therein.
Thus charge-transfer salts of the $\kappa$-(ET)$_{2}$X phase serve as a good platform to investigate the fermiology of quasi-2D systems. Furthermore, knowing precisely the relevant parameters that govern the Mott Physics, theoretical approaches can be employed to understand the fundamental aspects of the Mott transition, see e.g.\,\cite{Harald}.

\vspace*{0.2cm}
\vspace*{0.2cm}

Owing to the magnetic properties of the $\kappa$-(ET)$_{2}$X
family, several studies employing different experimental methods
have been carried out on the salt with X = Cu[N(CN)$_2$]Cl
\cite{Welp20i, Pinteric20j,Miyagawa20k}.

By means of resistance measurements, a magnetic-field-induced Mott MI transition was
observed by Kagawa \emph{et al.}
\cite{Kagawa20l}
by varying temperature, pressure and
the magnetic field.
From an
analysis of their NMR line shape, relaxation rate and
magnetization data, Miyagawa \emph{et al.} \cite{Miyagawa20k}
were able to describe the spin structure of this state. Below
$T= 26$ -- $27$\,K, they found a commensurate antiferromagnetic ordering
with a moment of (0.4 -- 1.0)\,$\mu_B$/dimer, as already mentioned
above. The observation
of an abrupt jump in the magnetization curves for fields applied
perpendicular to the conducting layers, i.e.\ along the
\emph{b}-axis, was discussed in terms of a spin-flop (SF)
transition. Furthermore, a detailed discussion about the SF
transition, taking into account the Dzyaloshinskii-Moriya exchange
interaction, was presented by 
Smith \emph{et al.}
\cite{Dylan20m, Dylan20n}. Similarly to that for the pressurized
chlorine salt, a magnetic-field-induced MI transition was also
observed in partially deuterated
$\kappa$-(ET)$_{2}$Cu[N(CN)$_{2}$]Br \cite{Kawamoto20o}. In
addition, a discussion on the phase separation and SF transition
in $\kappa$-D8-Br was reported in the literature
\cite{Miyagawa20g}.\newline

As already mentioned above, the organic charge-transfer salt
$\kappa$-(ET)$_2$Cu$_2$(CN)$_3$
has the peculiarity that the ratio of its hopping matrix elements $t'$ to $t$
is close to unity, more precisely
 0.83 \cite{Kandpal09,Nakamura09},
leading to a strongly frustrated
isotropic $S$ = 1/2 triangular lattice
with the coupling constant $J$ = 250\,K, where this coupling constant is
obtained by fitting the magnetic susceptibility using the
triangular-lattice Heisenberg model. Magnetic susceptibility and NMR
measurements revealed no traces of long-range magnetic ordering down
to 32\,mK \cite{Shimzu2003}. Based on these results, this system has
been proposed to be a candidate for the realization of a spin-liquid
state \cite{Anderson1973}.

\begin{figure}[!htb]
\begin{center}
\includegraphics[angle=0,width=0.46\textwidth]{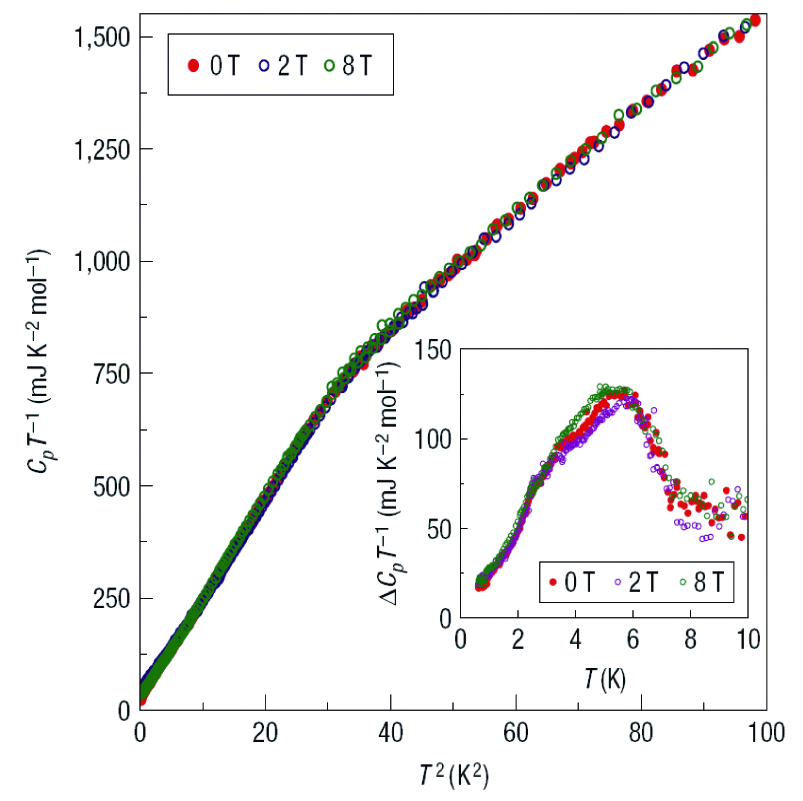}
\end{center}
\caption{\small Main panel: \textcolor{black}{Specific heat divided by $T$, i.e.\ $C_p/T$, data for the
$\kappa$-(ET)$_2$Cu$_2$(CN)$_3$ salt plotted against $T^2$
until 10\,K.} A broad hump anomaly around $T$ = 6\,K can be clearly
observed. Inset: Specific heat data for
$\kappa$-(ET)$_2$Cu$_2$(CN)$_3$,
assuming that the data of the superconductor
$\kappa$-(ET)$_2$Cu(NCS)$_2$ describe the phonon background.
Picture
extracted from \Ref{Yamashita2008}.}\label{yamshita4}
\end{figure}

Interestingly,
upon applying pressure (see
\Fig{SL-PD}), the system becomes a superconductor
\cite{Kurosaki2005}, i.e.\ superconductivity appears in the vicinity
of a spin-liquid state.
As shown in \Fig{yamshita4}, specific heat experiments revealed
the existence of a hump at 6\,K, insensitive to magnetic fields up
to 8\,T.

\begin{figure}[!htb]
\begin{center}
\includegraphics[angle=0,width=0.46\textwidth]{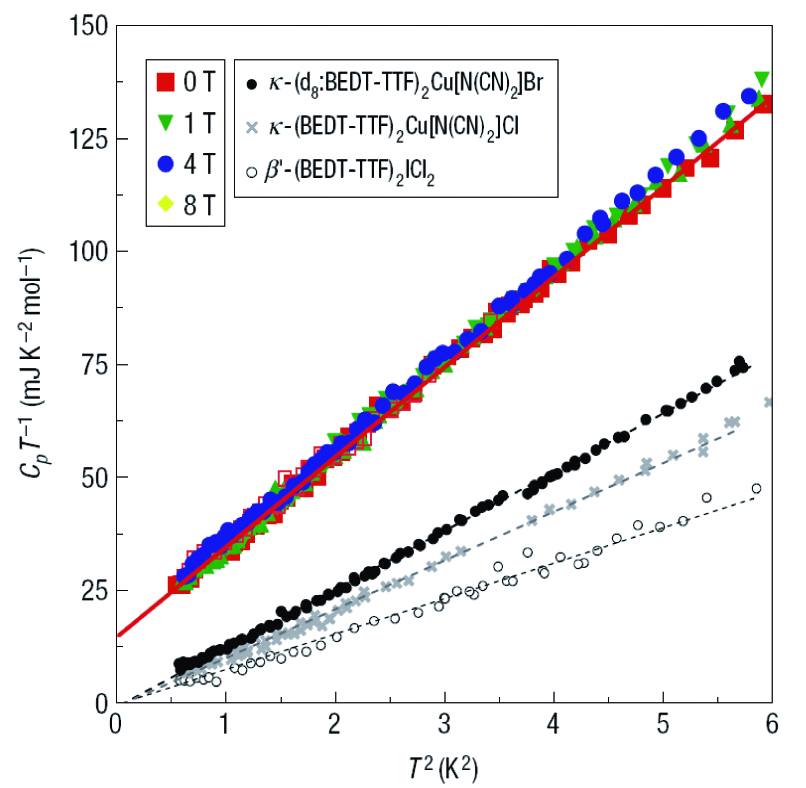}
\end{center}
\caption{\small Low-temperature specific heat data for the
$\kappa$-(ET)$_2$Cu$_2$(CN)$_3$ salt as well as
$\kappa$-D8-Br, $\kappa$-(ET)$_2$Cu[N(CN)$_2$]Cl and
$\beta'$-(ET)$_2$I\,Cl$_2$\textcolor{black}{, plotted is $C_p/T$ versus
$T^2$.}
The extrapolation of the red solid line towards zero temperature
indicates the existence of a $T$-linear contribution
for the spin-liquid \textcolor{black}{candidate} $\kappa$-(ET)$_2$Cu$_2$(CN)$_3$.
Data under magnetic fields for
$\kappa$-(ET)$_2$Cu$_2$(CN)$_3$ are also shown. Picture taken
from \Ref{Yamashita2008}.}\label{yamshita2b}
\end{figure}
This feature was assigned to a crossover from the paramagnetic Mott
insulating to the quantum spin-liquid state \cite{Yamashita2008}.
Such a ``crossover'' has been frequently referred to as ``hidden
ordering''. Below 6\,K, the specific heat presents a distinct
$T$-dependence, including a $T$-linear dependence in the temperature
window 0.3 -- 1.5\,K, as predicted theoretically for a spin-liquid
\cite{Anderson1987}. The spin entropy in this $T$-range is roughly
2.5\% of $R$\,ln\,2 \cite{Ramirez2008}, indicating that below 6\,K
only 2.5\% of the total spins contribute to the supposed
spin-liquid state. Extrapolating the \textcolor{black}{low-$T$} specific heat data, the
authors of \Ref{Yamashita2008}
found a linear specific heat coefficient $\gamma$ = (20
$\pm$ 5)\,mJ mol$^{-1}$ K$^{-2}$. \textcolor{black}{The latter} is sizable given the
insulating behavior of the material and contrasts with a vanishing
$\gamma$ value for the related compounds
$\kappa$-(ET)$_2$Cu[N(CN)$_2$]Cl and fully deuterated
$\kappa$-(ET)$_2$Cu[N(CN)$_2$]Br.
A finite $\gamma$ could indicate a spinon Fermi surface.

A gauge theory with an attractive interaction between spinons mediated via a non-compact $U(1)$ gauge field was proposed by Lee
\emph{et al.} \cite{Lee2007} to describe a possible spin-liquid
state in $\kappa$-(ET)$_2$Cu$_2$(CN)$_3$.
Within this model, a pairing
of spinons on the same side of the Fermi surface should occur
at low temperatures.
The attractive interaction between spinons with parallel momenta is quite analogous to  Amp\`ere's force law which states that two current carrying wires with parallel currents attract.
The authors deduce that
the pairing of spinons is accompanied by a
spontaneous breaking of the lattice symmetry, which in turn should
couple to a lattice distortion (analogously to the Spin-Peierls
transition) and should be detectable experimentally via X-ray
scattering.
 While Lee \emph{et al.} \cite{Lee2007} derive a specific heat which scales as $T^{2/3}$ and therefore contributes even stronger than linear at low $T$,
it should be kept in mind that the very-low temperature behavior of $\kappa$-(ET)$_2$Cu$_2$(CN)$_3$ is dominated by a diverging nuclear contribution to the specific heat.
\textcolor{black}{Indeed, the data set reported by Yamashita} \emph{et al.} \cite{Yamashita2008} is also reasonably well described by a $T^{2/3}$ behavior at intermediate temperatures \cite{Ramirez2008}.

Even though specific heat measurements by S.\ Yamashita \emph{et al.} \cite{Yamashita2008} are compatible with a spinon Fermi surface with gapless excitations, thermal conductivity measurements \textcolor{black}{reported by M.\ Yamashita \emph{et al.} in Ref.\,\cite{Yamashita09} show} a vanishing low-temperature limit of the thermal conductivity divided by temperature, indicating the absence of gapless excitations. The excitation gap was estimated to be $\Delta_\kappa \simeq 0.46$\,K.
Applying a theory based on $Z_2$ spin liquids,
Qi \emph{et al.} argued that thermal transport properties should be dominated by topological vison excitations \cite{Qi09}.  The gap found by M.\ Yamashita \emph{et al.} \cite{Yamashita09} should therefore be interpreted as a vison gap.
It was pointed out by Ramirez \cite{Ramirez2008} that before a final declaration of $\kappa$-(ET)$_2$Cu$_2$(CN)$_3$ as a spin-liquid is made
some points should be clarified, among them the anomaly observed
in the specific heat at 6\,K.
We will come back to the \textcolor{black}{system} $\kappa$-(ET)$_2$Cu$_2$(CN)$_3$ in Section \ref{SpinLiquidSection} where we will discuss its thermal expansion properties. \textcolor{black}{Regarding the fascinating properties of this salt, recently the existence of two magnetic field-induced quantum critical points was reported in the literature \cite{Blund}. Yet, based on a careful analysis of their muon spin rotation results, performed under extreme conditions, the authors of Ref.\,\cite{Blund} discuss a possible interpretation for the features observed in several physical quantities around 6\,K in terms of the formation of bosonic pairs emerging from a portion of the fermionic spins in $\kappa$-(ET)$_2$Cu$_2$(CN)$_3$.}

The remainder of this article is organized as follows:
In Section \ref{Experimental setup}, we discuss experimental aspects like sample preparation and high-resolution thermal measurements. Section \ref{TEMeasurementsonkappaD8} is dedicated to an exhaustive discussion on the thermal expansivity of fully deuterated salts of $\kappa$-(ET)$_2$Cu[N(CN)$_2$]Br, while in Section \ref{Criticality arount Tstar} the Mott criticality is comprehensively reviewed. Section \ref{SpinLiquidSection} is devoted to the discussion of the thermal expansivity of the spin-liquid candidate $\kappa$-(ET)$_2$Cu$_2$(CN)$_3$. Finally, conclusions and perspectives are presented in Section \ref{Conclusions and Perspectives}.

\section{Experimental setup}\label{Experimental setup}

\subsection{Sample Preparation}

As thermal expansion measurements under external hydrostatic pressure are extremely challenging, using compounds of different chemical composition currently represents the most practical way to probe the phase diagram of the $\kappa$-(ET)$_2$X family via dilatometry.
Deuterated (98\%) bis(ethylenedithiolo)-tetrathiofulvalene (D8-ET)
was \textcolor{black}{grown} according to Hartke {\it et al.} \cite{Hartke20a1} and
Mizumo {\it et al.} \cite{Mizuno20b}, \textcolor{black}{making the} reduction of carbon
disulfide with potassium in dimethylformamide and subsequent
reaction with deuterated (98\%) 1.2-dibromoethane. The intermediate
thione C$_5$D$_4$S$_5$ was \textcolor{black}{mixed with} triethylphosphite in an inert
atmosphere \textcolor{black}{kept at} $1200^{\circ}$C and recrystallized several times from
chlorobenzene. By investigating the CH$_2$ and CD$_2$
stretch vibration modes,
no signs of any CH$_2$ or CDH vibrations
could be detected, cf.\, \Ref{Gartner20c}. Hence, the grade of
deuteration of $\kappa$-(D8-ET)$_2$Cu[N(CN)$_2$]Br is at least
98\,\% \cite{Griesshaber20e}. As discussed in detail in
\Ref{Strack20d}, by employing the preparation technique
described above, hydrogenated single crystals of the same substance
resulted in samples whose resistivity revealed reduced inelastic
scattering and enhanced residual resistivity ratios. In general, the
crystals have bright surfaces in shapes of distorted hexagons with \textcolor{black}{typical}
dimensions of approximately 1 $\times$ 1 $\times$ 0.4 mm$^3$. Due to
the small size and fragility of the crystals, the experimental
challenge lies in assembling them in the dilatometer. The samples of
the deuterated and hydrogenated variants of
$\kappa$-(ET)$_2$Cu[N(CN)$_2$]Br
and $\kappa$-(ET)$_2$Cu$_2$(CN)$_3$ salts used for measurements
discussed in this work are listed in Table \ref{Amostras}. The fully
deuterated (hydrogenated) salts of $\kappa$-(ET)$_2$Cu[N(CN)$_2$]Br
will be subsequently referred to as $\kappa$-D8-Br ($\kappa$-H8-Br).
Samples of $\kappa$-(ET)$_{2}$Cu$_2$(CN)$_3$ studied here were
prepared according to \Ref{Geiser1991}.

\vspace{0.4cm}

\begin{table}[htb]
\centering \small
\begin{tabular}{|c|c|c|c|c|}
\hline\hline \textbf{Anion X} & \textbf{Crystal Number} & \textbf{Direction} & \textbf{Batch} \\
\hline\hline
  D8-Br & 1  & \emph{a}, \emph{b}, \emph{c} & A2907$^*$\\ \hline
  D8-Br & 2 & \emph{a}, \emph{b} & A2995$^*$ \\ \hline
  D8-Br & 3  & \emph{a}, \emph{b}, \emph{c} & A2907$^*$\\ \hline
  D8-Br & 4  & - & A2907$^*$\\ \hline
  H8-Br & 7  & \emph{b} & A2719$^*$\\ \hline
  H8-Cu$_2$(CN)$_3$ & 1 & \emph{a} & SKY 1050$^{**}$\\ \hline
  H8-Cu$_2$(CN)$_3$ & 1 & \emph{c} & KAF 5078$^{**}$\\ \hline
  H8-Cu$_2$(CN)$_3$ & 2 & \emph{b} & KAF 5078$^{**}$  
\\ \hline\hline
\end{tabular}
\caption{\small Samples of the organic conductor
$\kappa$-(ET)$_2$X on which thermal expansion measurements were
performed. $^*$ refers to samples provided by
D.\ Schweitzer (University of Stuttgart) and $^{**}$ to samples
provided by
J.\ Schlueter (Argonne National Laboratory). Crystal\,\#4 was used to perform preliminary studies of
Raman spectroscopy. In \Ref{Mariano20f1} crystal\,\#2 is
referred to as crystal \#3.}\label{Amostras}
\end{table}

\vspace{0.4cm}

 \subsection{Thermal Expansion Measurements}\label{TEMeassurements}
As already mentioned
in the introduction,
thermal expansion
coefficient measurements are a very useful and powerful tool to
explore phase transitions in solid state physics research
\cite{Barron9f, Mariano2008, Mott, Thesis, SL,
RSI,Bartosch,Lunke,Matej,Anton1,Anton2,MarianoP,Anton3,Haghi,Review2013,Field2012}.
\textcolor{black}{For instance, combining specific heat and thermal expansion data one can employ the Ehrenfest relation to determine the pressure dependence
of a second-order phase transition temperature. Such an analysis
may serve also as a check of its direct
measurement as a function of hydrostatic pressure. Furthermore,
contrasting with specific heat which is an isotropic property,
anisotropic effects may be studied via thermal expansion
measurements.}
Regarding the charge-transfer salts of the $\kappa$-phase, to our knowledge, the first high-resolution thermal expansion experiments along the three crystallographic directions were carried out about twenty years ago by Kund and collaborators \cite{Kund}.

The measurements presented in the present work
were textcolor{red}{carried out
by making use of an ultra-high-resolution capacitance dilatometric cell, built
after \Ref{Pott9e}, with a maximum resolution of $\Delta
l/l = 10^{-10}$, in the temperature range 1.3 (pumping of Helium bath)
to 200\,K under magnetic fields up to 10\,T.} In order to avoid
external vibrations, the cryostat is equipped with shock absorbers.
A detailed description of the dilatometer  used in this work has
been presented in several works \cite{LangPhD,Jens9f1,Andreas9h}.
The principle of measurement is described in the following. As
sketched in
\Fig{Cell for TE measurements},
the cell, which is entirely made of high-purity copper to ensure
good thermal conductivity and covered with gold is constituted
basically of a frame (brown line) and two parallel pistons (orange),
the upper one being movable. As a matter of fact, the gold layer
should work like a protection, avoiding oxidation of the cell.
Nevertheless, after some years in use, two striking anomalies have
been observed at $T$ $\simeq$ 212\,K and 230\,K. These anomalies are
assigned to the formation of copper-oxide (CuO) \cite{Loram1989} in
the body of the cell. Due to this, experimental data taken in this
$T$ window are not reliable.

\begin{figure}[!htb]
\begin{center}
\includegraphics[angle=0,width=0.49\textwidth]{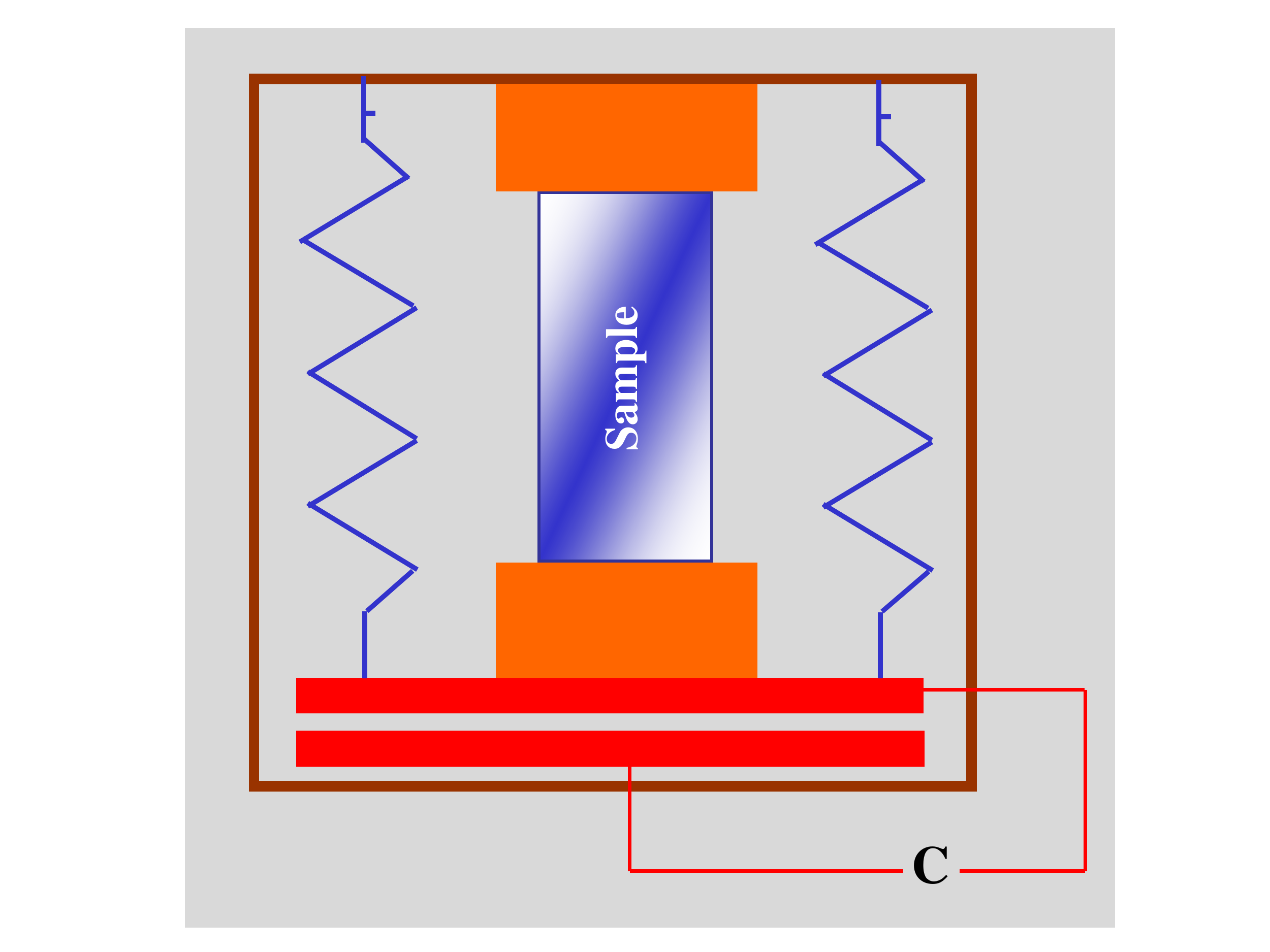}
\end{center}
\caption{Schematic representation of the cell
used for the thermal expansion measurements. Details are discussed
in the main text.
Picture adapted from \Ref{Thesis}.}
\label{Cell for TE
measurements}
\end{figure}

The sample is placed between these two pistons and by moving the
upper piston carefully, the starting capacitance is fixed. The
lower piston is mechanically linked to a parallel plate capacitor,
as schematically represented by springs in
\Fig{Cell for TE measurements}. The variation of the sample
length, i.e.\ contraction or expansion, as the temperature is
lowered or increased, together with cell effects corresponds exactly
to the change of the distance between the plates of the capacitor
and consequently to a change of the capacitance, so that very tiny
length changes can be detected. In this construction, the distance
between the plates of the capacitor is  about 100\,$\mu$m. \textcolor{black}{For the sake of avoiding stray electric fields, the parallel plate capacitor} is surrounded by guard
rings \cite{Pott9e}. The most remarkable characteristic in this
capacitance dilatometer, however, is its high resolution ($\Delta
l/l=10^{-10}$), corresponding to absolute length changes of $\Delta
l$ = 0.01\,\AA\,
\textcolor{black}{for a specimen of length 10\,mm}. This resolution
is roughly five orders of magnitude higher than that of conventional
methods like neutrons or X-ray diffraction \cite{Barron9f} and is
mainly due to the high resolution of the capacitance bridge and the
high quality of the plate capacitor in the dilatometer cell, which
allows the detection of very small changes, i.e.\,changes of about
10$^{-7}$ pF can be detected in the capacitance of the system. Two
different high precision capacitance bridges have been used in these
experiments: (i) Andeen Hagerling - Model 2500A 1\,kHz and (ii)
General Radio - Model 1616. However, the above-mentioned resolution
holds only until $T \simeq $ 40\,K, where a precise PID temperature
control becomes difficult due to the large time constant involved in
the experiment. As deduced in \Ref{Pott9e}, the sensitivity
of the measurements is proportional to the square of the starting
capacitance ($C^2$). This means that the higher the starting
capacitance, the higher the sensitivity.
However, \textcolor{black}{specimens of the molecular conductors investigated here are quite sensitive to the pressure
applied by the dilatometric cell} so that one cannot set a starting
capacitance too high since this would consequently lead to a break
in the sample. Hence, the starting capacitance for measurements
under normal pressure (see Subsection \ref{TEunderP}) was limited to
$\sim$18\,pF. Just for completeness, it is worth mentioning that the
empty capacitance of the system reads 16.7\,pF.

For measurements under magnetic fields, a magnet power supply (model
PS120-3) \textcolor{black}{supplied by} Oxford Instruments was used. In all performed
measurements under magnetic fields reported here the field was
applied parallel to the measured direction. For the
direction-dependent
thermal expansion measurements, the alignment
of the crystal orientation was made using an optical microscope and
 guaranteed with an error margin of $\pm 5^\circ$.
\textcolor{black}{To obtain the intrinsic thermal expansivity of the investigated specimen  the thermal expansivity of copper of
the dilatometric cell body was subtracted  from the raw data. Apart from that, no further treatments like splines or  any other
kind of mathematical fittings were done.}

The linear thermal expansion
coefficient $\alpha_i$ in the direction $i$ at constant $P$ is defined as
\vspace*{0.4cm}
\begin{equation}
\alpha_{i} = \frac{1}{\emph{l}} \Bigl(\frac{\partial l(T)}{\partial
T}\Bigl)_P , \label{linear coeficient}
\end{equation}
\vspace*{0.2cm}

\noindent
where \emph{l} \textcolor{black}{refers to the sample length.} The physical quantity described
by \Eq{linear coeficient} will be frequently used in this
work.
From the length changes of the sample $\Delta l(T)=l(T)-l(T_0)$
($T_0$ is a fixed temperature), which is the physical quantity
measured, the linear thermal expansion coefficient [\Eq{linear coeficient}]
was approximated by the differential quotient as

\begin{equation}
\alpha(T) \approx \frac{\Delta l(T_2)-\Delta l(T_1)}{l(300
\textrm{K})\times (T_2-T_1)} ,
\end{equation}

\vspace*{0.4cm}

\noindent
where $T = ( T_1 + T_2 )/2$, and $T_1$ and $T_2$ are two subsequent
temperatures of data collection.
Unless otherwise stated, in order to guarantee thermal equilibrium
in all thermal expansion measurements reported here, \textcolor{black}{a very low
temperature sweep rate ($\pm$\,1.5\,K/h) was employed.}

\subsection{Thermal Expansion Measurements Under Quasi-Uniaxial Pressure}\label{TEunderP}
One of the challenges in the context of thermodynamic measurements
is the realization of high-resolution thermal expansion measurements
under hydrostatic pressures, see e.g.\,\Ref{TEP}.

However, by changing the starting
capacitance and taking into account
the contact area between the dilatometer and the sample,
one can perform quasi-uniaxial thermal
expansion experiments under pressures up to $\sim 100$\,bar
employing the above-described setup, see e.g.~\Ref{Koeppen}. The term ``quasi-uniaxial'' is
used here in the sense that one may have a pressure-gradient on the
sample volume.
The dependence between
the force ($F$) exerted by the springs and the capacitance change
($\Delta C$) can be easily measured and is given by

\begin{equation}
\frac{\Delta F}{\Delta C} = 0.124 .
\frac{\mbox{N}}{\mbox{pF}}\label{TEpressure}
\end{equation}

\noindent
Making use of \Eq{TEpressure}, \textcolor{black}{the pressure applied by the
dilatometric cell on the sample can be estimated.} For example, taking
a contact area of
0.2 $\times$ 10$^{-6}$\,m$^2$ and a capacitance change of 2\,pF and
using the definition of pressure $P=F/A$ results in a
quasi-uniaxial pressure $P$ of roughly 12\,bar.

Hence, a slight change of the contact area and/or starting
capacitance results in a significant change of the pressure exerted
by the dilatometer on the sample. The crucial point of such
experiments lies in the assembling and disassembling of the sample
holder from the cryostat without damaging the sample, as such
procedure is necessary to set the new starting capacitance (new
pressure). \textcolor{black}{As depicted in \Fig{Fournier2},} for the
(BEDT-TTF) based charge-transfer salts a hydrostatic pressure of a
few hundred bar is enough to make a significant sweep across a wide
range of the pressure-temperature phase diagram. Measurements under
quasi-uniaxial pressure, to be discussed in Section\,\ref{Quasi-UP},
will show that a quasi-uniaxial  pressure of a few tenth of bar
changes the shape of the thermal expansion coefficient curves
dramatically, indicating therefore that the properties of this
material are altered.

\section{Thermal Expansion Measurements on Fully Deuterated Salts of $\kappa$-(ET)$_2$Cu[N(CN)$_2$]Br
  (``$\kappa$-D8-Br'')}
\label{TEMeasurementsonkappaD8}

\begin{figure}[!htb]
\begin{center}
\includegraphics[angle=0,width=0.48\textwidth]{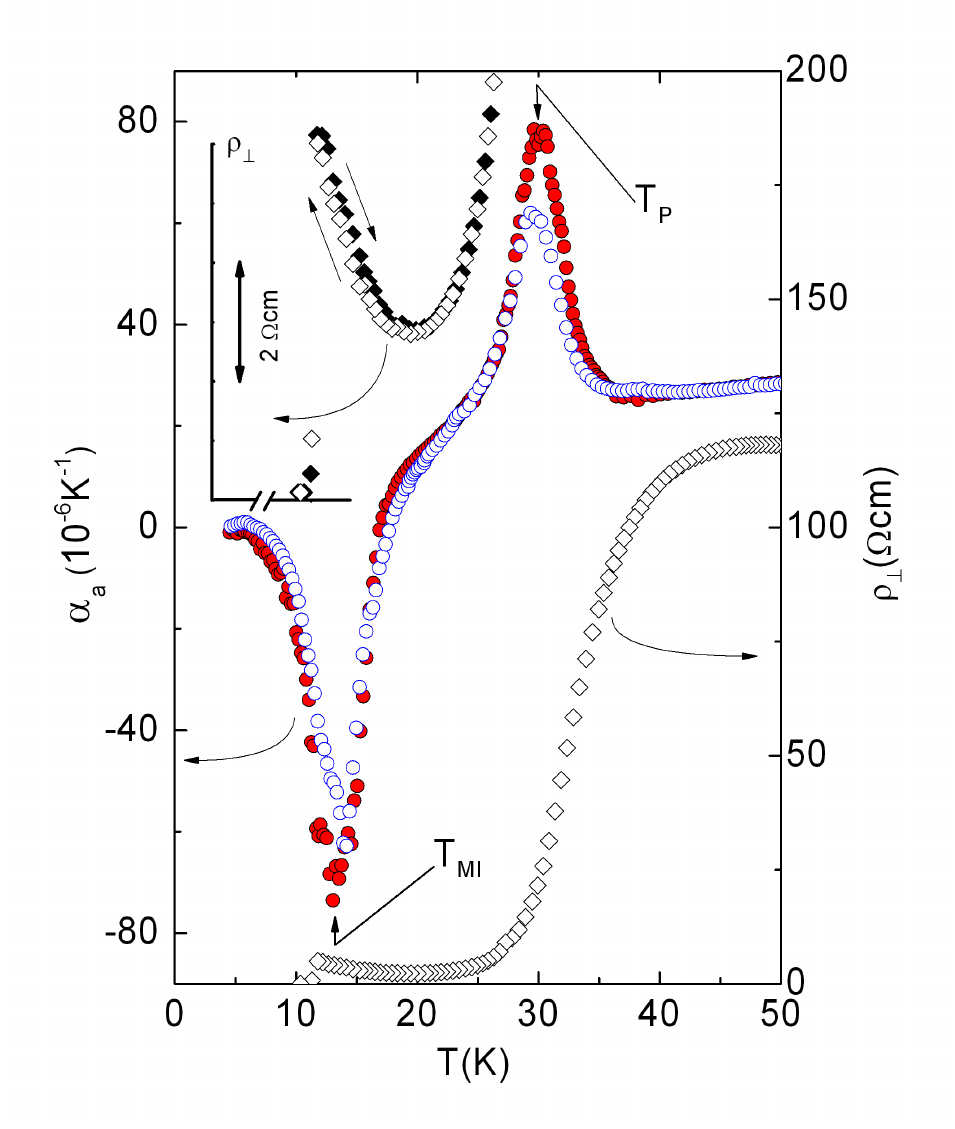}
\end{center}
\caption{\small \textcolor{black}{Left scale: thermal expansion coefficient along
the in-plane $a$-axis, $\alpha_{a}(T)$, for crystals \#1 (red) and
\#2 (blue); right scale: out-of-plane electrical resistivity,
$\rho_{\perp}$, for crystal \#1 of the $\kappa$-D8-Br salt. The upper inset depicts
a zoom of the low temperature $\rho_{\perp}(T)$ data employing the
temperature scale depicted in the main panel}. Picture taken from \Ref{Mariano20f1}.}
\label{K-d8-1.pdf}
\end{figure}

As otherwise stated, \textcolor{black}{in order to reduce cooling-speed dependent
effects, the temperature was decreased by a rate} of $\sim$\,$-$3\,K/h through the so-called
glass-like transition around 80\,K \cite{footnote1} in
all experiments reported here. \Fig{K-d8-1.pdf} \textcolor{black}{depicts the
linear thermal expansivity} (upper part)  along the
in-plane \emph{a}-axis for crystals\,\#1 and \#3 together with the
out-of-plane (\emph{b}-axis) electrical resistivity \cite{Strack} $\rho_{\perp}$
(lower part) for
crystal\,\#1. \textcolor{black}{As depicted in \Fig{K-d8-1.pdf}, upon cooling, $\rho_{\perp}$  achieves a broad maximum
centered at $\sim$45\,K, then abruptly drops and levels out at $\sim$30\,K.
The electrical resistivity remains metallic, i.e.\, d$R$/d$T$ $>$ 0, down to $\sim$20\,K, below which
$\rho_\perp$ increases again (cf.\,upper inset in
\Fig{K-d8-1.pdf}), indicating a phase transition
to an insulating
state. A comparable $\rho_{\perp}$ behavior was observed for the single
crystal\,\#3 as well as for the crystal investigated} in \Ref{Lang20f}, except
for small differences around the maximum and some details in the
insulating regime. Interestingly enough, for all three crystals studied, $\rho_{\perp}$ vanishes
below  \textcolor{black}{$\sim$11.5\,K. A zero electrical resistivity followed by a very small
feature in the thermal expansivity $\alpha_{i}(T)$ suggests percolative
superconductivity coexisting with an
antiferromagnetic insulating phase for $\kappa$-D8-Br}
\cite{Miyagawa20g}, cf.\,the phase diagram presented in \Fig{Phase
Diagram}. In \Ref{Kawamoto1997-20}, Kawamoto \emph{et al}.\
performed a.c. magnetic susceptibility measurements on $\kappa$-D8-Br \textcolor{black}{under two distinct conditions: (i)
cooling the system slowly, namely employing a cooling-rate of 0.2\,K/min; and (ii) using a high cooling-rate of 100\,K/min. From the
measurements carried out after slow cooling, below the onset of
superconductivity at $T_c$ = 11.5\,K, they did not observe a jump in
the real part of the a.c. magnetic susceptibility, as expected for bulk
superconductivity, but a gradual transition towards low
temperatures. From the latter behavior they estimated a percolative
superconducting volume fraction of 20\,\% . In the case of
measurements after rapid cooling, according to the authors, the
superconducting volume fraction is reduced to 1 -- 2\,\%.}

\vspace*{0.1cm}

\begin{figure}[h]
\begin{center}
\includegraphics[angle=0,width=0.46\textwidth]{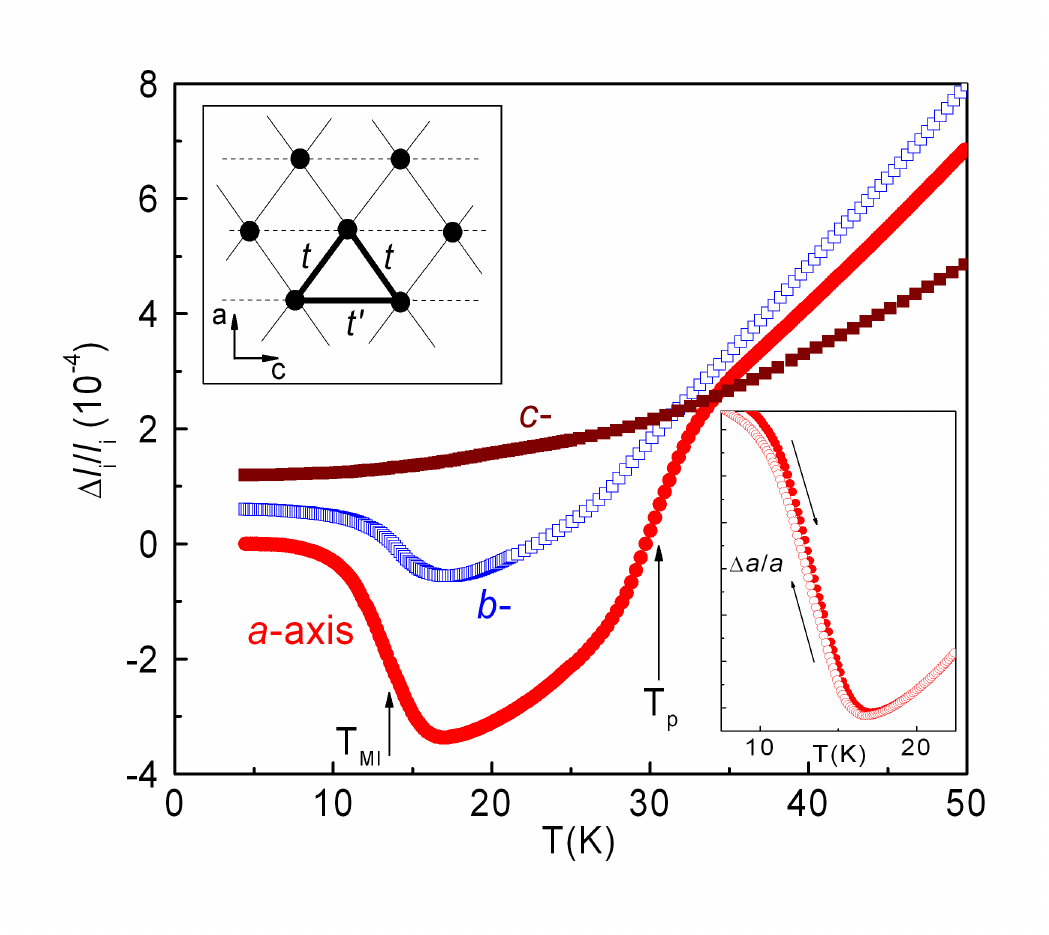}
\end{center}
\caption{\small \textcolor{black}{Main panel: relative length changes data $\Delta l_{i}(T)/l_{i}$, with $i = a, b, c$ for crystal
\#1 of the  $\kappa$-D8-Br salt along the in-plane $a$- and $c$- and out-of-plane $b$-axis.
For clarity, the data set have been offset. Lower inset: hysteresis in $\Delta l_{a}(T)/l_{a}$ associated with the Mott MI transition, measured under a very
low temperature sweep-rate of $\pm$1.5\,K/h. Upper inset: 2D
triangular-lattice dimer model with hopping terms $t$ and
$t'$, the dots represent dimers of ET
molecules.} Picture taken from \Ref{Mariano20f1}.}
\label{Lengthchanges.pdf}
\end{figure}

As can be seen in \Fig{K-d8-1.pdf}, some features observed in the electrical resistivity manifest themselves
in the coefficient of
thermal expansion as well.
Indeed, the rather abrupt drop of $\rho_{\perp}$ is
accompanied by a pronounced maximum in
$\alpha_a (T)$
centered at a
temperature $T_{p}$ $\simeq 30$\,K.
As will be discussed in more detail below, this behavior can be assigned to a
crossover phenomenon in the vicinity of a nearby critical point.
Upon cooling the system further, $\alpha(T)$ reveals a huge
negative peak, indicating
a phase transition,
see the phase diagram depicted in \Fig{Phase Diagram}.
The concomitant change in $\rho_{\perp}$ from metallic to insulating
behavior indicates
that the peak in the expansivity data, $\alpha _{a}(T)$, is due to the Mott MI transition \cite{C}.
Note that a similar behavior of $\alpha _{a}(T)$ is observed for crystal \#2,
albeit with somewhat reduced ($\sim 20\%$) peak anomalies at
$T_{p}$ and $T_{MI}$. This reduction might be related to small
misalignments of the crystal as well as to different pressures
exerted by the dilatometer cell on the crystals, approximately 4\,bar for
crystal \#1 and 6\,bar for crystal \#2. More information on the
nature of \textcolor{black}{such transitions can be obtained via investigation of possible anisotropic lattice effects, namely by measuring and analysing the
relative length changes along the three crystallographic directions, i.e.\ $\Delta l_{i}(T)/l_{i}$ = $(l_{i}(T)$ $-$
$l_{i}$(0\,K))/$l_{i}$(300 K)
(with $i = a, b, c$), as depicted in
\Fig{Lengthchanges.pdf}.}

\textcolor{black}{As can be seen in \Fig{Lengthchanges.pdf}, the most pronounced effects
appear along the $a$-axis (in-plane), i.e.\ parallel to the polymeric (Cu[N(CN)$_2$]Br)
anion chains, see \Fig{Anion}.}
Of special interest is the ``\emph{S}-shaped'' anomaly near $T_p$ which corresponds to the peak at $T_p$ in $\alpha_a(T)$ (see \Fig{K-d8-1.pdf}) and does not show any trace of hysteresis upon cooling or warming, at least within the resolution
of the present experiments.
This is
a generic feature of strong fluctuations in the vicinity of a critical end point.
Upon further cooling
through $T_{MI}$, the length of the crystal along
the $a$-axis shows a \textcolor{black}{distinct increase of
roughly $\Delta l_a/l_a = 3.5 \cdot 10^{-4}$
within a very limited $T$-window,
consistent with
a broadened first-order phase transition.
The assumption of a first-order phase transition is also supported by
the observation of a hysteresis around $T_{MI}$ of $\sim$0.4\,K, as depicted in the lower inset of
\Fig{Lengthchanges.pdf}, which nicely agrees with the hysteresis observed in
$\rho_{\perp}(T)$ (see upper inset of \Fig{K-d8-1.pdf}).
The associated anomalies observed
along the  $b$-axis (out-of-plane) are clearly less pronounced. Notably,
for the second in-plane direction ($c$-axis), anomalous behavior in
 $\Delta l_c/l_c$
can neither be detected at $T_{MI}$ nor at $T_{p}$}. The
reliability of the previous results is supported by the same
anisotropy observed for crystal\,\#3, investigated in
\Ref{Lang20f}. Based on the quasi-2D electronic structure of
the present material,
the observed anisotropic lattice effects at the
Mott \textcolor{black}{MI} transition are very intriguing
and deserve to be analyzed in more detail. The pronounced effects observed in the expansivity data along the \textcolor{black}{in-plane
\emph{a}-axis, along which dimer-dimer overlap is absent, agree
with the hypothesis that the diagonal hopping transfer integrals, $t$, play a crucial role in this process.
Considering that these transfer integrals carry a net
component along the in-plane \emph{c}-axis, which is expected to be softer
than the out-of-plane \emph{b}-axis and second in-plane
\emph{a}-axis, a noticeable effect along the \emph{c}-axis would
then be naturally expected at $T_{MI}$. According to the authors of
\Ref{Chasseau1991-20ga}, the uniaxial isothermal compressibilities
$k_i = l_i^{-1} (dl_i /dP)$  for the superconducting compound
$\kappa$-(ET)$_2$Cu(NCS)$_2$ are robustly anisotropic with $k_1$ :
$k_2$ : $k_3$ = 1 : 0.53 : 0.17 (ratio estimated under 1\,bar),
where $k_1$ refers to the isothermal compressibility along the in-plane
\emph{c} direction, $k_2$ along the second in-plane \emph{b}-axis
and $k_3$ along the direction parallel to a projection of the long
axis of the ET donor molecules, i.e.\ along the out-of-plane
\emph{a}-axis for this salt, since its crystallographic structure is monoclinic \cite{Taka}. Note that along the \emph{c}-axis a direct
dimer-dimer interaction $t'$, as indicated in the upper inset of
\Fig{Lengthchanges.pdf}, exists, making a zero-effect along
this axis even more surprising. A possible explanation for this feature
would be a cancellation of counteracting effects
related to} the transfer integrals $t$ and $t'$, which seems to be improbable.
Furthermore, it is not clear how these in-plane interactions may
produce comparatively strong effects in the out-of-plane
\emph{b}-axis.

\textcolor{black}{These findings provide strong evidence that other degrees of freedom, namely the lattice, should be somehow coupled to the $\pi$ electrons,
thus indicating that the Mott MI transition in the present material cannot be completely understood solely by taking into account a purely 2D electronic
scenario \cite{Mariano20f1}.}

\subsection{Anisotropic Lattice Effects in $\kappa$-D8-Br and
Rigid-Unit Modes}
\label{subsection:RUM}

\begin{figure}[!htb]
\begin{center}
\includegraphics[angle=0,width=0.49\textwidth]{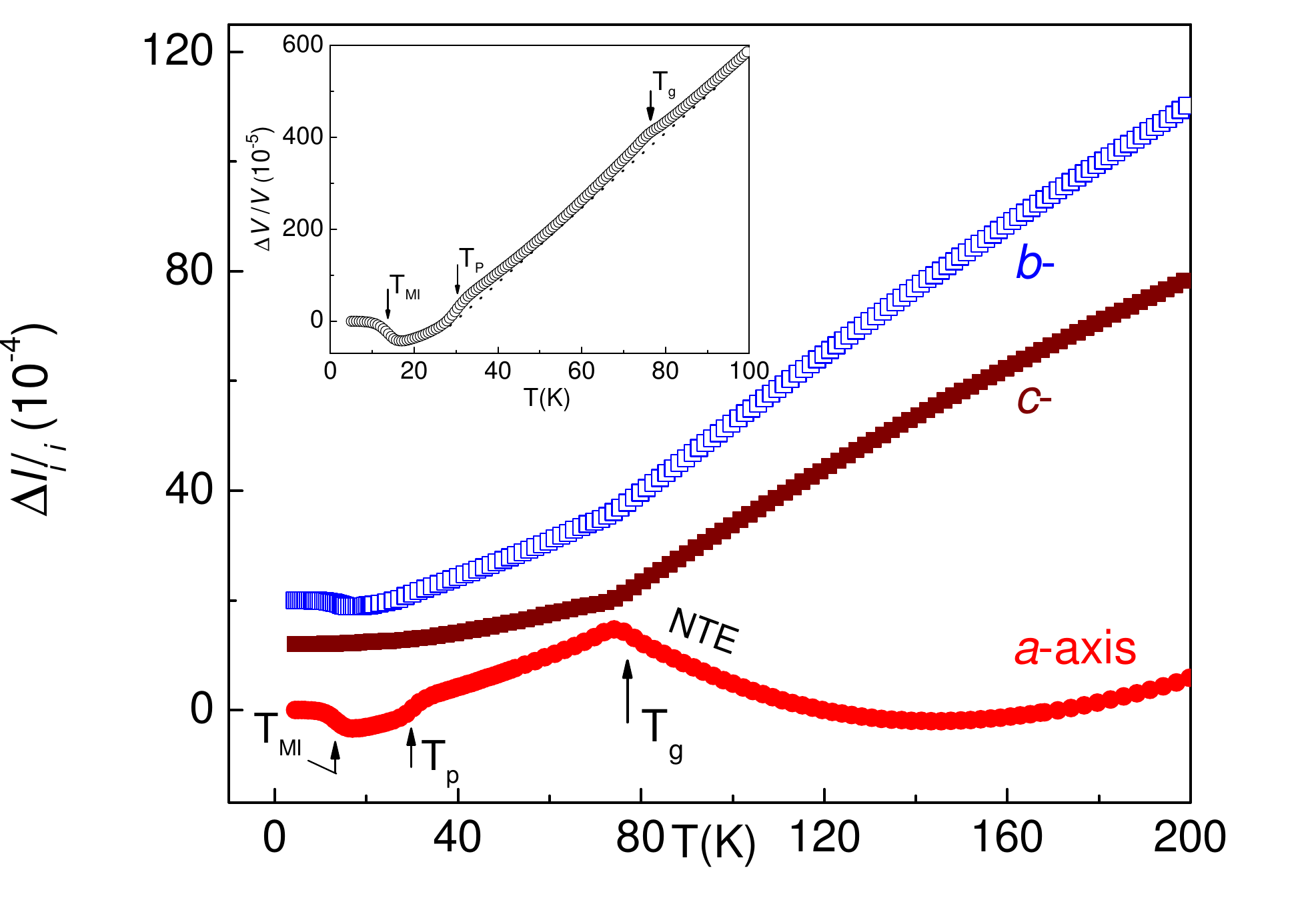}
\end{center}
\caption{\small Main panel: \textcolor{black}{Relative length changes along the
 \emph{a}- and \emph{c}-axis (in-plane), and
\emph{b}-axis (out-of-plane) for $\kappa$-D8-Br crystal\,\#1 up to 200\,K. Data
along the $b$- and $c$-axes are shifted for clarity. Position of
$T_g$ estimated according to \Ref{Jens2002-20}. NTE indicates
negative thermal
expansion along the \emph{a}-axis (in-plane) in the
$T$-range  $T_g$ $<$ $T$ $\lesssim$ 150\,K}. Inset: Relative volume
change
 $\Delta V/V = \Delta l_a/l_a + \Delta l_b/l_b + \Delta l_c/l_c$
versus
$T$. Low temperature data will be used for estimating the entropy
change associated with the Mott MI transition (Section
\ref{Entropy changes at the Mott transition}). Picture after \Ref{Thesis}.}
\label{TE_abc}
\end{figure}

\textcolor{black}{In \Fig{TE_abc} \cite{TEref} the relative length changes in the $T$-range 4.2\,K $< T < $ 200\,K for the
\emph{a}-, \emph{b}- and \emph{c}-axes are shown.} In the following, a possible
explanation for the unusual anisotropic lattice effects observed in
$\kappa$-D8-Br is presented.

\begin{figure}[!htb]
\begin{center}
\includegraphics[angle=0,width=0.46\textwidth]{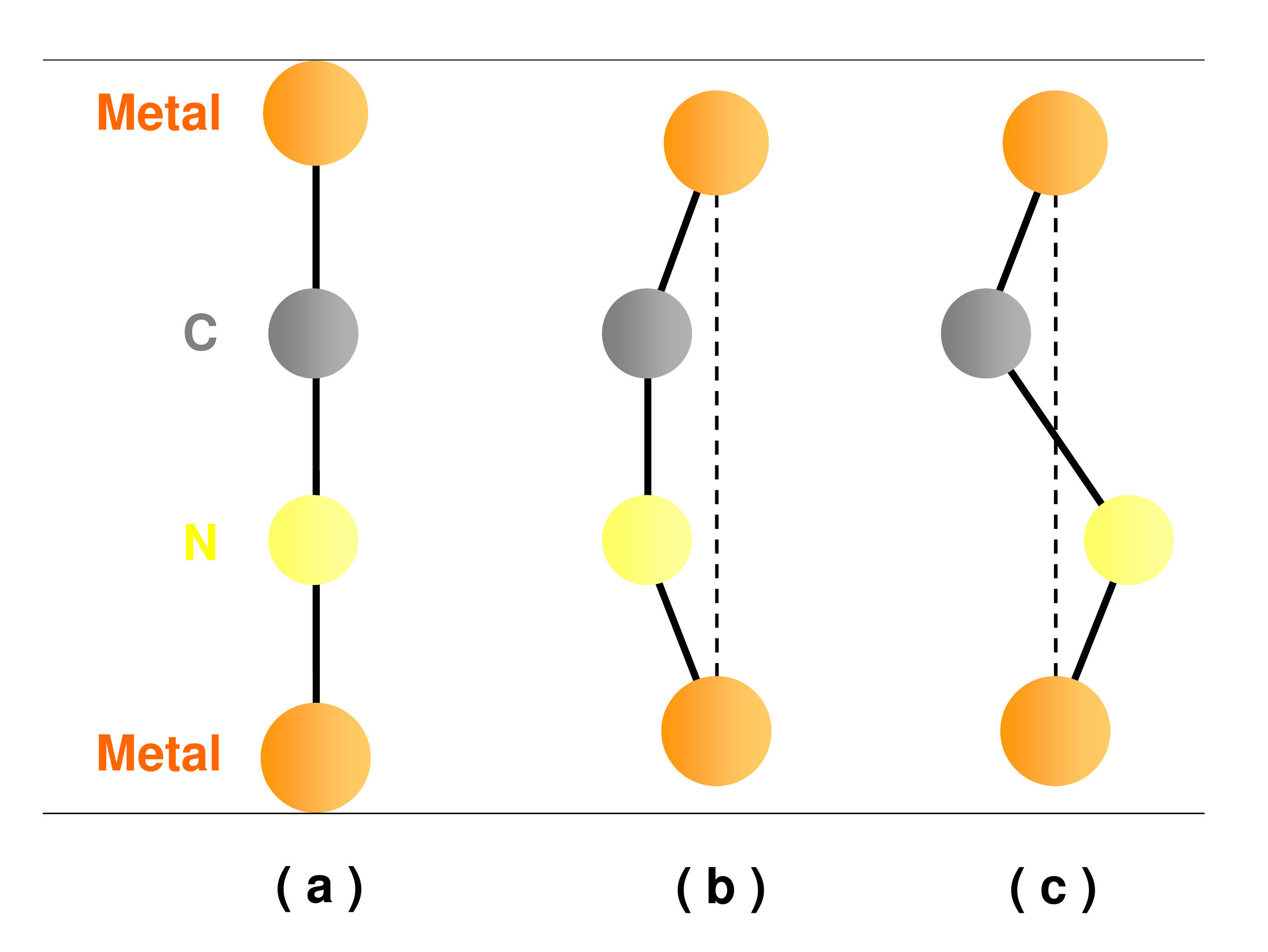}
\end{center}
\caption{\small Schematic representation of the M-CN-M linkage. a)
Linear configuration. b) \textcolor{black}{Shift of the carbon (C) and nitrogen (N) atoms aside
from the metal-metal axis in the same direction. c) Shift}
in the opposite direction. Note that the total length along the
metal-metal axis is reduced due to the displacement of the C-N
atoms. This is a mechanism giving rise
to negative thermal
expansion. Picture after \Ref{Goodwin2005}.}
\label{URM}
\end{figure}

Upon cooling, strongly anisotropic effects are observed in the whole temperature range.
The more pronounced effects are observed along the in-plane
\emph{a}-axis, which is marked by a minimum around 150\,K and an
abrupt change in slope at $T_g$. This feature will be discussed in
more detail below. \textcolor{black}{An interpretation in terms of disorder due to the
freezing-out of the ethylene groups of the ET molecules either in the staggered or in the eclipsed configuration was proposed by M\"uller \emph{et al.}
\cite{Jens2002-20}.} By substituting the terminal (CH$_2$)$_2$
groups by (CD$_2$)$_2$, one observes a shift of $T_g$ to higher
temperatures. A feature that is expected \textcolor{black}{because of the} higher mass of
the (CD$_2$)$_2$ in comparison to the mass of the (CH$_2$)$_2$
groups, which in turn implies a higher relaxation time at a given
temperature. Hence, the model proposed by M\"uller \emph{et al.}
is supported by the so-called isotope effect.
However, it does not
explain the \textcolor{black}{negative expansivity along the \emph{a}-axis (in-plane)} observed above $T_g$, namely in the interval $T_g$ $<$
$T$ $\lesssim$ 150\,K.
More recently, high-resolution synchrotron X-ray
diffraction experiments \cite{Wolter20g1} have provided indications
that such an anomaly is not solely linked to the \textcolor{black}{ethylene groups freezing-out. Essentially,  the} claims of the authors are based on the
comparison between the estimated number of the disordered ethylene
groups (staggered configuration) at 100\,K ($\sim$11 \%) with those
at extrapolated zero temperature (3 $\pm$ 3 \%). They conclude that
such a transition is most likely related to structural changes,
probably involving the insulating polymeric anion chains. Upon
further cooling, the anomalies at $T_p$ and $T_{MI}$, already
discussed in detail in the previous section, show up. From
\Fig{TE_abc} it becomes clear that the lattice effects at
$T_g$, $T_p$ and $T_{MI}$ are more pronounced along the polymeric
anion chain \emph{a}-axis, with negative thermal expansion (NTE),
i.e.\ a decrease of length with increasing temperature, observed
below $T_{MI}$ (due to charge carrier localization) and also in the
temperature window $T_g$ $<$ $T$ $\lesssim$ 150\,K. It should be
noted that the in-plane anisotropy at both $T_{MI}$ and $T_g$
corresponds to an orthorhombic distortion. Recently, NTE was also
observed by Goodwin \emph{et al.} in other materials containing
cyanide bridges in their structures, cf.\,\Refs{Goodwin2008-20g1a,
Goodwin2005} and references therein. These authors proposed a simple
model \cite{Goodwin2005} to explain the origin of this unusual
behavior.  In this model, the transverse vibrational displacement of
the CN units, away from the metal-metal axes (\Fig{URM} b) and
c)), is assigned to be the mechanism responsible for NTE in these
materials. As a matter of fact, NTE is also observable in other
materials. For example, the NTE of liquid water below $4^\circ$C is
connected with a breaking of the tetrahedral H bonding. Below
$4^\circ$C, NTE is required to overcompensate the increase of entropy
due to such a structural transition \cite{NTE}. Another example can be
found in ZrW$_2$O$_8$. This material exhibits NTE over a wide
temperature window of 50 -- 400\,K \cite{Mary1996}. The origin of
this anomalous behavior is assigned to its structural arrangement,
which is \textcolor{black}{built up by WO$_4$ tetrahedra and} rigid ZrO$_6$ octahedra.
Rigid-unit modes in this material show up due to the fact that each
WO$_4$ unit has one vertex not shared by another unit.

\textcolor{black}{The displacement of
the C and N atoms aside from the metal-metal axis has the effect of
decreasing the distance between the metal atoms as the temperature
is increased, cf.\,\Fig{URM} b) and c).} According to the authors of \Ref{Goodwin2005},
in polymeric crystalline materials, these vibrational modes will not
occur separately, but rather they will affect the vibration modes of
other metal-CN-metal linkages, giving rise to the NTE phenomenon.
Nevertheless, the deformation of the metal-cyanide bridge will carry
a high-energy penalty. Hence, assuming that the metal coordination
geometries are preserved, the vibration modes illustrated in
\Fig{URM} b) and c) can be observed only if they are coupled
with the crystal lattice in the form of phonon modes. Yet, according
to Goodwin \emph{et al.}, such modes are referred to as rigid-unit
modes (RUM) and can be experimentally observed in the low-energy
window  0 -- 2\,THz, i.e.\ in the range of typical phonon
frequencies \cite{Roth}. \textcolor{black}{Since the
Cu[N(CN)$_2$]Br$^-$ (\Fig{Anion})  counter anion has a polymeric nature}, the above-described model
is a good candidate to explain  why NTE  is observed only along the
in-plane anion chain \emph{a}-axis. The latter might have their
origin in the RUM of dicyanamide
[(NC)N(CN)]$^{-}$ ligands along the
anion polymeric chain. Note that the contraction of the lattice with
growing $T$ above $T_g$ along the \emph{a}-axis is accompanied by an
expansion along the \emph{b}- and \emph{c}-axes. In
addition, preliminary Raman studies (not shown here), carried out on
$\kappa$-D8-Br (crystal\,\#4),
revealed the appearance of a double peak structure in the frequency
window 50 -- 200\,cm$^{-1}$ at 20\,K (below $T_g$), not observed at
130\,K (above $T_g$, but below the onset of NTE) \cite{Raman}.
Further systematic
studies are required to verify if such splitting shows up above or
below $T_p$. Note that this frequency window fits roughly into the
above-mentioned energy window predicted for RUM. Interestingly
enough, the double peak structure is not observed at 5\,K (below
$T_{MI}$), indicating that vibration modes in this energy window are
no longer active below $T_{MI}$.
Hence, these results
support the idea of strong coupling between lattice and electronic
degrees of freedom at the Mott transition. Still owing to the
anomalous lattice effects above $T_g$, based on the above
discussion, \textcolor{black}{it seems that the freezing-out of the
degrees of freedom of the ethylene end-groups of the ET molecules in the staggered/eclipsed configuration alone
cannot be
responsible for a NTE along the \emph{a}-axis.}
Indeed, as pointed out
in \Ref{Jens2000-9c}, the glass-like anomaly is not observed
in the organic superconductors
$\beta''$-(ET)$_2$SF$_5$CH$_2$CF$_2$SO$_3$ ($T_c$ = 5\,K, large
discrete anion) and in $\kappa$-(ET)$_2$I$_3$ ($T_c$ = 3.5\,K,
linear anion) as well as in the nonsuperconducting
$\alpha$-(ET)$_2$KHg(SCN)$_4$ (polymeric anion). For such materials, a smooth
Debye-like behavior along the three crystallographic directions up
to 200\,K is observed, i.e.\ no glass-like transition is observed.
It is appropriate to mention here that the glass-like
anomaly is not observed in the $\kappa$-(ET)$_2$Cu$_2$(CN)$_3$ salt,
to be discussed in Section\,\ref{SpinLiquidSection}.
Thus, a possible scenario
to explain this feature
for the  $\kappa$-(ET)$_2$Cu[N(CN)$_2$]Z salts
would be the existence of a complicated
entwinement between the ethylene end groups and the vibration modes
of the polymeric anion chain, not knowingly reported as yet in the
literature. According to this idea, the ethylene end groups delimit
cavities, where the anions are trapped \cite{Geiser1991a}, so that
below $T_g$ an anion ordering transition occurs in a similar way to
the anion ordering transition observed in the quasi-1D
(TMT\emph{C}F)$_2$X charge-transfer salts \cite{Pouget96}. Further
systematic Raman and infra-red studies are necessary to provide more
information about the nature of the glass-like transition.

\subsection{Entropy Change Associated with the Mott MI
Transition} \label{Entropy changes at the Mott transition}
More information about the \textcolor{black}{Mott physics in the present materials can be obtained by estimating the amount of entropy linked to the Mott MI transition.
Such entropy change is directly associated with the slope of the first-order
phase transition line in the \emph{universal} $P$-$T$ phase diagram \Fig{Phase Diagram}} and the
volume change associated with the Mott MI transition by the
Clausius-Clapeyron equation, which reads
\vspace*{0.4cm}
\begin{equation}
\frac{d T_c}{d P} = \frac{\Delta V}{\Delta S} . \label{CCequation}
\end{equation}
\vspace*{0.4cm}

\noindent
\textcolor{black}{Here, $\Delta V = ( V_{I} - V_{M})$ ($\Delta S = (S_{I} - S_{M})$)
refer to the difference in volume (entropy)
between the insulating (I) and metallic (M) states.}
Employing $dT_{MI} /dP = (-2.7 \pm 0.1) $\,K/MPa, as obtained from the
slope of the \emph{S}-shaped line at $T_{MI}$ in \Fig{Phase
Diagram} and $\Delta V/V$ = $(4.2 \pm 0.5)$ $\times$ 10$^{-4}$,
 as estimated from the inset of \Fig{TE_abc},
one obtains $\Delta S =
-0.074$\,Jmol$^{-1}$K$^{-1}$. This small entropy change at $T_{MI}$
represents \textcolor{black}{a subtle fraction of the full entropy related to the
metallic state of the $\kappa$-H8-Br salt at $T \simeq$ 14\,K of $S
= \gamma \cdot T \simeq$ 0.375\,Jmol$^{-1}$K$^{-1}$, using the
Sommerfeld coefficient $\gamma$ = 0.025\,Jmol$^{-1}$K$^{-2}$
\cite{Elsinger20h}. Electronic specific heat data
by Y. Nakazawa \emph{et al.} \cite{Nakazawa20h1} revealed that by means of gradual
deuteration it is possible to tune the system from the Mott insulating state to the metallic regime of the phase diagram.}
As observed in these experiments, $\gamma$ decreases
towards zero in the insulating phase. Based on these literature
results, $\gamma$ of the fully deuterated salt
discussed in this work
should
actually be somewhat
smaller than that of the fully hydrogenated salt. It follows that the small
entropy change of $-$0.074\,Jmol$^{-1}$K$^{-1}$,
estimated as above-described,
demonstrates that the spin entropy of the paramagnetic insulating state
at elevated $T$ must be almost completely removed. In
addition, this result is consistent with the N\'eel temperature
$T_{N}$ coinciding with $T_{MI}$ at this point in the phase diagram
\cite{Lang20h2}.

\subsection{Magnetic Field Effects on
$\kappa$-D8-Br}\label{Field-Effects}

\begin{figure}[!htb]
\begin{center}
\includegraphics[angle=0,width=0.48\textwidth]{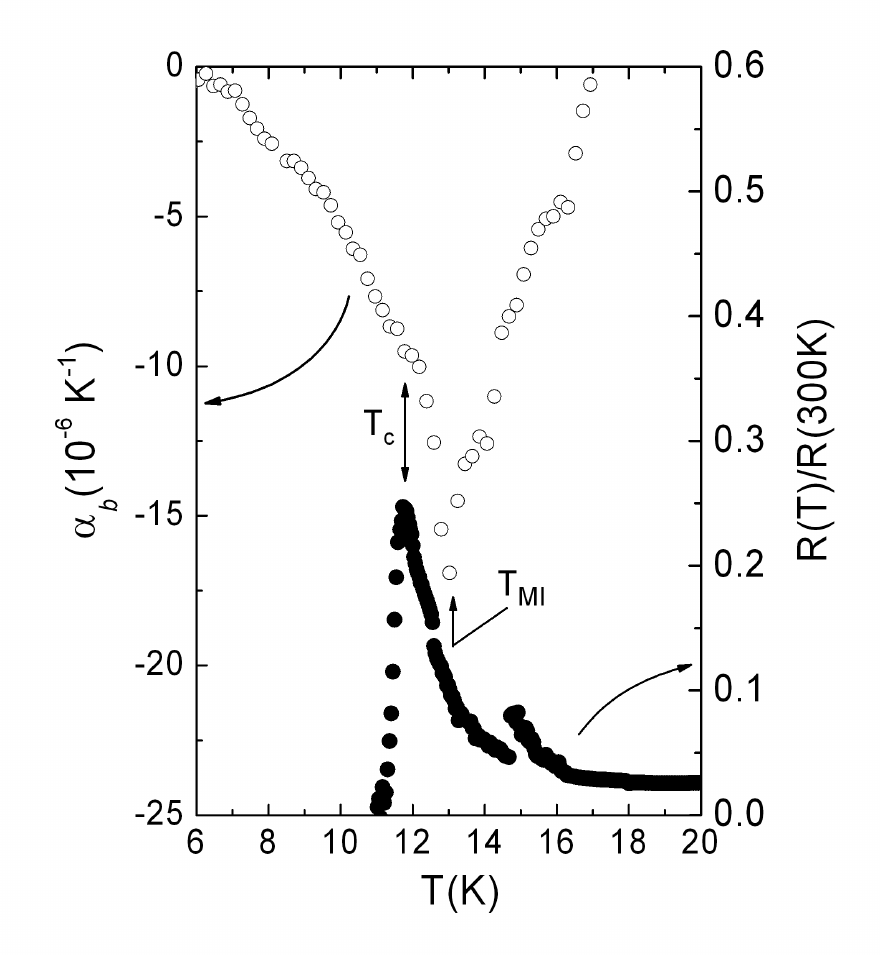}
\end{center}
\caption{\small \textcolor{black}{Left scale: thermal expansion coefficient along the
out-of-plane \emph{b}-axis. Right scale: electrical resistance data
normalized to its value at room temperature, being both data set for
$\kappa$-D8-Br crystal\,\#3. $T_{MI}$ denotes to the Mott MI transition
temperature and $T_c$ \textcolor{black}{refers to} the percolative SC critical
temperature.} Jumps in the resistance around 13\,K and 15\,K are most
likely related to the large expansivity of the material in this
temperature range, as discussed in the text. Picture taken from \Ref{Thesis}.}
\label{LowTdataRalpha}
\end{figure}

Having detected unusual features in the thermal expansivity of $\kappa$-D8-Br, additional information about the role played by
lattice degrees of freedom for the Mott MI transition in
$\kappa$-D8-Br can be obtained by carrying out thermal expansion
measurements under magnetic fields.
\textcolor{black}{In what follows we discuss the effects} of magnetic fields in the expansivity data at the immediate vicinity of the Mott transition.
However, before doing so,
let us consider first the low-temperature case at zero magnetic fields.
The thermal \textcolor{black}{expansivity along the \emph{b}-axis (out-of-plane) for $B$ = 0} is shown in \Fig{LowTdataRalpha} together with the electrical
resistance \cite{Strack} data normalized to its value at room temperature.
Upon cooling, a negative anomaly in the expansivity data at around $T$ = 13.6\,K
is observed.
The latter is linked to the Mott MI transition
and mimics the lattice response upon crossing the first-order \textcolor{black}{Mott MI} transition
line in the phase diagram, cf.\ the discussion in the previous sections.
Note that the
anomaly in $\alpha_b(T)$
is directly connected to an enhancement of
the electrical resistance around the same temperature, which corroborates the assumption that the
anomaly in the expansivity $\alpha_b(T)$ is
triggered by the Mott MI
transition.
The drop of the electrical resistance at $T_c$ = 11.6\,K,
accompanied by a very small kink (indicated by an arrow) in $\alpha_b(T)$,
are fingerprints of percolative SC in small portions of the sample
volume coexisting with the AF ordered insulating state, see
e.g.\,\Ref{Taniguchi20g2}.
In the temperature window 12\,K
$\lesssim$ $T$ $\lesssim$ 16\,K,
jumps in the electrical resistance are noticeable.
In fact, such jumps in $R$($T$) are
frequently observed in organic charge-transfer salts.
The huge expansivity of the crystal, which in turn is due to the softness of this class of materials, induces stress on the electrical
contacts, giving rise to such jumps. In general, the appearance of
jumps in resistivity data are attributed to cracks in the crystal. For the quasi one-dimensional (TMTTF)$_2$X and (TMTSF)$_2$X salts, in particular, such jumps in the resistivity data are assigned to stress caused by localized defects
and/or to mechanical twinning \cite{Ishiguro20}.

\begin{figure}[!htb]
\begin{center}
\includegraphics[angle=0,width=0.48\textwidth]{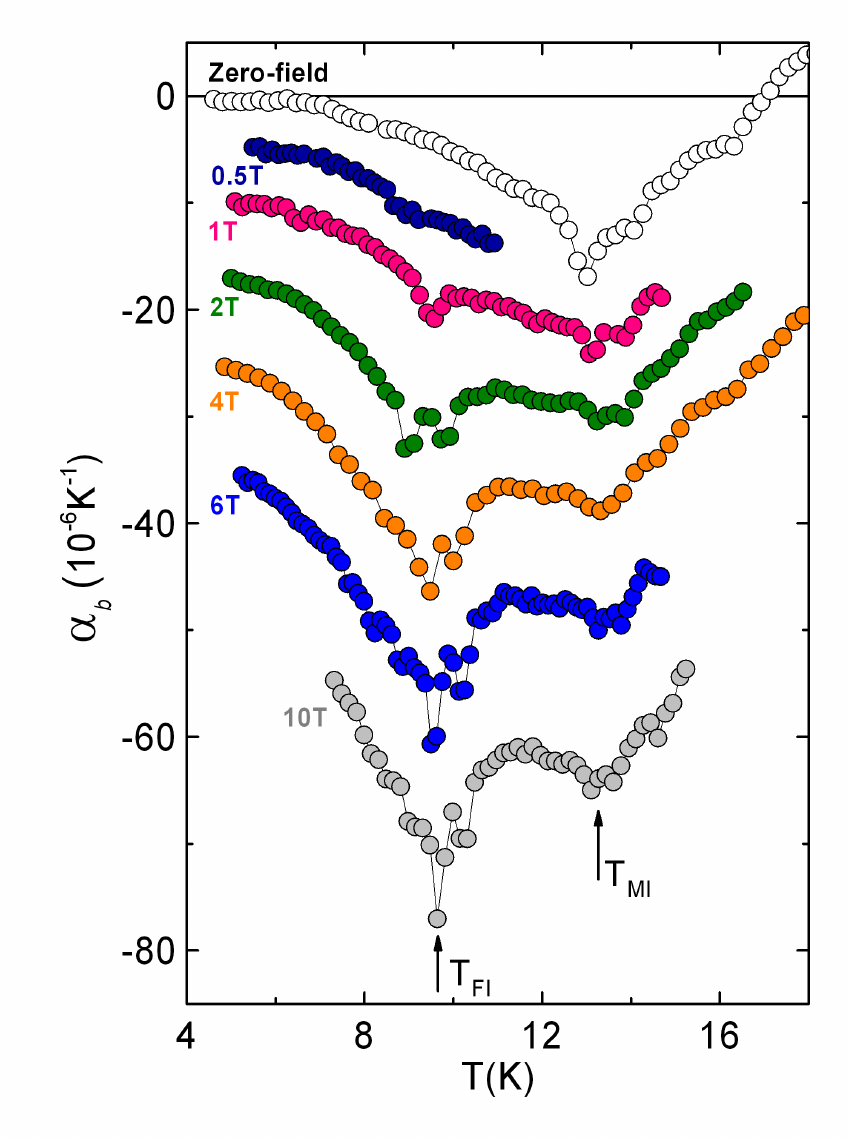}
\end{center}
\caption{\small Thermal expansion coefficient perpendicular to the
conducting layers under selected fields
for the $\kappa$-D8-Br crystal\,\#3. $T_{MI}$ refers to
the \textcolor{black}{Mott MI} transition temperature and $T_{FI}$ to the
field-induced phase transition temperature, as discussed in the
main text. Picture after \Ref{Field2012}.}
\label{Fig.3}
\end{figure}

The thermal \textcolor{black}{expansivity data along the
\emph{b}-axis (out-of-plane)  at low-$T$ under external magnetic fields of 0, 0.5, 1, 2, 4, 6, and 10\,T \cite{Field2012} is depicted in \Fig{Fig.3}.
From this data one can directly see that upon reducing $T$ under a magnetic field
up to 10\,T, the Mott MI transition temperature $T_{MI}$ stands essentially unaffected. Nevertheless, upon reducing $T$ further under a low
magnetic field of 0.5\,T,} a second \emph{field-induced}
anomaly at around $T_{FI}$ = 9.5\,K
is observed.
Increasing the magnetic field, this anomaly becomes more and more pronounced, until it saturates at around $\sim$4\,T. \textcolor{black}{A careful analysis of  the} data shown in \Fig{Fig.3}
reveals the presence of a double-peak structure for magnetic fields
higher than 1\,T. However, further experiments are necessary
to gain more insights into this feature. No such \textcolor{black}{magnetic
field-induced effects were observed along the in-plane} \emph{a}- and
\emph{c}-axis (not shown). As known from the literature, see e.g.\,\Ref{Blundell},
orientational field-dependent magnetic phenomena may indicate a
spin-flop (SF) transition. Such a transition takes place when the system
is in an AFI state and a magnetic field, exceeding a certain
critical value (critical field), is applied parallel to the so-called
easy-axis. In the present case, this field dependence of the thermal expansivity $\alpha_b(T)$
is \textcolor{black}{suggestive} of a SF transition with strong magneto-elastic coupling.
As can be directly seen from
\Fig{Fig.3}, the magnetic critical field $H_c$ is smaller than 0.5\,T
for $\kappa$-D8-Br, which is consistent with
$H_c$ $\sim$ 0.4\,T for the $\kappa$-(ET)$_2$Cu[N(CN)$_2$]Cl salt, as obtained by Miyagawa \emph{et al}. in \Ref{Miyagawa20k}. The physical explanation why the anomaly in the expansivity
$\alpha_b(T)$ at $T$ = $T_{FI}$ becomes more evident for
magnetic fields higher than $H_c$, however, cannot be \textcolor{black}{based} solely in terms of a SF transition, so that  a possibly different mechanism might be related to this feature.
It was pointed out by Ramirez \emph{et al.} in \Ref{Ramirez99} that the
emergence of pronounced anomalies under finite magnetic fields in thermodynamic quantities, like the specific heat or thermal expansion, have only been observed in a few systems and can be considered as unusual physical phenomena.
Let us discuss two possible distinct scenarios to explain this
unusual feature:

\emph{i}) \textcolor{black}{First of all, electrical resistance measurements carried out on partially  deuterated $\kappa$-(ET)${_2}$Cu[N(CN)$_{2}$]Br salts under magnetic field sweeps, being the magnetic field applied along the out-of-plane \emph{b}-axis and the temperature fixed at $T$ = 5.50\,K and 4.15\,K for 50\%
and 75\% salts, respectively, unveiled a sudden change from SC to high resistive-states for magnetic fields of about 1\,T \cite{Taniguchi1,Field2012}.
Such a state was found to
transform into a high-resistive state via jump-like increases in the
electrical resistance upon further increasing the applied magnetic field to 10\,T
\cite{Taniguchi1}. The authors
of \Ref{Taniguchi1} interpreted this behavior as a magnetic field-induced first-order SC-to-Insulator transition and related it to the theoretical description based on the SO(5) symmetry for
superconductivity and antiferromagnetism, as proposed for the high-$T_c$ cuprates by Zhang
\cite{Zhang}. Moreover, upon tuning the $\kappa$-(ET)${_2}$Cu[N(CN)$_{2}$]Cl salt  close to the
Mott MI transition by hydrostatic pressure \cite{Kagawa2004-20} a pronounced
$B$-induced increase in the electrical resistance was observed also for this
salt. In the latter case, it was proposed that minor metallic/superconducting phases near
the Mott MI transition undergo a magnetic field-induced localization transition
cf.\ theoretical predictions, see, e.g.\,
\Ref{Georges96} and references cited therein. Thus, based
on these findings, it is natural to expect that also for the fully deuterated $\kappa$-(ET)${_2}$Cu[N(CN)$_{2}$]Br
salt, located on the verge of the Mott MI transition line in the phase diagram, the electronic states may undergo drastic changes upon
increasing the strength of the applied magnetic field. Based on the above discussion,  the growth of the anomaly in $\alpha_b$(\emph{T}) with increasing fields $H$ $>$ $H_c$ may be related with the sensitivity of the electronic channel to applied magnetic fields in this particular region of the phase diagram. In fact,
the initial rapid growth and the tendency to saturation for field above about
4\,T is quite similar to the evolution of the electrical resistance, i.e.\, the
increase of $R(B, T = const.)$ with field observed in the
above-mentioned transport studies \cite{Taniguchi1,Kagawa2004-20}. It is worth mentioning that no such magnetic field-induced effects were observed for fields applied parallel
to the in-plane \emph{a}- and \emph{c}-axis (not shown), supporting thus the hypothesis of a SF transition.
Indeed, it was pointed out in \Ref{Dylan1} that the type of the
inter-layer magnetic ordering strongly depends on the direction of the applied
magnetic field. In particular, based on a detailed analysis of NMR
and magnetization data, taking into account the
Dzyaloshinskii-Moriya interaction, the atuhors of Refs.\,\cite{Taniguchi1,Kagawa2004-20} found that inter-plane antiferromagnetic
ordering  can be observed \emph{only} for magnetic
fields exceeding $H_c$ applied along the out-of-plane \emph{b}-axis. We stress that the term \emph{"inter-plane afm
ordering"} is used here as defined in Figs.\,4 and 5 of
\Ref{Dylan1}. Thus, given the absence of lattice effects at
$T_{FI}$ for magnetic fields applied along the in-plane \emph{a}- and
\emph{c}-axis, the present results suggest a close relation between
inter-plane antiferromagnetic ordering and the lattice response
observed at $T_{FI}$.
Employing the same
notation used by the authors of
\Ref{Dylan1}, the magnetization of the $+$($-$) sublattice at
the layer $l$ is $\textbf{M}_{+(-)l}$., being the staggered and
ferromagnetic moments given by $\textbf{M}^\dagger_l$ =
($\textbf{M}_{+l}$ $-$ $\textbf{M}_{-l}$)/2 and $\textbf{M}^F_l$ =
($\textbf{M}_{+l}$ $+$ $\textbf{M}_{-l}$)/2, respectively. For
fields above $H_c$ applied along the out-of-plane $b$-axis, $\textbf{M}^F_l$ is
along the $b$-axis and $\textbf{M}^\dagger_l$ lies in the $a$-$c$ plane.
$\textbf{M}^\dagger_A$ and $\textbf{M}^\dagger_B$ are antiparallel,
giving rise to an inter-plane antiferromagnetic ordering.
The pronounced negative peak anomaly in
$\alpha_b(T)$ observed at $T$ = $T_{FI}$ suggests that in order to obtain this particular spin
configuration, the increase in exchange energy forces the (ET)$^+_2$ layers
to move apart from each other.}

\emph{ii}) A possible second
interpretation for the sharp peaks in $\alpha_b (T)$
induced by a
magnetic field is that upon exceeding the critical field $H_c$,
percolative superconductivity is destroyed, the remaining electron
spins do not couple with the applied magnetic field due to the
correlated motion among them. These field-decoupled spins are
strongly coupled to the lattice and give rise to the minimum at
$T_{FI}$. The term field-decoupled spins is used in
\Ref{Ramirez99} to refer to a similar effect (double peak
structure) observed in specific heat measurements \textcolor{black}{under  magnetic
field for the ``spin ice'' compound Dy$_2$Ti$_2$O$_7$. According to
the authors of \Ref{Ramirez99}, for magnetic fields
applied parallel to the [100] crystallographic direction, due to
correlated motion among the spins,  half the spins have their Ising-axis} orientated perpendicular to the magnetic field. The latter are
called decoupled-field spins. It was observed that magnetic fields
exceeding a certain critical value lead to the ordering of these
field-decoupled spins. Monte Carlo calculations, also reported by
the authors in \Ref{Ramirez99}, support the proposed model.
The term ``spin ice'' is used by the authors to refer to the spin
orientations in analogy with the degeneracy of ground states
observed in ice (H$_2$O in solid phase), where hydrogen atoms are
highly disordered and give rise to a finite entropy as $T$
$\rightarrow$ 0. \textcolor{black}{This is because the oxygen atoms in water form a well
defined structure, while the hydrogen atoms remain disordered
as a result of the two inequivalent O-H bond lengths, as first pointed by
Pauling \cite{Pauling}.} The present results do not enable us to
determine the exact orientation of these field-decoupled spins.

\begin{figure}[!htb]
\begin{center}
\includegraphics[angle=0,width=0.48\textwidth]{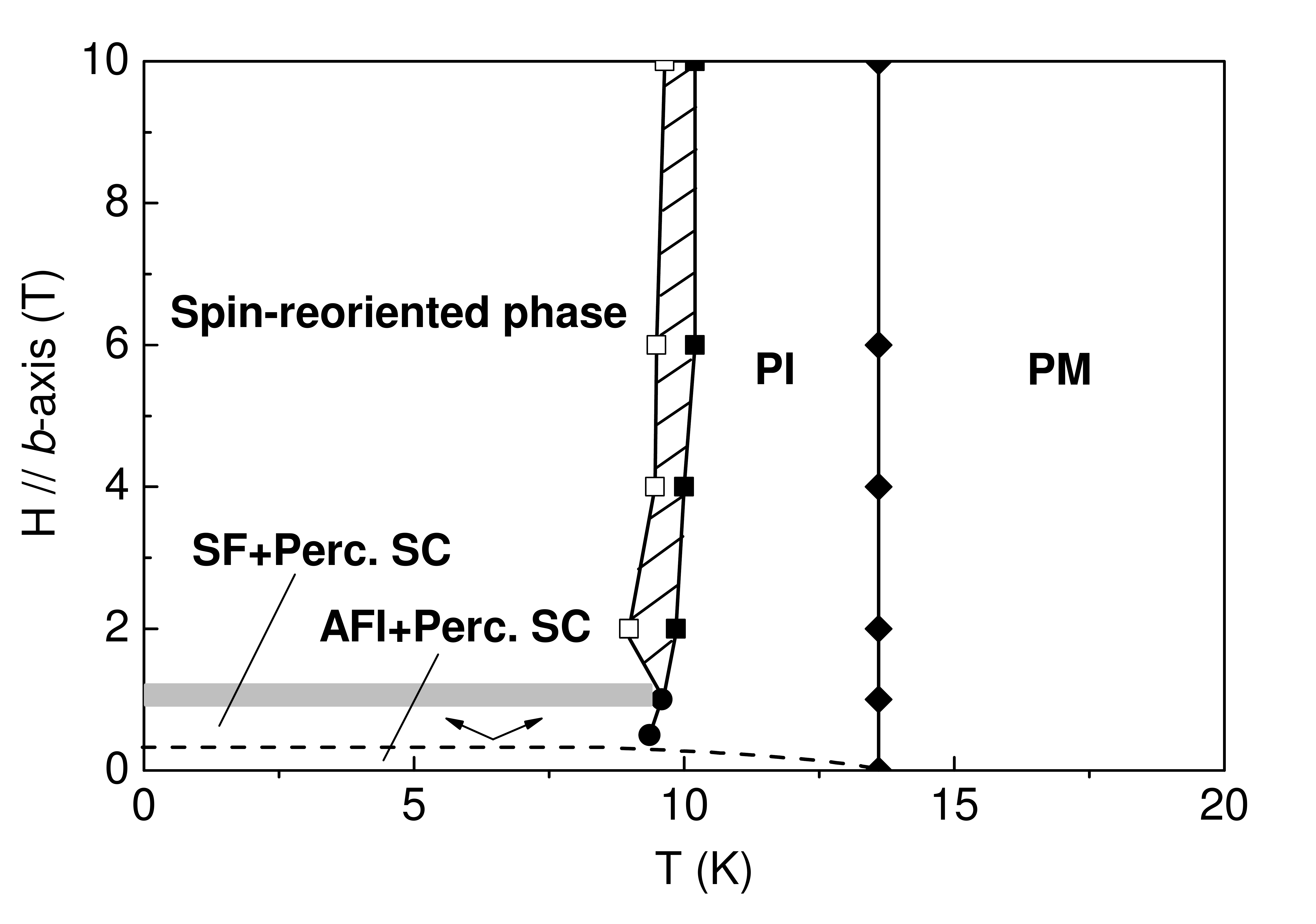}
\end{center}
\caption{\textcolor{black}{Schematic \emph{$H$}-\emph{$T$} diagram for $\kappa$-D8-Br for a magnetic field applied along the
out-of-plane \emph{b}-axis. Symbols indicate the peak anomaly observed in the expansivity $\alpha_b(T)$, while lines refer to
the various phase transitions, cf.\ the discussion in the main text. PI and PM
stand, respectively, for the paramagnetic insulator and the paramagnetic metal.
The hatched area marks the temperature interval in which a double-peak structure
in $\alpha_b(T)$ is observed. Spin-reoriented phase denotes the
region where $\textbf{M}^\dagger_A$ and $\textbf{M}^\dagger_B$ are
antiparallel, giving way to an inter-plane antiferromagnetic ordering, cf.\ the discussion in the main text.
Picture after \Refs{Thesis, Field2012}.}
}\label{Fig.4}
\end{figure}

It is worth mentioning that the above-discussed scenarios should be seen as
``possible scenarios'' and that the concrete physical origin of the
features observed at $T_{FI}$ remain elusive.
Another particularity in the data presented in \Fig{Fig.3} is that for
magnetic fields $H$ $>$ $H_c$, in a
temperature range between $T_{MI}$ and $T_{FI}$, an intermediary
phase appears, probably paramagnetic, as depicted in the schematic
phase diagram in \Fig{Fig.4}.
It is tempting  to speculate
that this feature can be seen as a separation of the $T_N$ and
$T_{MI}$ lines in the phase diagram. A possible physical description
to this would be related to the energetic competition between the
AFI ordered phase
and the paramagnetic phases after crossing the MI
first-order line, resulting thus in a decrease of $T_N$ with
increasing magnetic field.  Similar experiments were \textcolor{black}{performed
in two other $\kappa$-D8-Br crystals (crystals\,\#1 and \,\#2). However, for
both crystals,  in contrast to the pronounced anomaly observed at $T_{FI}$
$\approx$ 9.5\,K under magnetic fields (\Fig{Fig.3}),
the application of magnetic fields results in a smooth change of the out-of-plane expansivity
$\alpha_b(T)$ around the same temperature. In this regard, two factors should be
considered as a possible explanation for the absence of the above-described effects:}

i) As described in e.g.\,\Ref{Pankhurst20g4}, the SF transitions are
very sensitive to the alignment between the applied magnetic field and
the easy-axis. A subtle misalignment between the magnetic field and the easy-axis can therefore give rise to a suppression of the transition;

ii) The absence of a sharp transition can also be due to sample
inhomogeneities and defects.
Sample inhomogeneities would imply, for instance, that portions of
percolative SC may vary from sample to sample, reflecting therefore
differences in their magnetic properties \cite{Kawamoto1997-20}.

The present findings are compiled in the schematic  $T$ versus $H$ diagram
shown in \Fig{Fig.4}. The dashed line around $\sim$0.5\,T
defines the separation boundary between the AFI and the SF phases. The thick line indicates
the suppression of percolative SC giving place to
field-decoupled and/or flopped spins, here referred to as mixed states.

Concluding this section, the thermal expansion results on
$\kappa$-D8-Br along the out-of-plane \emph{b}-axis under a magnetic field show that the Mott MI transition in this salt is largely field-independent under magnetic fields up to at least 10\,T. Such findings are consistent with the picture of a Mott insulator formed by a set of
hole localized in dimers formed by two ET molecules \cite{Lang4}.
At $T_{FI}$ = 9.5\,K a phase transition induced by a magnetic field is observed, indicative of a SF transition with strong magneto-elastic coupling, accompanied by an enhancement of $\alpha_b(T)$ due to the suppression of the percolative SC under
magnetic fields above 1\,T. Further experiments, like
magnetostriction measurements above and below $T_{FI}$ are highly desired to
shed more light on the above-discussed magnetic field-induced lattice effects.

\subsection{Thermal Expansion Under Quasi-Uniaxial
Pressure}\label{Quasi-UP} 
\begin{figure}[!htb]
\begin{center}
\includegraphics[angle=0,width=0.48\textwidth]{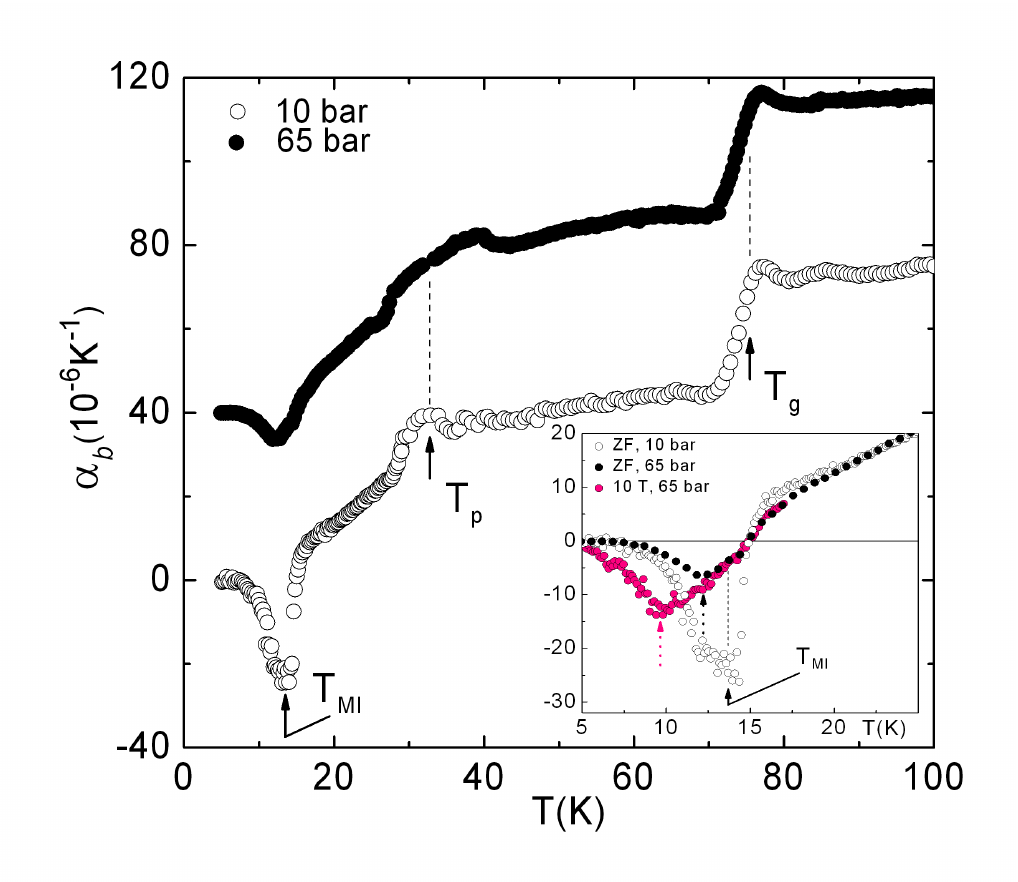}
\end{center}
\caption{\small Main panel: Thermal expansion coefficient along
the out-of-plane \emph{b}-axis for $\kappa$-D8-Br crystal\,\#2
under ambient (10\,bar) and quasi-uniaxial pressure of 65\,bar.
Dashed lines are guide for the eyes. \textcolor{black}{For clarity, data are shifted.}
Inset: blowup of the low-$T$ $\alpha_b(T)$ data together with
measurement under pressure (65 bar) and magnetic field of 10\,T.
ZF refers to zero magnetic field. The black dashed arrow indicates the
peak position under quasi-uniaxial pressure of 65\,bar, while the pink
dashed arrow highlights a shift of the peak position to lower $T$
when a magnetic field of 10\,T is applied. Picture after \Ref{Thesis}.}
\label{QUP}
\end{figure}

Given the high anisotropy of the
$\kappa$-(ET)$_2$X charge-transfer salts, thermal expansion
measurements under quasi-uniaxial pressure could provide more
insights to better understand the transitions at $T_{MI}$, $T_p$
and $T_g$. The effect of quasi-uniaxial pressure on the sample was
studied for crystal\,\#2 as shown in \Fig{QUP}. In these
experiments, a pressure of (65 $\pm$ 5)\,bar was applied along the
out-of-plane \emph{b}-axis. Roughly speaking, pressure application
along the \emph{b}-axis means a change of the contact between the
polymeric-anion chains and the ET molecules and/or enhancement of
the tilt of the ET molecules.

\textcolor{black}{As shown in the main panel of \Fig{QUP}, the shape of
the expansivity curves at $T_{MI}$ and $T_p$ is dramatically changed by
applying a quasi-uniaxial pressure of 65\,bar.
These findings reveal that the anomalies observed in the out-of-plane expansivity
$\alpha_b(T)$ at $T_{MI}$
and $T_p$ are strongly affected upon applying pressure, while $T_g$
remains practically unaffected. Since $\kappa$-D8-Br is located on
the boundary of the Mott MI
transition (\Fig{Phase
Diagram}), this feature might be associated with a slight shift of its
position from the insulating to the metallic side of the phase
diagram. Assuming the peak position $T_{MI}$ = 13.6\,K as the
transition temperature under ambient pressure (actually under
a pressure of roughly 10\,bar)
and $T_{MI}$  = 11.8\,K as the transition temperature under
quasi-uniaxial pressure (65\,bar),
one obtains $dT_{MI}/ d P_b$
$\simeq$ $-$33\,K/kbar. Such a value is roughly one order of magnitude
smaller than the hydrostatic pressure dependence of $T_{MI}$ ($d
T_{MI}/ d P$ $\simeq$ $-$380\,K/kbar), estimated from the slope of
the Mott MI transition line in \Fig{Phase Diagram} at $T$ = 11.8\,K.} Such
discrepancy might reflect the anisotropy in the uniaxial-pressure
effects. However, the shift of $T_p$ ($T_{MI}$) to high (low)
temperatures is in perfect agreement with a positive (negative)
pressure dependence of $T_p$ ($T_{MI}$) in the phase diagram
(\Fig{Phase Diagram}). Applying a magnetic field of 10\,T
(pink curve in the inset of \Fig{QUP}), the peak position of
the transition, indicated by black and pink dashed arrows, shifts to
lower temperatures. As $T_{MI}$ is insensitive to magnetic fields up
to 10\,T (Section\,\ref{Field-Effects}), this observation suggests
that the anomaly in
$\alpha_b(T)$ under quasi-uniaxial pressure should
be triggered by superconductivity. The present thermal expansion
results under quasi-uniaxial pressure can be seen as a starting
point for experiments under hydrostatic pressure, which will provide
important information about the physics in the vicinity of the Mott
MI transition region in the phase diagram.

\subsection{Influence of the Cooling Speed on $T_g$ in $\kappa$-D8-Br}
\begin{figure}[!htb]
\begin{center}
\includegraphics[angle=0,width=0.48\textwidth]{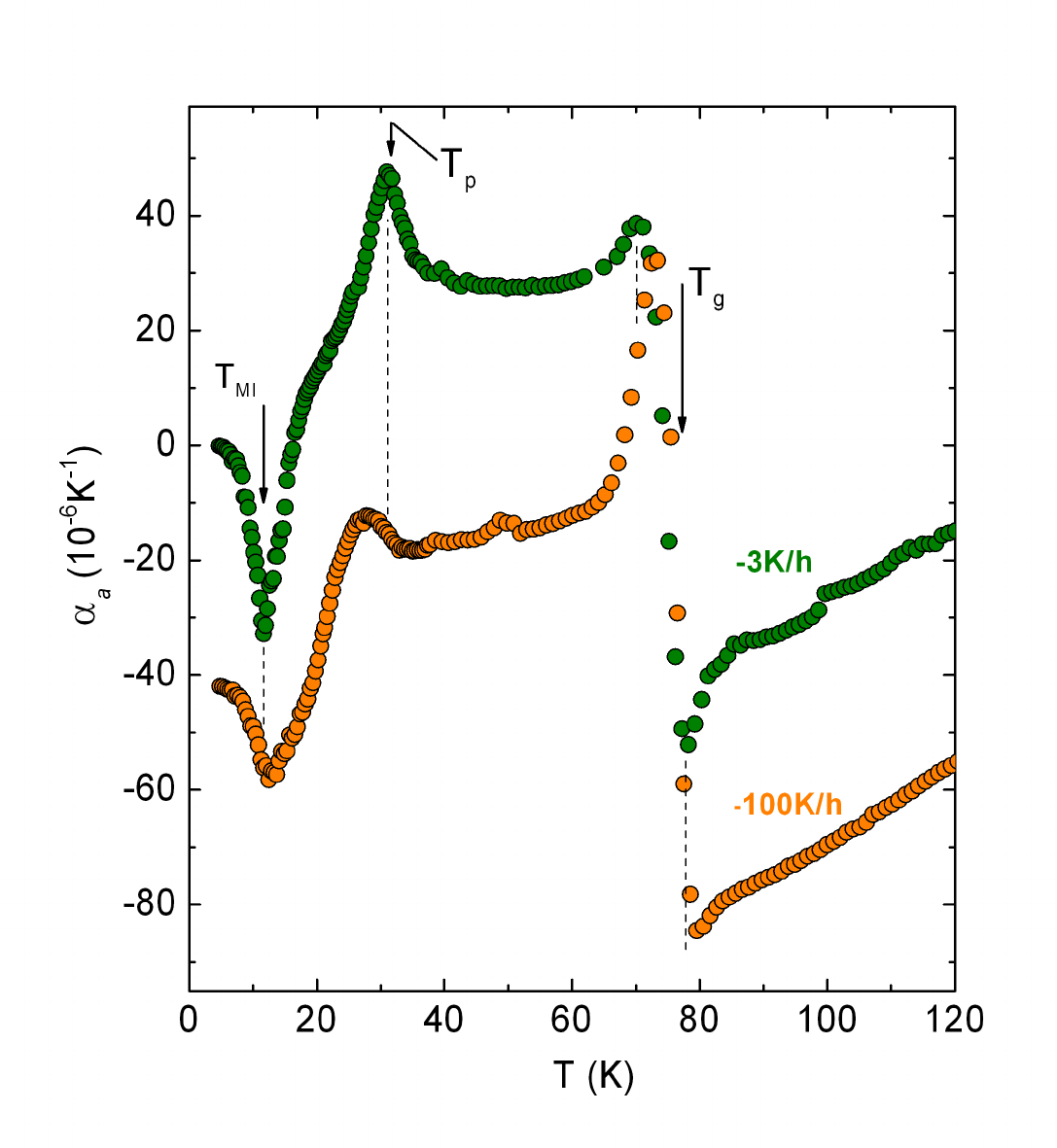}
\end{center}
\caption{\small Thermal expansion coefficient as a function of
temperature for $\kappa$-D8-Br crystal\,\#3 along the in-plane
\emph{a}-axis. Measurements taken on warming after employing
different cooling speeds ($-$3\,K/h and $-$100\,K/h). Data are
shifted for clarity. Arrows indicate the respective
transition temperatures
 for the slowly cooled crystal. Dashed lines are used
to show the shift of $T_{MI}$ (towards higher temperatures), $T_p$
(towards lower temperatures) and $T_g$ (towards higher
temperatures) under fast cooling across $T_g$. Picture after \Ref{Thesis}.}
\label{Cooling}
\end{figure}

In this section, the influence of the cooling speed below the
glass-like transition at $T_g$ $\approx$ 77\,K
on $T_{MI}$ and
$T_p$ is discussed. In fact, the role of the cooling speed on the
physical properties of fully and partially deuterated salts of
$\kappa$-(ET)$_2$Cu[N(CN)$_2$]Br has been intensively discussed in
the literature, see e.g.\ \Refs{Taniguchi20g2, Su1998-20g}.
As shown in \Fig{Cooling}, the \textcolor{black}{expansivity}
along the in-plane \emph{a}-axis for a crystal of $\kappa$-D8-Br
was measured after cooling the sample through the so-called
glass-like phase transition by employing two distinct cooling
speeds ($-$3\,K/h and $-$100\,K/h).

A broadening of the transition, accompanied by a reduction in the
size of the peak anomaly at $T_p$ $\simeq$ 30\,K, which is related
to the critical \textcolor{black}{end-point of the first-order Mott MI transition} line in the phase
diagram, and the Mott MI temperature $T_{MI}$ $\simeq$ 13.6\,K, is
observed for fast cooling ($-100$\,K/h). Another remarkable feature is
that the peak position of the $T_p$ = 30\,K anomaly shifts toward
lower temperatures, while the anomaly at $T_{MI}$ = 13.6\,K shifts
slightly to higher temperatures. Based on these observations, one
can conclude that the increase of the cooling speed through $T_g$
results in an opposite effect to that observed upon applying
quasi-uniaxial pressure (Section\,\ref{Quasi-UP}). The observed
shift of $T_g$ towards higher temperatures under fast cooling is in
agreement with literature results \cite{Jens2002-20}. Resistivity
measurements on $\kappa$-D8-Br salts
\cite{Griesshaber20e,Griesshaber99}, synthesized by employing the
same method as that used for the salts studied in this work,
revealed that
under a
fast cooling speed \cite{TEref} the resistance
increases dramatically below $T_g$ $\simeq$ 80\,K, superconductivity
is suppressed, giving way to a residual resistance ratio of $\sim$5.
This feature together with the observed  enhancement of the anomaly
in $\alpha_a$ on fast cooling (\Fig{Cooling}) can be assigned
to an increase of the scattering provoked by the randomly
distributed  potential of the polymeric Cu[N(CN)$_2$]Br$^-$ anion. A
similar situation is encountered in the quasi-1D (TMTSF)$_2$ClO$_4$
superconductor salt, where by rapid cooling ($>$50\,K/min)
superconductivity is destroyed, giving way to a first-order
anion-ordering transition at $T$ = 24\,K, \textcolor{black}{which  is
accompanied by a spin-density wave transition around 6\,K \cite{Ishiguro20}.}

\section{Mott Criticality}\label{Criticality arount Tstar}
Before discussing the Mott criticality in detail, let us first present a general discussion about critical behavior and universality classes.

\subsection{Critical Behavior and Universality Classes}\label{Critical Behavior and Universality Classes}

Continuous phase transitions involve fluctuations on all length scales,
leading to the remarkable phenomenon called universality.
While details of the interaction do not matter, only robust properties
of the system like its effective dimensionality, the absence or presence of
long-range interactions or the symmetry of relevant low-energy
fluctuations determine the low-energy behavior.
As a consequence, it is possible to classify continuous phase
transitions according to their universality classes.

In real materials, upon approaching
the critical temperature, fluctuations are no longer negligible.
The more the critical
temperature is approached, the stronger are the fluctuations.
Finally, fluctuations succeed in destroying the original phase and a
new phase, often with distinct symmetries, emerges.
To distinguish the two different phases, one usually tries to introduce an
order parameter as the expectation value of some field which vanishes in one phase but not the other.

The transition is, in many cases,
accompanied by broadening effects due to crystal defects or
inhomogeneities.
Amazingly, around $T_c$ some physical quantities tend to obey power
laws in the
reduced temperature $t = (T-T_c)/T_0$,  which measures the
relative distance to the transition temperature.
As a consequence of universality, the divergence of the compressibility
of water in the vicinity of its critical point is described by exactly the same power-law dependence as the divergence of the susceptibility of a uniaxial magnet in the vicinity of the Curie temperature.
In such a uniaxial magnet, the
spins can only align themselves parallel or anti-parallel to a given
crystal axis.
As the magnetization is finite in the symmetry-broken state below
$T_c$ and vanishes above $T_c$, the expectation value of the magnetization may
serve as an order parameter.

Concerning thermodynamic properties,
one usually defines the four critical
exponents $\tilde \alpha$, $\beta$, $\gamma$, and $\delta$, which in the
case of magnets are related
to the specific heat (at constant pressure)
$C_p(T)$, the spontaneous magnetization $M_s(T)$, and the
susceptibility $\chi(T)$.
More generally, $M_s(T)$ and $\chi(T)$ describe the order parameter and the order parameter susceptibility.
Usually, $\alpha$ is used in the literature to refer to the specific heat
critical exponent. In order to avoid any confusion with the linear
thermal expansion coefficient $\alpha(T)$, here, $\tilde{\alpha}$ is
used to refer to the specific heat critical exponent.

Close to criticality, one expects the following power-law behavior,
\begin{equation}
C_p (T) \sim A^{\pm}
\frac{|t|^{-\tilde{\alpha}}}{\tilde{\alpha}},
\label{Specific Heat Div}
\end{equation}
\begin{equation}
M_{s}(T) \sim B|t|^{\beta},
\\\\\\ t \leq 0 ,
\end{equation}
\begin{equation}
\chi(T) \sim C^{\pm}|t|^{-\gamma},
\end{equation}
\begin{equation}
M_s(H) \sim DH^{1/\delta},
\\\\\\ t = 0 .
\end{equation}
While the amplitudes $A^{\pm}$, $B$, $C^{\pm}$, and $D$ are non-universal and the coefficients $A^+$ and $C^+$ governing the behavior for $t>0$ are generally different from the corresponding coefficients $A^-$ and $C^-$ for $t<0$, the ratios $A^+/A^-$ and $C^+/C^-$ are in fact universal, i.e. they are not material-specific and only depend on the underlying universality class.
In fact, not all of the above exponents are independent.
Using concepts of scaling or the renormalization group, the following identities can be derived \cite{Huang1b,GoldenfeldBook,CardyBook,KardarBook,KopietzBook},
\begin{equation}
\tilde{\alpha} + 2\beta + \gamma = 2\label{Universallaw} ,
\end{equation}
\begin{equation}
\gamma=\beta(\delta-1)\label{Universallaw1} .
\end{equation}
These two relations are
frequently called the Rushbrooke and Widom identities.
It is striking to note that in the study of critical behavior
near
a second-order phase transition, materials displaying completely different
crystal structures as well as quite different subsystems obey the
same critical behavior
near
$T_c$, giving thus rise to the
universality classes. The theoretical  values for the critical
exponents of different universality classes accompanied by  an
example of a phase transition are listed in Table\,\ref{universality}.
It should be noted that sufficiently far away from the transition,
fluctuations can largely be ignored and
mean field behavior prepails. More precisely, corrections to mean field theory
can only be expected to become important for temperatures satisfying
\begin{equation}
  \label{eq:GinzburgRegime}
  | t | \lesssim t_G ,
\end{equation}
\vspace{0.2cm}
where
\begin{equation}
  \label{eq:GinzburgScale}
  t_G = C \left(\frac{k_B}{\Delta c_v \xi_0^d}\right)^{2/(4-d)} .
\end{equation}
is the Ginzburg scale.
Here, $d$ is the effective dimensionality of the system, $\Delta c_v$ is the jump of the specific heat
across the phase transition, and $\xi_0$ is the bare coherence length.
In three dimensions, the universal dimensionless proportionality constant $C$
introduced here is given by $C= 1/(32\pi^2)$ \cite{Levanyuk59,Ginzburg60}, which is much smaller than 1.

Quite generally, a crossover is expected near $t_G$ from mean field to non-mean field behavior.
If, however, the bare coherence length is sufficiently large, the non-mean field regime can turn out to be
unmeasurably small.
In conventional superconductors, for example, the size of the bare coherence length can be one thousand the size of the lattice spacing,
leading to $t_G \approx 10^{-18}$.
As a consequence, mean field theory is essentially exact.

\begin{table*}[htb]
\centering \small
\begin{tabular}{|p{2.3cm}|c|c|c|c|p{3.4cm}|}
\hline\hline \textbf{
Universality class} & \textbf{$\tilde{\alpha}$} & \textbf{$\beta$} & \textbf{$\gamma$} & \textbf{$\delta$} & \textbf{Examples of phase transition}\\
\hline\hline

Mean-field & 0 & 0.5 & 1.0 & 3.0 &
Superconducting transition in conventional superconductors or
Mott MI transition in
(V$_{1-x}$Cr$_x$)$_2$O$_3$ \cite{Limelette03-9d3}\\ \hline

2D Ising  & 0 & 0.125 & 1.75 & 15 & Preroughening transition in
GaAs \cite{Labella2000-9d}\\ \hline

3D Ising  & 0.110(1) & 0.3265(3) & 1.2372(5) & 4.789(2) & Liquid-gas transition \\ \hline

3D XY & -0.0151(3) & 0.3486(1) & 1.3178(2) & 4.780(1) & Superfluid transition in
${}^4$He \cite{Lipa96}
\\ \hline

3D Heisenberg & -0.1336(15) & 0.3689(3) & 1.3960(9) & 4.783(3) & Ferromagnetic
transition in a clean and isotropic ferromagnet \\ \hline

Unconventional Criticality & $-$1 & 1 & 1 & 2 & Mott MI transition
in
$\kappa$-(BEDT-TTF)$_2$Cu[N(CN)$_2$]Cl \cite{Kagawa2005_20}\\
\hline\hline
\end{tabular}
\caption{\small Different universality classes with their respective critical
exponents, accompanied by a proposed example of the phase
transition.
The theoretical estimates for the critical exponents
are from
\Ref{Pelissetto02} ($3D$ Ising),
\Ref{Campostrini06} ($3D$ $XY$), and
\Ref{Campostrini02} ($3D$ Heisenberg).
The mean-field values are derived in any textbook on critical phenomena
(see e.g.~\Refs{GoldenfeldBook,CardyBook,KardarBook,KopietzBook}),
values quoted for the $2D$ Ising model are the exact values from Onsager's famous solution \cite{Onsager44}.
}
\label{universality}
\end{table*}

\vspace{0.2cm}

From the experimental point of view, the estimate of critical
exponents
can be a hard task.
For instance, owing to the specific heat
critical exponent $\tilde{\alpha}$, a reliable estimate of the
phonon background
can be one of the crucial points.
 In addition, for a
reliable estimate of the critical behavior of a system,  fine
measurements close to $T_c$
are
necessary.
Accurate measurements of e.g. the specific heat are
required over several orders of magnitude of $t$.
For real materials, a broadening of the
transition due to inhomogeneities (impurities or crystal defects) is
frequently observed.
Due to this, $T_c$ cannot be
measured directly, but is rather obtained indirectly via self
consistent fittings.

\subsection{Scaling ansatz for the Mott metal-insulator transition}

In 2005, in a stimulating article entitled
``Unconventional Critical Behaviour in a Quasi-2D Organic
Conductor'', Kagawa and
collaborators reported on the criticality at the pressure-induced
Mott transition in the organic
$\kappa$-(BEDT-TTF)$_2$Cu[N(CN)$_2$]Cl charge-transfer salt \cite{Kagawa2005_20}.
In this
study, the authors made use of the isothermal pressure-sweep
technique, using helium as a pressure
transmitting medium, to explore the critical behavior
of this organic salt
through conductance measurements.
The pressure-sweep technique had been previously applied by Limelette and collaborators \cite{Limelette03-9d3}
to study the Mott critical behavior of Cr-doped V$_2$O$_3$, which is
now recognized as a canonical Mott insulator system and behaves mean-field like.
Very close to the transition the authors also claim Ising universality.
In contrast to $\kappa$-(BEDT-TTF)$_2$Cu[N(CN)$_2$]Cl, chromium-doped
vanadium sesquioxide is a truly three-dimensional material.
In the study of the quasi-two-dimensional charge-transfer salt
$\kappa$-(BEDT-TTF)$_2$Cu[N(CN)$_2$]Cl,
the critical behavior
of the conductance data at the critical endpoint was analyzed in the
framework of the scaling theory of the liquid-gas transition
\cite{Kadanoff9d1}.
For simplicity, in the so-called non-mixing approximation,
the rescaled pressure and the rescaled temperature
were used as two independent scaling variables
to obtain the critical
exponents $\beta = 1$, $\gamma = 1$, and $\delta = 2$, as listed in
Table\,\ref{universality}.
Substituting these values in the
Rushbrooke relation one obtains $\tilde{\alpha}$ = $-$1.
As pointed
out by the authors, the obtained critical exponents do not fit in
the well-known universality classes, indicating the
experimental discovery of a new
universality class.
A possible explanation of the experimentally observed exponents of unconventional criticality was given by Imada and collaborators \cite{Imada05,Imada10}.

The appearance of unconventional critical exponents came as a surprise to many who expected Ising universality.
This expectation was based on an argument by Castellani \emph{et al.} \cite{Castellani79} who argued that, even though there is no symmetry breaking, the double occupancy of lattice sites can serve as an order parameter.
While doubly occupied (or empty) sites are localized on the insulating side of the phase transition, they do proliferate on the metallic site. The order parameter was therefore expected to be of an Ising type.
The Mott criticality was also studied in the framework of dynamical mean-field theory of the Hubbard model by Kotliar \emph{et al.} \cite{Kotliar}. These authors confirmed the statement that the Mott transition should lie in the Ising universality class.
Having a phase transition between a low-density gas of localized doubly occupied or empty lattice sites to a high-density fluid of unbound doubly occupied or empty lattice sites is quite analogous to the well-known liquid-gas transition which lies also in the Ising universality class.

An attempt to reconcile the experimentally observed unconventional critical exponents $\beta =1$, $\gamma = 1$, and $\delta =2$ with the well-established two-dimensional Ising universality class was made by
Papanikolaou and collaborators \cite{Papanikolaou2008_20}.
These authors pointed out that the critical exponents derived from transport measurements do not necessarily coincide with the thermodynamic critical
exponents which are defined in terms of an order parameter.
According to Papanikolaou \emph{et al.} \cite{Papanikolaou2008_20},
the singular part of the conductivity $\Delta\Sigma$ does not only depend on the
singular part of the order parameter, but also depends on the related energy density.
As a consequence, $\Delta\Sigma$ is given by
\begin{equation}
  \label{eq:DeltaSigmaDensity}
  \Delta\Sigma = A m + \text{sgn}(m) B |m|^\theta .
\end{equation}
Here, $A$ and $B$ are non-universal coefficients,
$m$ is the order parameter
 and $\theta = (1- \tilde\alpha)/\beta$. In an extended regime not too close to the critical point the first term on the r.h.s.\ can be much smaller than the second one such that
\begin{equation}
  \label{eq:DeltaSigmaDensity0}
  \Delta\Sigma \propto \text{sgn}(m) |m|^\theta .
\end{equation}
It then follows that the critical exponents as obtained from the conductivity can be expressed in terms of the thermodynamic critical exponents as follows \cite{Papanikolaou2008_20}:
$\beta_\sigma = \theta \beta$, $\delta_\sigma = \delta/\theta$, and $\gamma_\sigma = \gamma + \beta (1 - \theta)$. In particular, for the 2D Ising universality class one obtains with $\theta = 8$ the exponents $\beta_\sigma = 1$, $\delta_\sigma = 7/4$, and $\gamma_\sigma = 7/8$ which within the experimental resolution are consistent with the exponents obtained by Kagawa \emph{et al.} \cite{Kagawa2005_20}.

To explain the lattice response in the vicinity of the MI finite temperature critical end point and explain the corresponding peak in the thermal expansivity a description in terms
of critical exponents is not sufficient unless the pressure is tuned to its critical value.
Instead, the usage of a scaling theory was proposed in \Ref{Bartosch}.
While such an approach is extremely general and can be applied to any universality class, it was shown in \Ref{Bartosch} that the expansivity data of \Ref{Mariano20f1} is in fact consistent with 2D Ising criticality:
Within the scenario of two-parameter scaling, the Gibbs free energy can be written as
\begin{equation}
  \label{eq:FreeEnergy}
  f(t,h) = |h|^{d/y_h} \Phi \left( t/|h|^{y_t/y_h} \right) .
\end{equation}
Here, $\Phi(x)$ is a universal scaling function which only depends on the universality class, $t$ and $h$ are the two (suitably normalized) scaling variables, and $y_t$ and $y_h$ are their corresponding RG eigenvalues.
These are the two relevant eigenvalues of the underlying universality class and determine all critical exponents.
For example, $\tilde\alpha = 2 - d/y_t$ or $\beta = (d-y_h)/y_t$ (see e.g. \Refs{GoldenfeldBook,CardyBook,KardarBook,KopietzBook}).
Conventionally, $t=(T-T_c)/T_0$ and $h=(P-P_c)/P_0$ can be thought of as temperature- and pressure-like quantities, but more generally there are also linear mixing terms such that
$t=(T-T_c- \zeta (P-P_c))/T_0$ and $h=(P-P_c- \lambda (T-T_c))/P_0$.
In fact, very close to the transition Hooke's law of elasticity breaks down and one should expect compressive strain to enter linearly the scaling variables instead of pressure, see below and \Ref{Zacharias}.

The appearance of the mixing terms is a consequence of the experimental fact that the first-order transition line ending in the critical point $(P_c,T_c)$ is not parallel to the $T$ axis, see \Fig{Schematic-PD}.
While the linear mixing terms turn out to be important for a quantitative description of the expansivity data \cite{Bartosch}, let us for simplicity focus here on the much simpler case of no mixing.
For the $2D$ Ising universality class, $y_t = 1$ and $y_h = 15/8$. Having one integer-valued RG eigenvalue, logarithmic corrections to scaling should be expected \cite{CardyBook}.
As it turns out, the scaling ansatz for the $2D$ Ising universality class requires a correction term and reads \cite{Fonseca03}
\begin{equation}
  \label{eq:fIsing}
  f(t,h) = \frac{t^2}{8\pi} \ln t^2 + |h|^{d/y_h} \Phi \left( t/|h|^{y_t/y_h} \right) .
\end{equation}
Differentiating twice with respect to $t$ and setting $h$ equal to zero one then obtains the well-known logarithmic divergence of the specific heat $\alpha_{\text{sing}}$. Assuming no mixing terms, the singular part of the thermal expansivity
is given by \cite{Papanikolaou2008_20}
\begin{equation}
  \label{eq:expansivitypc}
  \alpha_{\text{sing}} \Big|_{h= \pm 0}  \propto \frac{\partial^2 f}{\partial h\, \partial t} \Big|_{h= \pm 0}  \propto \text{sgn} (h)\, (-t)^{-1+\beta} , \quad t < 0 .
\end{equation}
As a consequence, the Gr\"uneisen ratio
$\Gamma_{\text{sing}} =  \alpha_{\text{sing}} / c_{\text{sing}}$
diverges for $t < 0$ as \cite{Bartosch}
\begin{equation}
  \label{eq:Grueneisencriticalendpoint}
  \Gamma_{\text{sing}} = \left. \frac{\alpha_{\text{sing}}}{c_{\text{sing}}} \right|_{h=\pm 0} \propto \text{sgn}\,(h) \, (-t)^{-1+\tilde\alpha+\beta} .
\end{equation}

Let to emphasize here that the true divergence of the Gr\"uneisen parameter is a consequence of neglecting linear mixing terms.
Including such mixing terms effectively leads to a saturation of the Gr\"uneisen parameter at a very large value. It was estimated in \Ref{Bartosch} that the crossover scale for the D8-Br crystals is of order 0.1\,bar and thus
for most practical purposes beyond experimental resolution. The smallness of this crossover scale can be related via the Clausius-Clapeyron equation to comparatively small entropy differences between the metallic and insulating phases.

The Gr\"uneisen ratio is also known to diverge at a quantum critical point \cite{Zhu03}.

In the case $\tilde\alpha < 0$ there is no divergence in the specific heat, implying that the specific heat is bounded and can safely be replaced by a constant. This has interesting consequences for the universality class of unconventional criticality for which $\tilde\alpha = -1$ and $\beta = 1$. In this case both the expansivity and the specific heat stay finite such that there is no divergence of the Gr\"uneisen ratio.
Still, there is no reason to expect the expansivity to be proportional to the specific heat, i.e.\ Gr\"uneisen scaling breaks down.\cite{Bartosch}

\begin{figure}[b]
\centering
  {\includegraphics*[width=0.8\columnwidth]{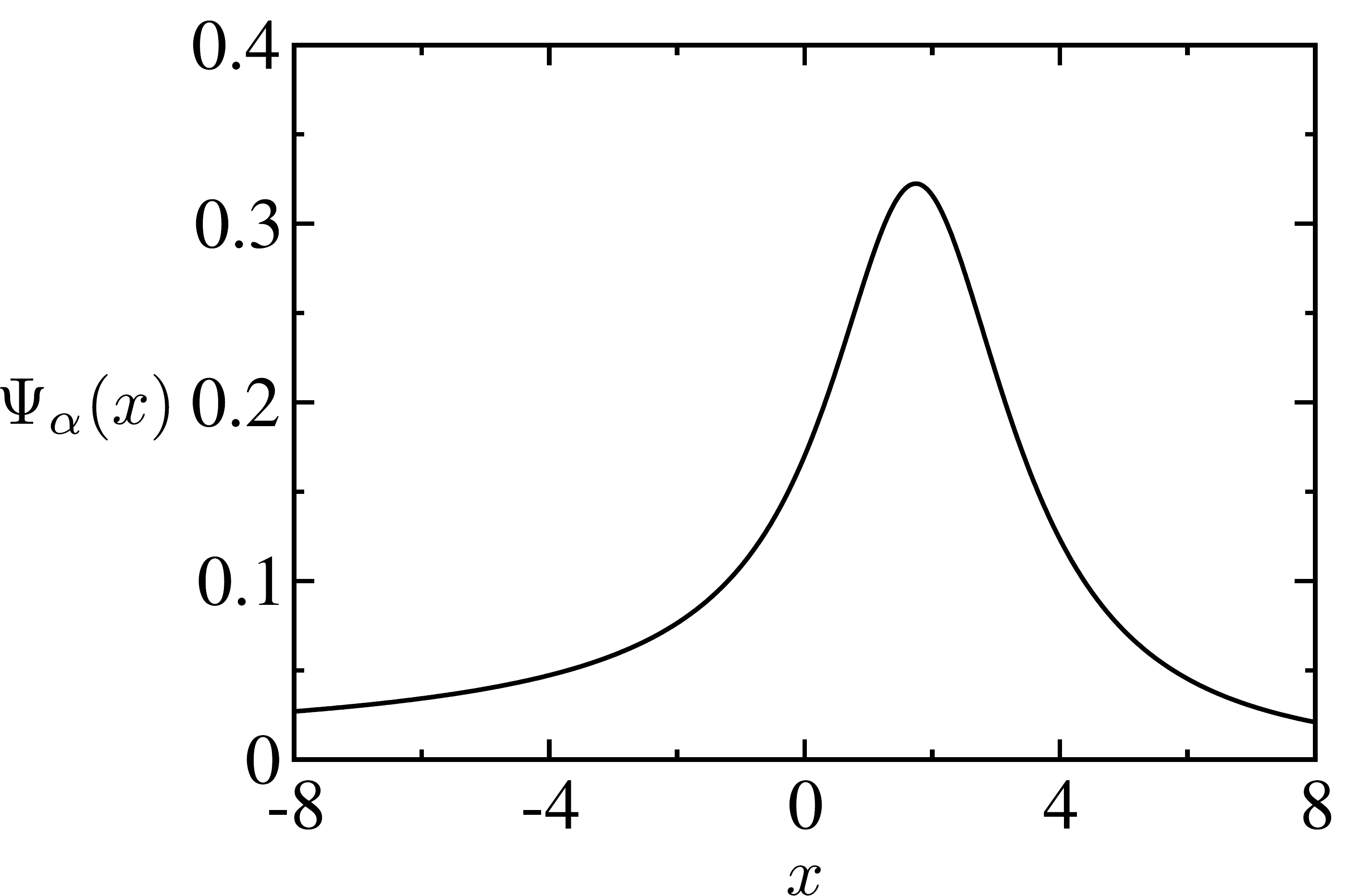} \quad}
  \caption{Plot of the scaling function $\Psi_\alpha(x)$ of the thermal expansivity for the $2D$ Ising universality class. Figure taken from \Ref{Bartosch}.
}
  \label{FigScalingFunction2D}
\end{figure}

\begin{figure}[tb]
  \includegraphics*[width=0.8\columnwidth]{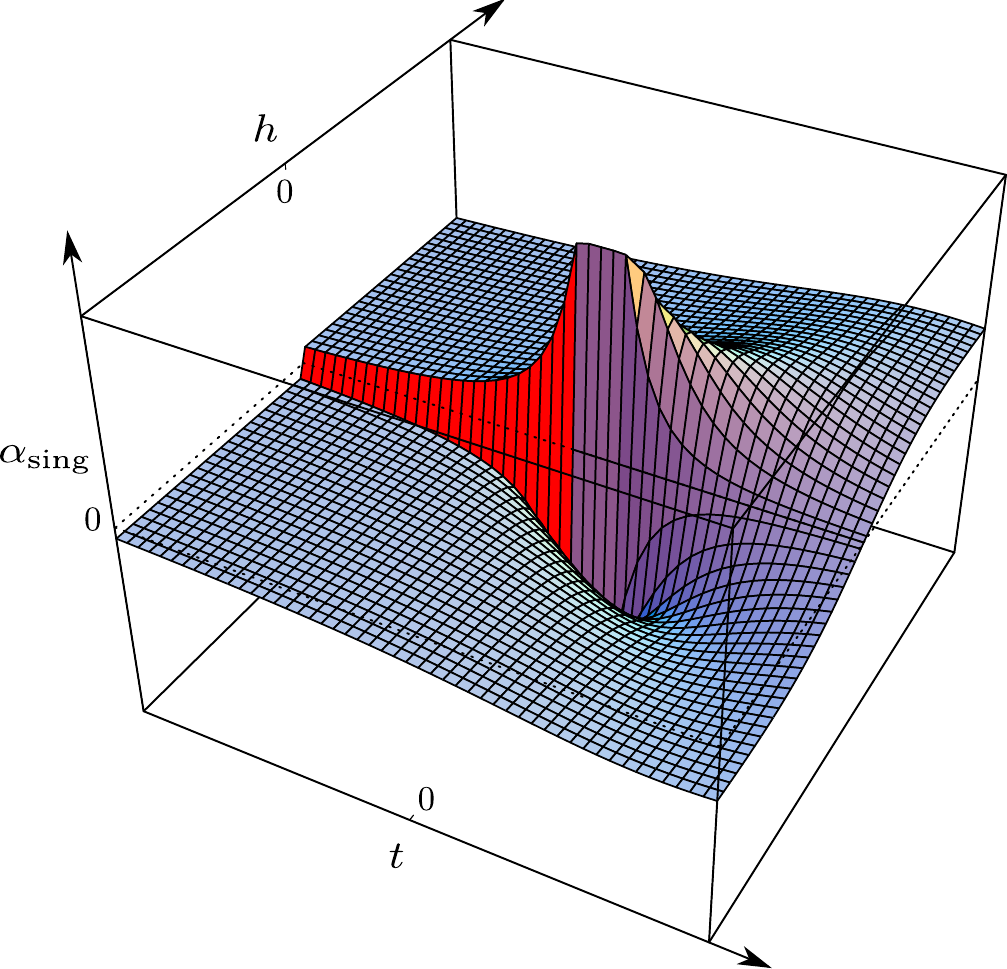}
  \caption{
Plot of the singular part of the thermal expansivity $\alpha_{\text{sing}}$ as a function of the scaling variables $t$ and $h$.
The first-order transition line along the negative $t$-axis manifests itself by a jump in the expansivity which for $t \to 0$ diverges as
$\Delta \alpha_{\text{sing}} \propto (-t)^{-1+\beta} = (-t)^{-7/8}$, see \Eq{eq:expansivitypc}.
It should be noted that every cut at constant $h$ is just given by the (rescaled) scaling function $\Psi_\alpha(x)$, as graphically depicted in \Fig{FigScalingFunction2D}. Figure taken from \Ref{Bartosch}.
}
  \label{FigScalingFunction3D}
\end{figure}

\begin{figure}[tb]
  \includegraphics*[width=0.8\columnwidth]{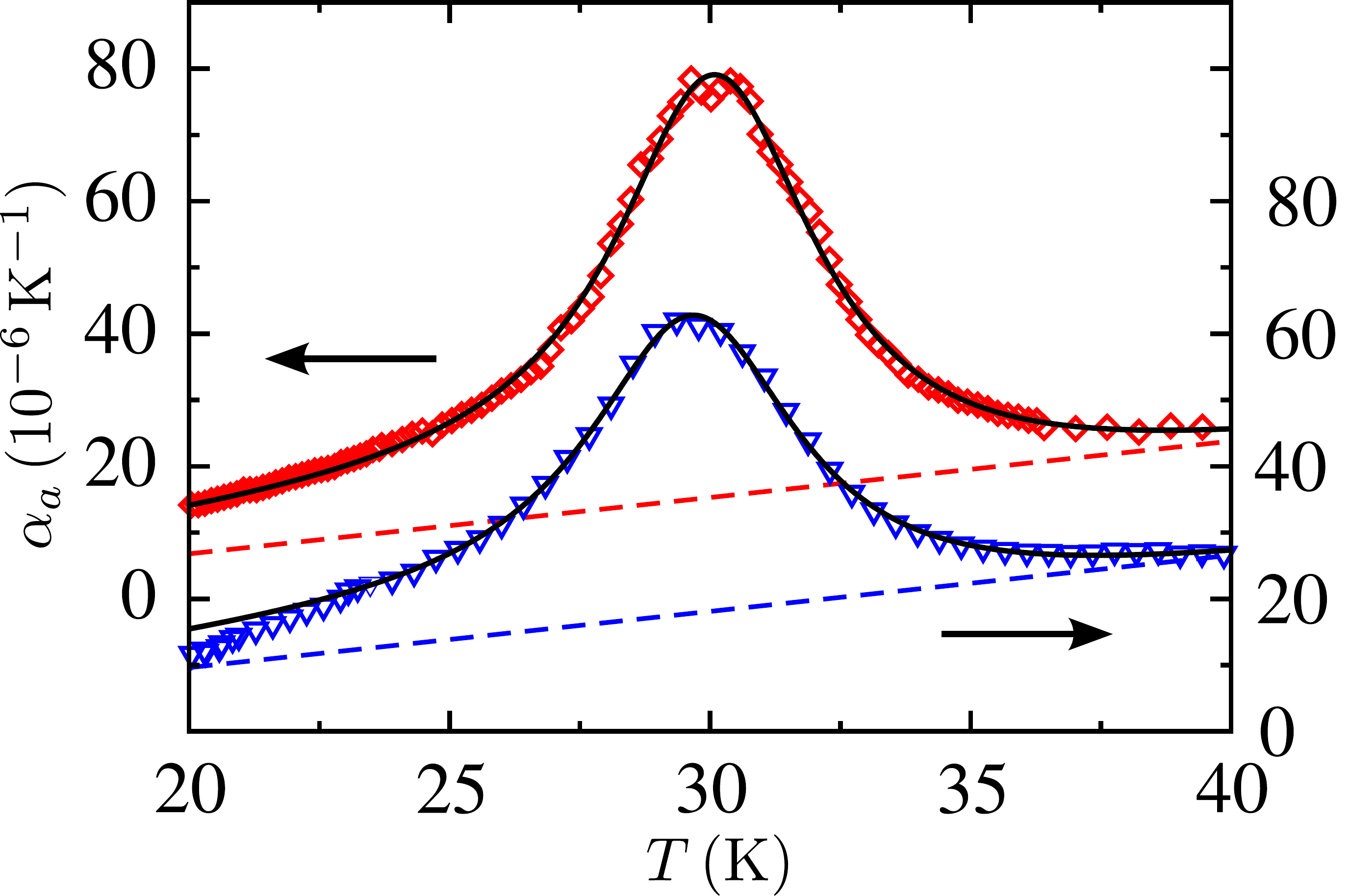}
  \caption{
Fit of the expansivity data of the two $D8$-Br crystals \#~1 (diamonds $\Diamond$) and \#~3 (triangles $\nabla$) from \Ref{Mariano20f1} to the scaling theory outlined in the text.
In total, there are six fitting parameters involved. Two fitting parameters determine the height and width of the peak, two fitting parameters determine the orientation of the temperature and pressure axes
(and thereby describes linear mixing terms), and finally two fitting parameters are necessary to describe the analytic background contribution, as described by the dashed line. Figure taken from \Ref{Bartosch}.
}
  \label{fig:fit}
\end{figure}

To describe the expansivity of the $\kappa$-ET salts away but close to criticality, i.e.\ for $h \neq 0$, we can express the singular part of the expansivity in terms of derivatives of the scaling function $\Phi(x)$. Differentiating \Eq{eq:fIsing} with respect to $t$ and $h$, one obtains \cite{Bartosch}
\begin{equation}
  \label{eq:expansivityoftandh}
  \alpha_{\text{sing}}(t,h) \propto \text{sgn} (h)\, |h|^{-1+(d-y_t)/y_h} \Psi_\alpha(t/|h|^{y_t/y_h}) ,
\end{equation}
with the scaling function $\Psi_\alpha(x)$ given by
\begin{equation}
  \label{eq:Psi}
  \Psi_\alpha(x) = \frac{d-y_t}{y_h} \Phi'(x) -\frac{y_t}{y_h} x \Phi''(x).
\end{equation}
Even though the 2D Ising model can be solved exactly, there is no known exact expression for the scaling function $\Phi(x)$ and hence no exact expression for $\Psi_\alpha(x)$.
Nevertheless $\Phi(x)$ [and hence $\Psi_\alpha(x)$] can be calculated with very high accuracy numerically\cite{Fonseca03,Bartosch}, for a plot see \Fig{FigScalingFunction2D}.
With this scaling function at hand, the singular part of the thermal expansivity can now be plotted as a function of $t$ and $h$, see \Fig{FigScalingFunction3D}.
This singular part of the thermal expansivity corresponds to the critical region in the pressure-temperature phase diagram.
The first-order transition line is simply represented by the jump in $\alpha_{\text{sing}}(t,h)$ for $h = 0$ and $t < 0$ which describes the change in volume $\Delta V \propto (-t)^\beta$.
It should be noted that every cut of the graph for constant $h$ is just a rescaled version of the scaling function $\Psi_\alpha (x)$, as shown graphically in \Fig{FigScalingFunction2D}. Also, the change of sign of $\alpha_{\text{sing}}$ with a change of sign of $h$ can clearly be seen.
While a constant pressure corresponds to a constant $h$ in the non-mixing approximation, in the case of linear mixing terms a constant $P$ corresponds to cutting the graph in \Fig{FigScalingFunction3D} at an angle.
Fitting the thermal expansivity of the D8-Br crystals \cite{Mariano20f1} \#~1 and \#~3 by a scaling form of the thermal expansivity based on
\Eq{eq:expansivityoftandh} on top of a linear background including linear mixing terms gives indeed an excellent fit for the Ising universality class \cite{Bartosch}, see \Fig{fig:fit}.

Even though the Ising universality class leads to an excellent fit of the dilatometric data, we would like to stress here that this is by far no proof of Ising universality.
NMR mearsurements of $1/T T_1$ obey roughly $\Delta (1/T T_1) \propto (P-P_c)^{1/\delta}$, with $\delta \approx 2$, as also obtained from conductance measurements \cite{Kagawa2005_20}.
As far as we know, the NMR measurements have not found a natural explanation based on Ising criticality yet.

Notwithstanding the analogy between the Mott MI and the liquid-gas transition, differences do exist and the compressible MI with its long-range shear forces should rather be classified as a
solid-solid transition \cite{Zacharias}.
Hopping matrix integrals do depend on the distance between neighboring atoms, such that the Mott MI transition is sensitive to elastic strain and couples only indirectly to the applied stress,
e.g.\ compressive pressure.
As described in detail in \Ref{Zacharias} and similarly also in the context of the compressible Hubbard model in \Ref{Hassan2005-20}, the homogeneous component $E$ of the strain can be obtained by minimizing the effective potential density
\begin{equation}
  \label{eq:strainPotential}
  V(E) = \frac{K_0}{2} E^2 - (P-P_0) E + f_{\text{sing}}(t(T,E),h(T,E)) ,
\end{equation}
where both $t(T,E)$ and $h(T,E)$ are linear functions of $T$ and $E$, $K_0$ is the bare bulk modulus, and $P_0$ is an offset pressure.
For simplicity, we have replaced all tensorial quantities by scalars.
It was shown in \Ref{Zacharias} that while not too close to the critical end point Hooke's law holds and our above treatment was legitimate, the feedback of the electronic subsystem back on the lattice drives the effective bulk modulus to zero even above the phase transition described by $f_{\text{sing}}$ alone, thereby leading to a violation of Hooke's law and a preempted phase transition.
This phase transition was explicitly shown to be governed by Landau criticality with mean-field exponents.

While there is no final answer on the universality class underlying the Mott transition in the $\kappa$-ET charge-transfer salts yet, we hope that future experiments will shed more light on this issue.

\vspace{5mm}

\section{Thermal Expansion Measurements on $\kappa$-(ET)$_2$Cu$_2$(CN)$_3$}\label{SpinLiquidSection}
\begin{figure}[!htb]
\begin{center}
\includegraphics[angle=0,width=0.46\textwidth]{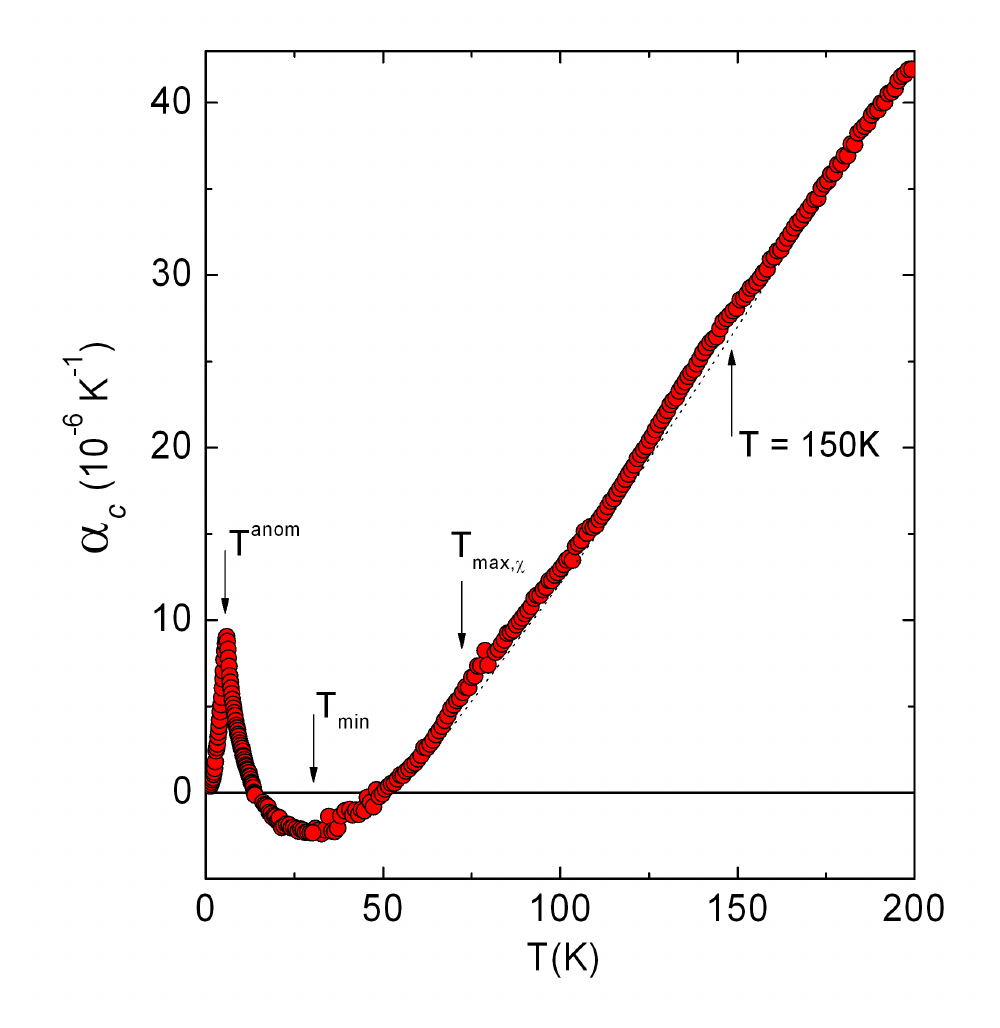}
\end{center}
\caption{\small Expansivity along the in-plane \emph{c}-axis for a
single crystal of $\kappa$-(ET)$_2$Cu$_2$(CN)$_3$. $T^{\text{anom}}$
indicates the temperature at which, according to
\Ref{Yamashita2008}, the crossover (hidden ordering) to a
spin-liquid state occurs. The dashed line is used to indicate a
hypothetical linear background.  Broad hump anomalies at $T$
$\simeq$ 150\,K and $T_{\text{max},\chi}$ $\simeq$ 70\,K are indicated by
the arrows, cf.\
the discussion in the text. Figure taken from \Ref{Thesis}.}
\label{TE-SpinLiquid}
\end{figure}

\begin{figure}[!htb]
\begin{center}
\includegraphics[angle=0,width=0.48\textwidth]{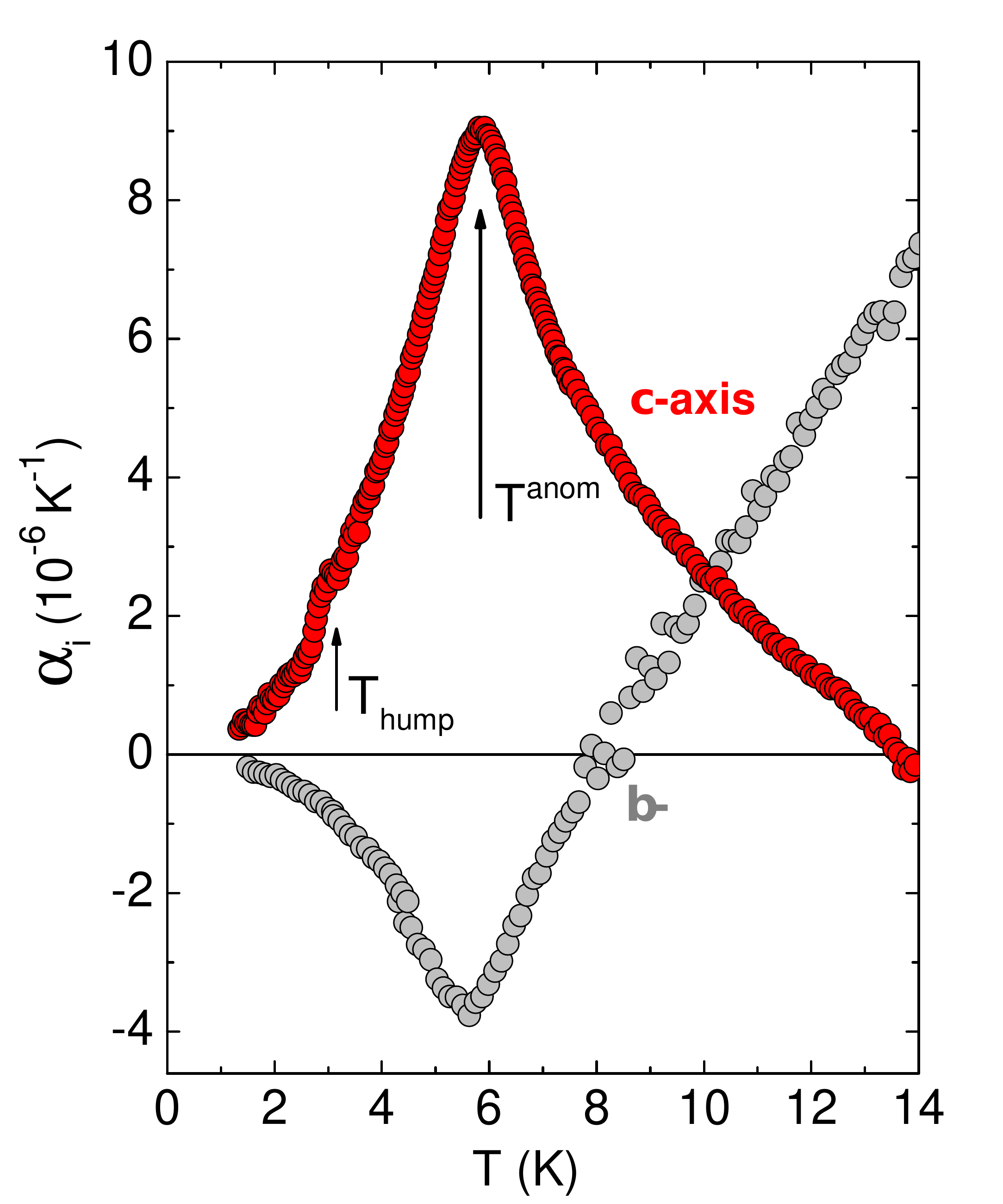}
\end{center}
\caption{\small
Blow-up of the low-temperature expansivity
data along the in-plane \emph{c}- (crystal \#1) and \emph{b}-axis
(crystal \#2) of $\kappa$-(ET)$_2$Cu$_2$(CN)$_3$ on expanded
scales, showing the sharpness of the transition at $T^{\text{anom}}$
$\simeq$ 6\,K and a hump in $\alpha_{c}$ at $T_{hump}$ $\simeq$
2.8\,K along the \emph{c}-axis.
Figure taken from \Ref{Thesis}, see also \Ref{SL}.
}
\label{TE-SpinLiquidLowT}
\end{figure}

\textcolor{black}{The thermal expansion coefficient
along the in-plane \emph{c}-axis for crystal \#1 of the
salt $\kappa$-(ET)$_2$Cu$_2$(CN)$_3$ is shown in \Fig{TE-SpinLiquid}}. Recall that this is the spin-liquid candidate.
Upon cooling, $\alpha_{c}$ decreases monotonously down to
$T_{\text{min}}$ $\simeq$ 30\,K. Around $T$ = 150\,K, indications of a
small and broad hump are observed. In \Ref{Shimzu2003},
Shimizu and collaborators observed an enhancement of the spin
relaxation rate above 150\,K, which was attributed to the freezing
of the thermally activated vibration of the ethylene end groups.
Note that no traces of a glass-like anomaly around $T$ = 77 -
80\,K are observed.  This behavior is quite distinct from that
observed in $\kappa$-(ET)$_2$Cu[N(CN)$_2$]Cl (see \Ref{Jens2002-20})
and $\kappa$-D8-Br
(discussed in Section \ref{TEMeasurementsonkappaD8})
and $\kappa$-H8-Br \cite{Jens2002-20}, where clear signatures in the
thermal expansion coefficient show up at $T_g$ $\simeq$ 77\,K.
This discrepancy indicates that the lattice dynamic for the
$\kappa$-(ET)$_2$Cu$_2$(CN)$_3$ salt is different from that of the
above-mentioned compounds. In fact, according to the RUM model,
introduced in Section \ref{subsection:RUM},
 the absence of a glass-like
anomaly \textcolor{black}{in the $\kappa$-(ET)$_2$Cu$_2$(CN)$_3$ salt} can be understood in
the following way: the anion Cu$_2$(CN)$_3^-$
(\Fig{Anion-SL}) consists of a 2D network of Cu(I) and
bridging cyanide groups \cite{Geiser1991}. As the Cu$_2$(CN)$_3^-$
anion is arranged in a network fashion, which is quite different
from the polymeric arrangement of Cu[N(CN)$_2$]Cl$^-$ and
Cu[N(CN)$_2$]Br$^-$ (\Fig{Anion}), the vibration modes of
the CN groups are confined between nearest Cu(I) atoms, so that
they cannot propagate along the structure. Hence, the formation of
RUM in $\kappa$-(ET)$_2$Cu$_2$(CN)$_3$ is very unlikely and as a
consequence no signatures of the glass-like transition can be
observed. Cooling the system further, another broad hump is
observed at $T_{\text{max},\chi}$ $\simeq$ 70\,K. The latter coincides
roughly with a broad maximum observed in magnetic susceptibility
measurements \cite{Shimzu2003}. Below $T$ $\simeq$ 50\,K,
$\alpha_{c}$ assumes negative values. Cooling the system further,
a broad minimum is observed at $T_{\text{min}}$ $\simeq$ 30\,K. Below $T$
$\simeq$ 14\,K, $\alpha_{c}$ starts to assume positive values. A
possible scenario for explaining the negative thermal expansion in
the \textcolor{black}{temperature} range 14\,K $\lesssim$ $T$ $\lesssim$ 50\,K is discussed in
the following. In this temperature range, lattice vibration modes
(most likely from the anion) become soft, the Gr\"uneisen
parameter in turn assumes negative values
and the lattice expands
upon cooling.  Interestingly enough, negative thermal
expansion has been also observed in amorphous Y$_{100-x}$Fe$_{x}$
(x = 92.5 and 84) alloys over the temperature range in which
these systems go from the paramagnetic to the spin-glass state.
The latter feature is assigned to the \textcolor{black}{thermal dependence of the} spin
fluctuations \cite{Fujita1994}. Thus, for the spin-liquid candidate the hypothesis of
negative thermal expansion driven by spin degrees of freedom
cannot be ruled out. In fact, the connection between negative
thermal expansion and frustration has been reported in the
literature, but a theory able to describe these phenomena is still
lacking \cite{Schlesinger2008}. Upon further cooling, a
huge anomaly is observed at $T^{\text{anom}}$ $\simeq$ 6\,K. The latter
coincides with the hump anomaly observed in specific heat
measurements \cite{Yamashita2008}. This finding constitutes the
first observation of lattice effects associated with the
transition (or crossover/hidden ordering) at 6\,K. A further
analysis of the present data is difficult because the actual
phonon background cannot be estimated.  The low-temperature data
will be discussed in more detail in the following.
\Fig{TE-SpinLiquidLowT} shows the thermal expansion
coefficient below 14\,K on expanded scales.

Upon cooling below 6\,K, a hump at $T$ $\simeq$ 2.8\,K is
observed. Several runs were performed in order to check for
reproducibility. The latter feature has not been observed in
specific heat measurements \cite{Yamashita2008}, most likely due
to the lack of resolution of such experiments.

\begin{figure}[!htb]
\begin{center}
\includegraphics[angle=0,width=0.49\textwidth]{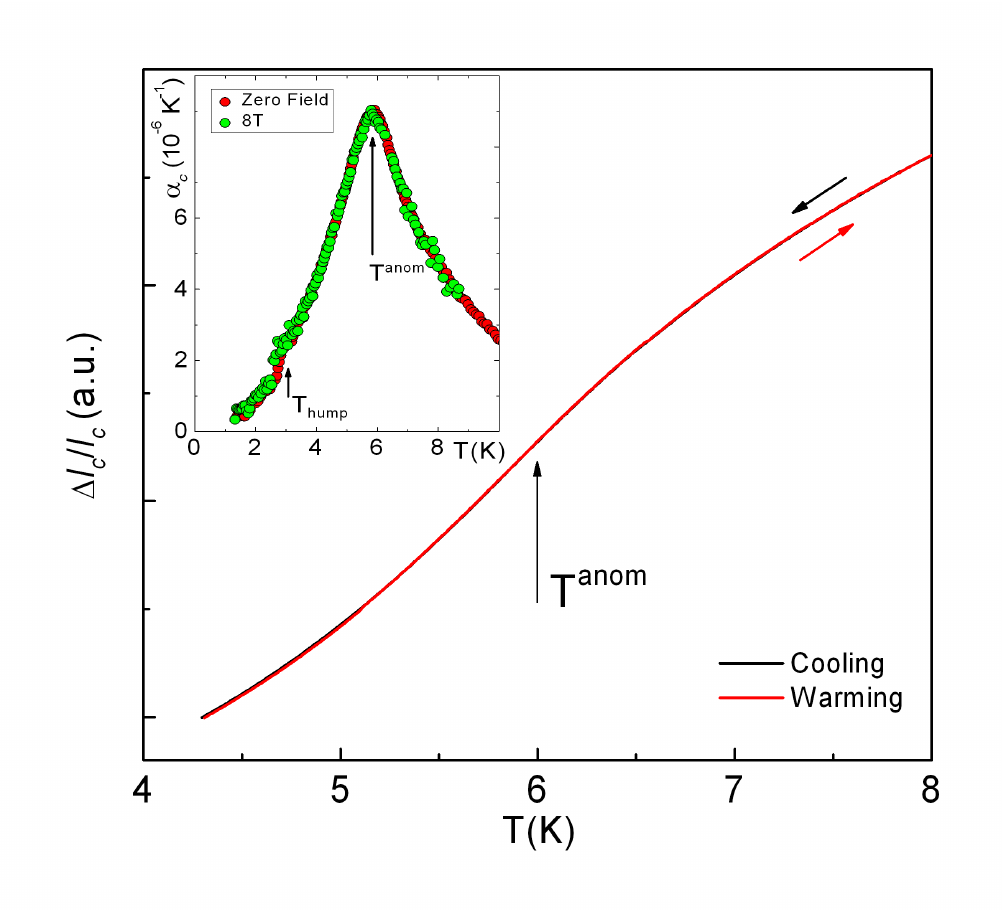}
\end{center}
\caption{\small Main panel: Relative length changes (in arbitrary
units) for $\kappa$-(ET)$_2$Cu$_2$(CN)$_3$ along the in-plane
\emph{c}-axis measured at very \textcolor{black}{low sweep-rate of} $\pm$1.5\,K/h,
showing the absence of hysteretic behavior. Inset: Thermal expansion
data under zero magnetic field and 8\,T. Such measurements were
taken on warming,
with the magnetic field applied at 1.3\,K. Figure taken from \Ref{Thesis}.
}\label{Sliquid3}
\end{figure}

The signatures observed in $\alpha_{c}$ have their direct
correspondence in the spin-lattice relaxation rate ($T_1^{-1}$)
and magnetic susceptibility ($\chi$) \cite{Shimzu2003}. Below
50\,K, both quantities decrease monotonously with temperature down
to 4\,K, where $T_1^{-1}$ starts to increase and shows a broad
maximum at 1\,K, while $\chi$ varies smoothly \cite{Shimzu2003}.
Thermal expansion measurements under magnetic fields (inset of
\Fig{Sliquid3}) revealed that under 8\,T, the peak position
around 6\,K as well as the hump at $T$ $\simeq$ 2.8\,K remain
unaltered, in agreement with specific heat measurements under
a magnetic field \cite{Yamashita2008}, indicating that the
anomaly at 6\,K is unlikely to be due to long-range magnetic
ordering, at least for magnetic fields applied parallel to the
\emph{c}-axis. At this point, it is worthwhile mentioning that for the
``spin ice'' system Dy$_2$Ti$_2$O$_7$ \cite{Ramirez99}, already
mentioned previously, the spins' disorder is suppressed when a
magnetic field is applied perpendicular to their Ising axis.
In addition, measurements on cooling and warming (main panel of
\Fig{Sliquid3}) revealed no traces of hysteretic behavior,
at least in the resolution of the present experiments, so that the
hypothesis of a first-order transition can be ruled out.

Hence, the actual origin of the anomaly in $\alpha_{c}$ at 6\,K
remains unclear. However, at first glance, the present thermal
expansion findings appear to fit in the model proposed by Lee
\emph{et al.} \cite{Lee2007},
which we described briefly at the end of the introduction.
According to this model,
spin pairing on the Fermi surface generates a spontaneous breaking of
the lattice symmetry, giving rise to a phase transition at finite
temperature, which in turn is coupled to a lattice distortion. In
other words, upon cooling, the spin entropy has to be (partially)
frozen. The only way for this, without long-range magnetic
ordering, is to introduce a lattice distortion. This process
presents some similarities with a classical Spin-Peierls
transition, where the formation of a singlet state requires that
the spin entropy of the triplet state goes to zero by introducing
a lattice distortion. In the present case, the amount of entropy
in the temperature range 0.3\,K $<$ $T$ $<$ 1.5\,K  corresponds to only
a few percent of $R$\,ln\,2 \cite{Yamashita2008}, indicating
therefore that only a minor part of the total number of spins
contribute to the spin-liquid phase. Interestingly enough, the
shape of the transition at 6\,K (\Fig{TE-SpinLiquidLowT}) resembles the shape of the anomaly in the thermal expansion
coefficient associated with the Spin-Peierls transition for the
quasi-1D organic conductors (TMTTF)$_2$X (X = PF$_6$ and AsF$_6$) \cite{Mariano2008}.
Hence, the sharp
lattice distortion at $T^{\text{anom}}$ $\simeq$ 6\,K seems to be
associated with a real phase transition and not with a crossover,
as proposed in \Ref{Yamashita2008}.   In fact,
thermal expansion measurements along
the second in-plane \emph{b}-axis revealed striking
anisotropic in-plane lattice effects \cite{SL}.
Interestingly enough, along
the in-plane \emph{b}-axis (see \Fig{TE-SpinLiquidLowT}) a
negative anomaly is observed at $T^{\text{anom}}$ $\simeq$ 6\,K,
indicating that below $T^{\text{anom}}$ a lattice expansion (shrinkage)
along the \emph{c}-axis (\emph{b}-axis) occurs. The observed
distinct in-plane anisotropy implies  that the hopping integral
terms $t$ and $t'$ are strongly affected.

It remains to be seen whether
$\kappa$-(ET)$_2$Cu$_2$(CN)$_3$ is in fact
in a spin-liquid state for $T \to 0$.
In this
sense, thermal expansion measurements
at very low temperatures
along the \emph{a}-, \emph{b}- and \emph{c}-axes are required in
order to achieve a better understanding of the physics of this
exciting material. Such experiments will reveal, for example,
whether below 1.3\,K $\alpha$/$T$ behaves linearly and remains
finite as $T$ $\rightarrow$ 0, as observed in specific heat
measurements \cite{Yamashita2008} or if $\alpha$/$T$ decreases
exponentially.
As already mentioned at the end of the introduction, it was more recently
reported \cite{Yamashita09}
that the thermal conductivity of
$\kappa$-(ET)$_2$Cu$_2$(CN)$_3$ is described by an activated
behavior with a gap $\Delta_\kappa$ $\simeq$ 0.46\,K. The latter
results are at odds with the specific heat measurements performed by
S.\ Yamashita \emph{et al.} \cite{Yamashita2008}.
Based on such results, one can see that a full picture of the fundamental aspects of the spin-liquid phase in $\kappa$-(ET)$_2$Cu$_2$(CN)$_3$ is still incomplete and requires further investigations.

\section{Conclusions and Perspectives}\label{Conclusions and Perspectives}

In this review, we have addressed one of the central questions in the field of strongly correlated electron systems,
whether lattice degrees of freedom play
a role for several physical phenomena, among them the Mott MI
transition. \textcolor{black}{Molecular conductors} of
the $\kappa$-(BEDT-TTF)$_2$X family have
been revealed as model systems for the study of the latter
phenomena. The main findings discussed in this review are summarized
below.

\vspace*{0.2cm}

\textcolor{black}{High-resolution thermal expansion measurements on fully deuterated salts
of $\kappa$-(BEDT-TTF)${_2}$Cu[N(CN)$_{2}$]Br revealed
pronounced lattice effects} at the Mott
metal-to-insulator transition temperature $T_{MI}$ = 13.6\,K,
accompanied by a striking anisotropy. While huge effects are
observed along the in-plane \emph{a}-axis, along which the polymeric
anion chains are aligned, an almost zero effect is observed along
the second in-plane \emph{c}-axis, along which the polymeric anion
chains are linked via Br-N weak contacts. Still more amazing is the
observation of pronounced lattice effects along the out-of-plane
\emph{b}-axis. These findings provide strong evidence that the Mott
transition for the present class of materials cannot be
fully understood
in
the frame of a purely 2D electronic triangular dimer model.
Even though there is still an ongoing debate about the universality class and possible crossover scales of the Mott transition, the scaling theory reviewed here can be used with any universality class.
Candidates suggested for the Mott transition in the $\kappa$-ET salts are unconventional criticality \cite{Kagawa2005_20,Imada05,Imada10}, 2D Ising \cite{Papanikolaou2008_20}, and a crossover to Landau criticality \cite{Zacharias}.
It was explicitly shown that the thermal expansivity data for the $\kappa$-D8-Br salt is consistent with 2D Ising criticality \cite{Bartosch}, which of course does not provide a proof of this universality class.
Finally, the experimentally observed critical exponent $\delta = 2$ was also argued to be due to subleading quantum effects preceding an asymptotic critical regime \cite{Semon12}.

In order to achieve a better understanding of the Mott
transition in the present material, thermal expansion measurements
were taken under \textcolor{black}{magnetic field, quasi-uniaxial pressure as well as
by employing different cooling speeds across the glass-like
transition. For example, measurements under magnetic field reveal
that the Mott MI insulator transition temperature is insensitive to magnetic fields
up to 10\,T, which is in agreement with the picture of a Mott
insulator with a hole localized on a dimer formed by an ET molecule. A magnetic field-induced
first-order phase transition at $T_{SF}$ $\simeq$ 9.5\,K for a
field applied along the out-of-plane \emph{b}-axis was observed, indicative of a
spin-flop transition with strong magneto-elastic coupling.
Measurements under quasi-uniaxial pressure applied along the
out-of-plane \emph{b}-axes revealed effects in contrast to those
observed upon increasing the cooling speed across the glass-like
transition. While quasi-uniaxial pressure shifts $T_{MI}$ towards
lower temperatures and $T_p$ to higher temperatures, the opposite
situation is observed upon fast cooling through the glass-like
transition. Interestingly enough, the so-called glass-like transition
temperature
$T_g$ remains unaffected under quasi-uniaxial
pressure.} Although by means of thermal expansion measurements one
has access to the macroscopic behavior of the sample studied, the
strong anisotropic lattice effects observed,  marked by a negative
thermal expansion above $T_g$, enable us to draw some conclusions at
the microscopic level. Hence, a model within the rigid-unit-mode
scenario, taking into account the low-energy vibration modes of the
cyanide ligands of the polymeric anions Cu[N(CN)$_{2}$]Br$^-$, was
proposed to describe the negative thermal expansion above the $T_g$
$\simeq$ 77\,K. In this model, low-energy vibration modes of the
cyanide groups of the polymeric anion chains become frozen upon
cooling down to $T_g$, shrinking thus the \emph{a} lattice parameter
upon warming the system above $T_g$.

\textcolor{black}{Additional experiments, including magnetostriction measurements both
above and below the magnetic field-induced phase transition temperature $T_{FI}$ as well as magnetic measurements with $B$
thoroughly aligned along the out-of-plane $b$-axis, will certainly help to achieve a better
understanding of the behavior of almost localized correlated
electrons in this interesting region of the phase diagram of the
$\kappa$-phase of (BEDT-TTF)$_2$X charge-transfer salts.  A theoretical
study of the stability of the Mott insulating state and the neighboring
superconducting phase, taking into account magnetic field-induced lattice
effects, is highly desired as well.}
We have here presented a phenomenological description of the Mott metal-insulator
transition within a scaling theory. While transport measurements revealed unconventional
critical exponents, it would be desirable to have direct access to thermodynamic quantities.
In particular, we hope that future expansivity measurements under pressure might give rise
to a better understanding.

We have also examined in this review
thermal expansion measurements on the
$\kappa$-(BEDT-TTF)${_2}$Cu$_2$(CN)$_{3}$ salt, a system which is
\textcolor{black}{proposed} as a candidate for the realization of a
spin-liquid ground state.
Measurements taken along the
in-plane \emph{c}-axis reveal some interesting
 aspects, which can be of
extreme importance for a wider understanding of the physical
properties of this material. For example, the absence of the
so-called glass-like transition in
$\kappa$-(BEDT-TTF)${_2}$Cu$_2$(CN)$_{3}$ is in line with the
above-mentioned model within the rigid-unit modes scenario. While
the anion Cu[N(CN)$_{2}$]Br$^-$ is arranged in form of chains weakly
linked  via Br-N contacts, the anions Cu$_2$(CN)$_3^-$ are arranged
in a 2D  network fashion, so that the cyanide groups are confined
between the Cu atoms and therefore the vibration modes associated
with the cyanide groups cannot propagate along the structure as it
occurs, for example, in fully deuterated salts of
$\kappa$-(BEDT-TTF)${_2}$Cu[N(CN)$_{2}$]Br, cf.\
the discussion
above. A
negative thermal expansion is observed \textcolor{black}{in the temperature range} 14 $<$ $T$ $<$ 50\,K. In
addition, the thermal expansion coefficient reveals  a huge anomaly,
indicative of a phase transition at $T$ = 6\,K, coinciding nicely
with a hump in the specific heat, which has been assigned to a
crossover to the quantum spin-liquid state. Last but not least, during the preparation of the final version of this article, a theoretical treatment for the thermal expansivity of the charge-transfer salts of the $\kappa$-(BEDT-TTF)${_2}$X family based on electronic correlations was proposed by Kokalj and McKenzie \cite{Mac}. These authors consider both the dependence of the system entropy on the Hubbard parameters $U$, $t$, $t'$  and the dependence with temperature of the double occupancy and bond-orders to describe the thermal expansion of these materials.

\vspace*{0.2cm}

\section*{Acknowledgements}

We would like to thank A.~Br\"uhl, M.~Dressel, R.M.~Fernandes, M.~Garst, A.A.~Haghighirad, K.~Kanoda, S.~K\"ohler, P.~Kopietz, R.S.~Manna, R.E.L.~Monaco, J.~M\"uller, R.~Paupitz, J.-P.~Pouget, B.~Powell, T.~Sasaki, J.A.~Schlueter, J.~Schmalian, D.~Schweitzer, A.C.~Seridonio, Ch.~Strack, R.~Urbano, B.~Wolf, M.~Zacharias
and in particular M.~Lang
for valuable discussions and collaborations along the years.
MdS acknowledges financial support from the DFG via SFB/TRR 49, S\~ao Paulo
Research Foundation -- Fapesp (Grants No. 2011/22050-4) and National Counsel of Technological and Scientific Development -- CNPq (Grants No. 308977/2011-4).
\vspace{10mm}\\


\begin{thebibliography}{99}
\bibitem{pressao} For the Mott insulator V$_2$O$_3$, for example, a pressure of a
few kbar is required to cross the first-order line in the $P$ - $T$
phase diagram \cite{Limelette03-9d3}.
\bibitem{footnote1} This cooling speed ($\sim$\,$-$3\,K/h) was employed
in the temperature range $\sim$85 - 65\,K. From room temperature
down to $\sim$85\,K as well as from 65\,K down to 4.2\,K,  an
average cooling speed of $-$40\,K/h was employed.
\bibitem{Strack} Resistance measurements shown in
\Fig{LowTdataRalpha} were carried out by C.~Strack.
\bibitem{C} In employing $\alpha$ = $-$80 $\cdot$ 10$^{-6}$\,K$^{-1}$,
obtained from the height of the anomaly in the thermal expansion
coefficient for crystal\, \#1 (\Fig{K-d8-1.pdf}), using the
entropy change $\Delta S = -0.074$\,Jmol$^{-1}$K$^{-1}$  at the Mott
MI transition, whose detailed deduction is presented  in
Section\,\ref{Entropy changes at the Mott transition}, results in a
jump of the specific heat of about 0.2\,Jmol$^{-1}$K$^{-1}$. Details
of this estimate are provided in Appendix 2. This value corresponds
roughly to 0.4\% of the background at $T_{MI}$ in the specific heat
(specific heat measurements on $\kappa$-D8-Br crystal\,\#1 were performed
by S. K\"ohler)
employing ac calorimetric method, so that no signatures in the
specific heat at $T_{MI}$ could be detected.
\bibitem{TEref} In \Ref{Griesshaber20e} Grie{\ss}haber refers to a cooling rate
of $\sim-$1\,K/min as a slow cooling, while the rapid cooling rate
is not mentioned.
\end{thebibliography}
\end{document}